\documentclass[a4paper,fleqn,usenatbib]{mnras}

\usepackage{newtxtext,newtxmath}

\usepackage[T1]{fontenc}
\usepackage{ae,aecompl}
\usepackage{color, soul}

\usepackage{etoolbox}
\makeatletter
\patchcmd\@combinedblfloats{\box\@outputbox}{\unvbox\@outputbox}{}{%
  \errmessage{\noexpand\@combinedblfloats could not be patched}%
}%
\makeatother

\usepackage{graphicx}	
\usepackage{subfig}
\usepackage{amsmath}	
\usepackage{amssymb}	

\newcommand{\ds}{$\,{\rm deg}^2$} 
\newcommand{\chone}{$[3.6\mu{\rm m}]$}
\newcommand{\chtwo}{$[4.5\mu{\rm m}]$}
\newcommand{\Zsun}{${\rm Z}_{\odot}$}

\title[The bright-end of the LF at z = $\mathit{8}$--$\mathit{10}$]{A lack of evolution in the very bright-end of the galaxy luminosity function from $\mathbf {z \simeq 8}$--$\mathbf{10}$}

\author[R. A. A. Bowler et al.]{R. A. A. Bowler,$^{1}$\thanks{E-mail: rebecca.bowler@physics.ox.ac.uk}
M. J. Jarvis,$^{1, 2}$
J. S. Dunlop,$^{3}$
R. J. McLure,$^{3}$
D. J. McLeod,$^{3}$
\newauthor N. J. Adams$^{1}$, B. Milvang-Jensen,$^{4, 5}$ H. J. McCracken$^{6}$
\\
$^{1}$Astrophysics, The Denys Wilkinson Building, University of Oxford, Keble Road, Oxford, OX1 3RH \\
$^{2}$Department of Physics, University of the Western Cape, Bellville 7535, South Africa\\
$^{3}$Institute for Astronomy, University of Edinburgh, Royal Observatory, Edinburgh, EH9 3HJ\\
$^{4}$Cosmic Dawn Center (DAWN)\\
$^{5}$Niels Bohr Institute, University of Copenhagen, Lyngbyvej 2, 2100 Copenhagen, Denmark\\
$^{6}$Sorbonne Universit\'{e}, CNRS, UMR 7095, Institut d'Astrophysique de Paris, 98 bis Boulevard Arago, F-75014 Paris, France
}

\pubyear{2020}

\begin{document}

\label{firstpage}
\pagerange{\pageref{firstpage}--\pageref{lastpage}}
\maketitle

\begin{abstract}
We utilize deep near-infrared survey data from the UltraVISTA fourth data release (DR4) and the VIDEO survey, in combination with overlapping optical and~\emph{Spitzer} data, to search for bright star-forming galaxies at $z \gtrsim 7.5$.
Using a full photometric redshift fitting analysis applied to the $\sim 6$\ds~of imaging searched, we find 27 Lyman-break galaxies (LBGs), including 20 new sources, with best-fitting photometric redshifts in the range $7.4 < z < 9.1$.
From this sample we derive the rest-frame UV luminosity function (LF) at $z = 8$ and $z = 9$ out to extremely bright UV magnitudes ($M_{\rm UV} \simeq -23$) for the first time.
We find an excess in the number density of bright galaxies in comparison to the typically assumed Schechter functional form derived from fainter samples.
Combined with previous studies at lower redshift, our results show that there is little evolution in the number density of very bright ($M_{\rm UV} \sim -23$) LBGs between $z \simeq 5$ and $z\simeq 9$.
The tentative detection of an LBG with best-fit photometric redshift of $z = 10.9 \pm 1.0$ in our data is consistent with the derived evolution.
We show that a double power-law fit with a brightening characteristic magnitude ($\Delta M^*/\Delta z \simeq -0.5$) and a steadily steepening bright-end slope ($\Delta \beta/\Delta z \simeq -0.5$) provides a good description of the $z > 5$ data over a wide range in absolute UV magnitude ($-23 < M_{\rm UV} < -17$).
We postulate that the observed evolution can be explained by a lack of mass quenching at very high redshifts in combination with increasing dust obscuration within the first $\sim 1 \,{\rm Gyr}$ of galaxy evolution.

\end{abstract}

\begin{keywords}
galaxies: evolution -- galaxies: formation -- galaxies: high-redshift
\end{keywords}



\section{Introduction}
The study of galaxies at ultra-high redshifts has the potential to answer fundamental questions in the field of galaxy formation and evolution.
As probes of the Universe at less than a billion years after the Big Bang, galaxies at redshifts $z > 7$ (as well as other probes such as quasars and gamma-ray bursts; e.g.~\citealp{Banados2018, Tanvir2018}) give an insight into the formation of the first stars, dust and super-massive black holes, as well as the process of reionization.
Over the past decade the study of galaxies at these extreme redshifts has become possible with the advent of deep imaging in the near-infrared from the~\emph{Hubble Space Telescope} Wide Field Camera 3 (\emph{HST}/WFC3) amongst other facilities.
By combining optical and near-infrared imaging it is possible to select star-forming galaxies at $z \gtrsim 7$ by identifying the strong Lyman-break in the spectral-energy distribution (SED) as it is redshifted beyond $\lambda_{\rm obs} = 1\mu{\rm m}$.
Galaxies discovered by applying this `Lyman-break technique' to deep~\emph{HST} survey data now number many thousands at $z > 4$ (e.g.~\citealp{Finkelstein2015, McLure2013}) with tens at $z > 8.5$ (e.g.~\citealp{Kawamata2018, Salmon2018, McLeod2016}).
These samples have allowed increasingly precise measurements of the rest-frame UV luminosity function (LF) back to $\sim 500$ Myrs after the Big Bang (e.g.~\citealp{Bouwens2019, Oesch2016}).
Using either an expanded colour-colour cut methodology (e.g.~\citealp{Ono2018, Bouwens2015}) or a photometric redshift fitting analysis (e.g.~\citealp{Finkelstein2015, Bowler2015}), the results of the past decade have revealed a rapid evolution in the LF of these galaxies at $z \simeq 4$--$10$.
While the majority of previous studies agree within the errors at the binned level, the exact form of the evolution, for example whether the normalisation ($\phi^*$) or characteristic luminosity ($L^*$) is the key driver, remains debated (e.g.~\citealp{Bowler2015, Bouwens2015}).
A change in shape of the LF at these redshifts is postulated to signal a change in fundamental galaxy properties due to the young age of the Universe~\citep{Silk2012}.
For example, a reduced AGN feedback efficiency, lack of dust content or inefficient star-formation (e.g.~\citealp{Bower2012, Paardekooper2013, Dayal2014, Clay2015}) can lead to a change in the relative number of bright and faint galaxies compared to lower redshifts.

A key component in determining the form of any evolution is having sufficiently large samples of bright galaxies to accurately constrain the position of the knee in the LF.
With~\emph{HST} alone this has been a challenge, even when a wide-area survey strategy is implemented, due to the small field-of-view of WFC3 (e.g. the Cosmic Assembly Near-infrared Deep Extragalactic Legacy Survey, CANDELS;~\citealp{Grogin2011, Koekemoer2011} covers only $\sim 0.2\,$\ds).
A powerful alternative approach has been to utilize the deep ground-based near-infrared data from the UK Infrared Telescope (UKIRT) and the Visible and Infrared Survey Telescope for Astronomy (VISTA).
The $YJHK$ imaging provided by these facilities has resulted in the first statistical samples of LBGs bright-ward of $M_{\rm} \lesssim -21.5$ at $z \simeq 7$~\citep{Bowler2014, Bowler2016} and more recently the detection of similarly bright $z\simeq 8$--$9$ galaxies~\citep{Stefanon2017, Stefanon2019}.
Selected over several square degrees, these samples probe the very bright-end of the LF and hence dramatically increase the dynamic range over which it can be constrained.
Prior to these data, the high-redshift LF was typically fitted with a Schechter function ($\phi\, {\rm d}L = \phi^*\,(L/L^{*})^{\alpha}\,e^{-L/L^*}\,{\rm d}L$), which tends to a power-law with slope $\alpha$ at faint luminosities, and has an exponential decline in the number of galaxies bright-ward of the characteristic luminosity at $L > L^*$.
Using a sample of extremely luminous galaxies at $z \simeq 7$,~\citet{Bowler2014} found evidence for an excess of galaxies compared to that expected from the previous best-fit Schechter function determined from fainter samples.
The $z \simeq 7$ LF was found to be better described by a double power-law (DPL), potentially indicating a lack of quenching or dust obscuration at these redshifts~\citep{Bowler2015}.
This deviation from a Schechter form has also been found to continue to brighter magnitudes as demonstrated by samples derived from deep Hyper-SuprimeCam (HSC) $y$-band photometry~\citep{Ono2018}. 
The DPL form ($\phi\, {\rm d}L = \phi^*\,/[(L/L^{*})^{-\alpha} + (L/L^{*})^{-\beta}]  \,{\rm d}L$) removes the requirement for an exponential decline in the number of bright galaxies, and instead the slope of the bright-end is governed by the power-law index $\beta$.
If there is a change in shape of the LF at high-redshift as a consequence of fundamental changes in astrophysical effects, it holds that a shallower decline in the number of very bright galaxies should also be seen at $z > 7$.
To-date however, it has not been possible to determine fully the shape of the function at $z \gtrsim 8$ due to a lack of dynamic range at the bright-end (\citealp{Stefanon2019}; hereafter S19).

\begin{figure*}

\includegraphics[width = \textwidth]{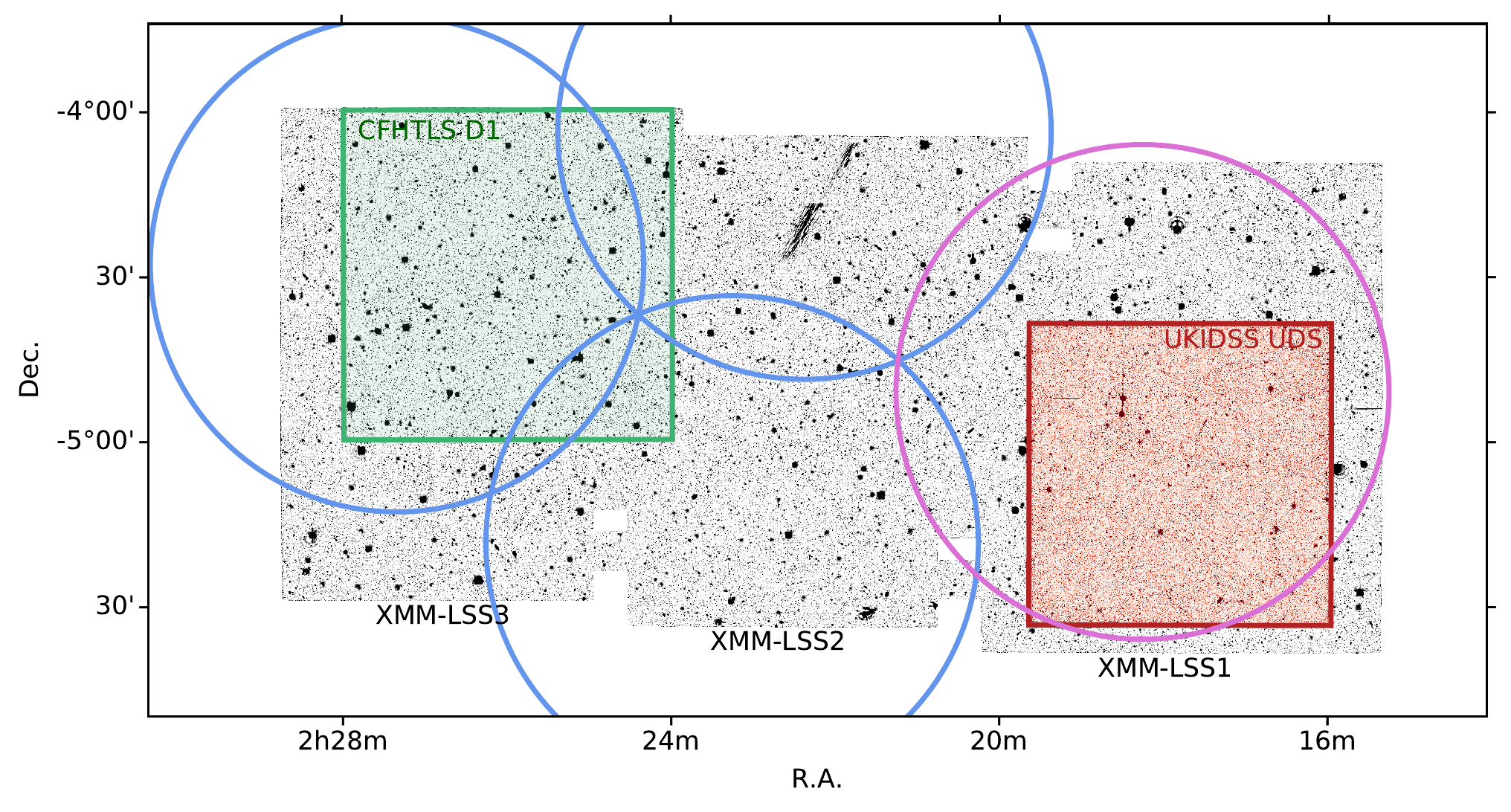}
\caption{The footprint of the optical/near-infrared datasets in the XMM-LSS field utilized in this study.
The VIDEO near-infrared data are shown as the background image, which covers an area approximately $3 \times$ that of the COSMOS field shown in Fig.~\ref{fig:cosmos}.
Near-infrared data that extends deeper than the VISTA VIDEO data exist in the UKIDSS UDS field, which sits within the XMM-LSS tile 1 on the right of the figure.
Optical coverage of the full field is provided by the HSC SSP using four overlapping circular pointings.
The three blue circles covering the middle and left part of the figure are part of the `deep' tier of the HSC SSP, whereas the right-most pointing illustrated by the purple circle is part of the `ultra-deep' tier.
In the XMM-LSS tile 3 there are additional optical data from the CFHTLS D1 field.
The maximal survey area searched corresponds to the overlap between the HSC data and the VIDEO footprint.
}\label{fig:xmm}

\end{figure*}

What shape do we expect the rest-frame UV LF to take at high-redshift?
The majority of cosmological hydrodynamic (e.g.~\citealp{Henriques2014, Genel2014}) and analytic/semi-analytic models (e.g.~\citealp{Cai2014, GonzalezPerez2013}) indeed show a power-law like decline at the bright-end, when galaxies are considered to be dust-free.
In addition to showing a different shape, these models tend to over-predict the number density of the brightest galaxies by an order of magnitude in some cases (see comparison in~\citealp{Bowler2015}).
It is only after the inclusion of significant dust obscuration that these predictions are brought into agreement with the normalisation and the steeper decline in the bright-end of the observations.
There is therefore considerable uncertainty in the expected rest-frame UV LF from simulations, as it depends sensitively on how dust attenuation is implemented.
The underlying power-law like LF predicted by the dust-free simulations more closely follows the underlying dark matter halo mass function.
As observations push to higher and higher redshifts, where galaxies are becoming progressively less dusty (e.g. as suggested by the rest-frame UV colours, e.g.~\citealp{Dunlop2013}), it is to be expected that a power-law will better describe the observed rest-frame UV LF at the high-luminosity end.

Prior to the launch of~\emph{Euclid} and~\emph{WFIRST}, the premier near-infrared data on the degree-scale comes from surveys conducted with VISTA and UKIRT.
While previously thought to be too shallow to find extremely high-redshift galaxies, the detection of very bright $z \simeq 7$ galaxies in UltraVISTA, in addition to the discovery of very bright LBGs in the small area~\emph{HST}/WFC3 surveys (e.g.~\citealp{Morishita2018,RobertsBorsani2015}), has highlighted the potential of this ground-based data in finding $z > 7$ galaxies.
In this work we present the results of the widest area search to-date for bright LBGs at $z =8$--$10$.
We use a total of $5.8\,$\ds~of ground-based near-infrared survey data from the UltraVISTA~\citep{McCracken2012}, the UKIRT Infrared Deep Sky Survey Ultra Deep Survey (UKIDSS UDS;~\citealp{Lawrence2007}) and the VISTA Deep Extragalactic Observations (VIDEO;~\citealp{Jarvis2013}) surveys, in addition to deep optical and mid-infrared data from HSC and~\emph{Spitzer}, to search for high-redshift LBGs.
The structure of this paper is as follows.
In Section~\ref{sect:data} we describe the ground-based datasets used in this analysis.
In Section~\ref{sect:selection} and Section~\ref{sect:sample} we present the selection procedure and the resulting $z > 7.5$ galaxy candidates.
The new samples allow us to compute the bright-end of the rest-frame UV LF at $z \simeq 8$--$10$ which we present in Section~\ref{sect:lf}.
We end with a discussion of these results in Section~\ref{sect:discussion} and our conclusions in Section~\ref{sect:conc}.
Throughout this work we present magnitudes in the AB system~\citep{Oke1974,Oke1983}.
The standard concordance cosmology is assumed, with $H_{0} = 70 \, {\rm km}\,{\rm s}^{-1}\,{\rm Mpc}^{-1}$, $\Omega_{\rm m} = 0.3$ and $\Omega_{\Lambda} = 0.7$.

\section{Data}\label{sect:data}

The two survey fields considered in this study are the COSMOS and the~\emph{XMM-Newton} - Large Scale Structure (XMM-LSS) fields.
These fields were chosen as they contain the deepest near-infrared ($YJHK_{s}$) photometric data on the degree-scale, in addition to other multi-wavelength data from the X-ray to the Radio.
The near-infrared data is essential to detect $z > 7 $ LBGs as their rest-frame UV emission is redshifted beyond the red-optical bands.
In addition, we require deep optical data to confirm the photometric redshift and to remove red contaminants such as cool brown dwarfs.
Photometric data in the mid-infrared from~\emph{Spitzer}/Infrared Array Camera (IRAC) are also key in the removal of low-redshift dusty galaxy contaminants, which typically have much redder near-to-mid IR colours than $z > 7$ galaxies.
The overlap of these various datasets leads to a variety of regions analysed in this work.
In Table~\ref{table:areas} we summarise these regions and their area.

\begin{table}\caption{The primary optical and near-infrared datasets utilized for each sub-field.
In the XMM-LSS field there is VISTA VIDEO data over the full region, with additional deeper near-infrared data from the UKIDSS UDS survey in Tile 1.
Deeper optical data from the CFHTLS D1 field is also available in Tile 3.
The COSMOS field consists of two depths of $YJHK_{s}$ data from UltraVISTA, `deep' and `ultra-deep'.
The total area of the survey data utilized, which corresponds to the overlap between the HSC and the VISTA data accounting for masked regions is 5.8 \ds.
}\label{table:areas}
\begin{tabular}{l c c l}
\hline
Field & Region & Area & Primary Datasets \\
& & /\ds &(near-infrared, optical) \\
\hline
XMM-LSS & UDS & 0.79 & UKIDSS UDS, HSC-UD \\
XMM-LSS & wide & 1.01 & VIDEO Tiles 1+2, HSC-UD \\
XMM-LSS & wide & 1.49 & VIDEO Tiles 2+3, HSC-D \\
XMM-LSS & wide/D1 & 0.97 & VIDEO Tile 3, CFHTLS \\
COSMOS & ultra-deep & 0.86 & UltraVISTA, HSC-UD \\
COSMOS & deep & 0.65 & UltraVISTA, HSC-UD \\
\hline
Total & & 5.77 &\\ 
\hline
\end{tabular}
\end{table}

\subsection{The XMM-LSS field}

The full XMM-LSS field has been imaged in the $YJHK_{s}$ bands as part of the VIDEO survey~\citep{Jarvis2013}.
As shown in Fig.~\ref{fig:xmm}, the full near-infrared mosaic is comprised of three completed tiles of the Visible and Infrared Camera (VIRCAM), and hence provides three times the area of the UltraVISTA data, albeit to shallower depths.
The XMM-LSS field has optical data from both the `deep' and `ultra-deep' tier of the HSC Subaru Strategic Program (SSP;~\citealp{Aihara2017}) in the $GRIZy$ filters.
In total the field is covered by four pointings of HSC, and we use the DR1 release of the SSP.
A sub-set of the XMM-LSS field has been imaged as part of the UKIDSS UDS~\citep{Lawrence2007}.
The UDS data, which sits within Tile 1 of the XMM-LSS VISTA imaging, reaches approximately $1\,$mag deeper than VIDEO in the $JHK$ bands.
The `ultra-deep' HSC SSP imaging pointing coincides with the UDS deep near-infrared data.
There exists deeper $z'$-band imaging from the Subaru/SuprimeCam (SC) in the UDS that we utilize in addition to the new HSC $Z$-band imaging~\citep{Furusawa2016}.
In addition, in `Tile 3' of the XMM-LSS field there exists deeper optical data in the $u^*griz$ bands from the Canada-France-Hawaii Telescope Legacy Survey (CFHTLS) D1 field.
Over the full XMM-LSS field there is~\emph{Spitzer}/IRAC imaging from the~\emph{Spitzer} Extragalactic Representative Volume Survey (SERVS;~\citealp{Mauduit2012}).
We also include deep IRAC imaging from the~\emph{Spitzer} Large-Area Survey with HSC (SPLASH;~\citealp{Steinhardt2014}).
 and the~\emph{Spitzer} Extended Deep Survey (SEDS;~\citealp{Ashby2013}) programs in XMM-LSS tile 1.

\subsection{The COSMOS field}

\begin{figure}
\includegraphics[width = 0.48\textwidth]{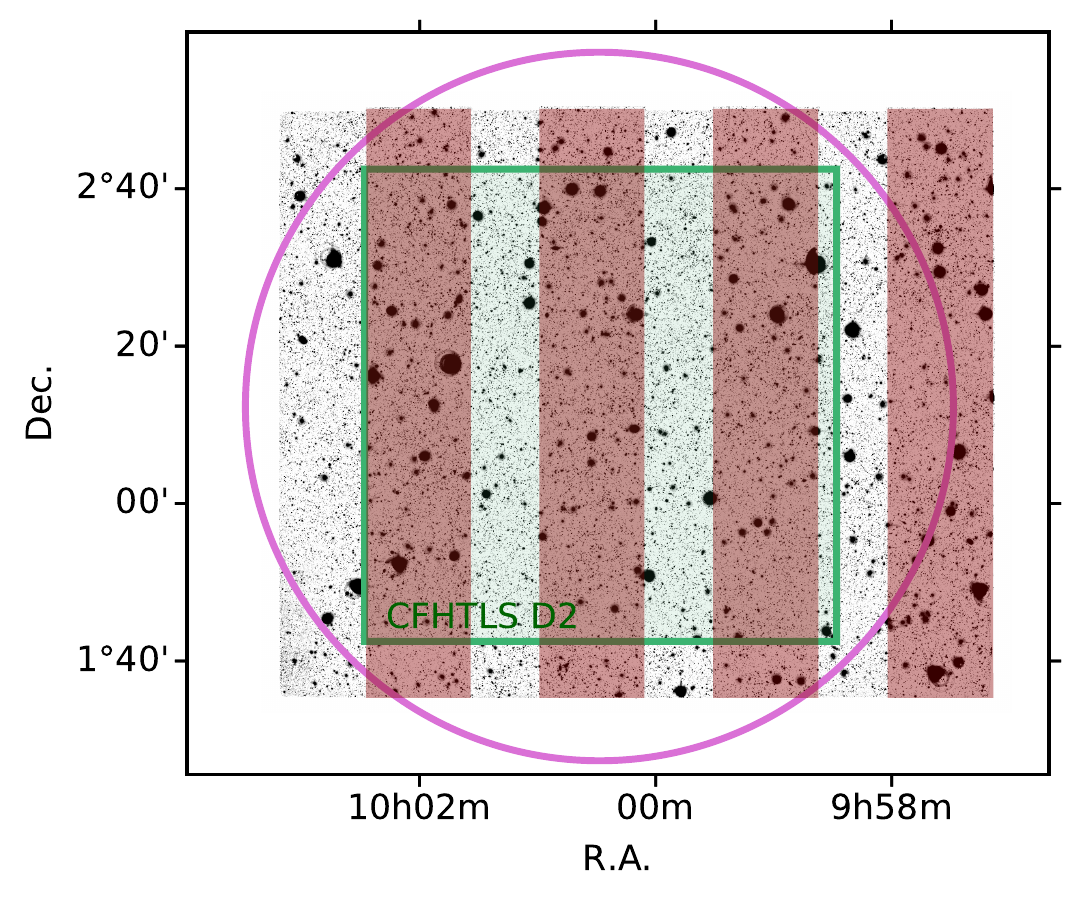}
\caption{The footprint of the optical/near-infrared datasets in the COSMOS field utilized in this study.
The background image shows the UltraVISTA near-infrared data, where the dark circles are haloes of bright stars.
The four `ultra-deep' stripes in the UltraVISTA data are shown as the shaded red regions.
Optical data from the CFHTLS is shown as the green square, and the purple circle denotes the extent of the HSC data (from the `ultra-deep' tier of the HSC surveys).
The full region of the COSMOS field that we search for $z > 7$ LBGs corresponds to the overlap between the UltraVISTA and HSC footprint.
}\label{fig:cosmos}
\end{figure}

Deep near-infrared imaging of the COSMOS field has been acquired as part of the ongoing UltraVISTA survey~\citep{McCracken2012}.
UltraVISTA consists of deep $YJHK_{s}$ imaging taken with the VISTA/VIRCAM over $1.5\,$\ds~over the COSMOS field~\citep{Scoville2007}.
The UltraVISTA survey has two tiers.
The `ultra-deep' tier consists of four deeper stripes covering approximately half the full area.
The remaining regions in the `deep' tier are approximately $1\,{\rm mag}$ shallower.
We use the fourth data release (DR4) of UltraVISTA in this work.
The footprints of the survey and the auxiliary optical data are shown in Fig.~\ref{fig:cosmos}.
The edges of the `ultra-deep' region were defined using the local depth map.
Optical data in the $GRIZy$ filters covering the majority of the field were provided by the `ultra-deep' tier of the HSC SSP DR1, where we also utilize the deeper data provided in the incremental data release~\citep{Tanaka2017}.
In addition deeper optical imaging in the $u^*griz$ filters from the CFHT D2 field was used in the central $1\rm$\ds.
In this central region we also used deep $z'$-band imaging from Subaru/Suprime-Cam~\citep{Furusawa2016}.
This dataset reaches $\sim 0.8\,{\rm mag}$ deeper than the HSC DR1 $Z$-band imaging.
We use~\emph{Spitzer}/IRAC data from several different programmes.
The shallowest imaging comes from SPLASH, which we supplement with deeper data from the~\emph{Spitzer} Matching Survey of the UltraVISTA ultra-deep Stripes survey (SMUVS;~\citealp{Ashby2018}) and SEDS.

\subsection{Image processing and catalogue creation}\label{sect:cats}
The imaging data presented above was matched to the astrometry and pixel scale of the VISTA near-infrared imaging.
This common pixel scale was $0.2\,$arcsec/pix in XMM-LSS and $0.15\,$arcsec/pix in COSMOS.
The astrometry of the UltraVISTA imaging in COSMOS is registered to the Gaia reference frame\footnote{\url{http://www.eso.org/rm/api/v1/public/releaseDescriptions/132}}.
The astrometry of the VIDEO imaging in the XMM-LSS field is registered to the Two Micron All Sky Survey (see~\citealp{Jarvis2013}).
Astrometric solutions and re-sampling of the data were performed using the {\sc SCAMP} and {\sc SWARP} packages respectively~\citep{Bertin2006, Bertin2002}.
We created inverse variance weighted stacks of the near-infrared data to increase our sensitivity to high-redshift sources.
Catalogues were produced using {\sc SExtractor}~\citep{Bertin1996} in `dual-image' mode with the $J+H$ and $H+K_s$ stacked data as the detection images.
We also created catalogues using the $J$, $H$ and $K_s$ bands as the detection images, however in practice the resulting objects were all detected in the stacked images.
The aperture photometry was measured in a 1.8 arcsec diameter circular aperture, and corrected to a total flux assuming a point-source correction derived from {PSFEx}~\citep{Bertin2013} in each band.
This correction ranged from $0.2$--$0.4\, {\rm mag}$ in the optical bands, to $0.4$--$0.6 \, {\rm mag}$ in the near-infrared and~\emph{Spitzer}/IRAC bands.
Photometry in the~\emph{Spitzer}/IRAC \chone~and \chtwo~bands was obtained via a deconfusion analysis using the {\sc T-PHOT} software~\citep{Merlin2015}.
The high-resolution image was taken as the VISTA $K_{s}$-band (or $K$-band in the UDS) and point-spread functions for this band and the IRAC dataset were derived using {PSFEx}.
The convolution kernel for T-PHOT was obtained using a Richard-Lucy deconvolution algorithm.
We used {\sc T-PHOT} to fit and subtract the neighbouring galaxies around each high-redshift candidate.
Aperture photometry was then obtained using a $2.8$ arcsec diameter aperture on this cleaned image, which was then corrected to total flux using a point-source correction.

\subsection{Image depths}

The $5\sigma$ limiting magnitudes for the imaging data utilized in this study are presented in Table~\ref{table:depths}.
Depths were calculated using empty aperture measurements on background subtracted images.
We use $1.8$ arcsec diameter circular apertures in this work as a compromise between optimising the signal-to-noise and the robustness of aperture measurements given the pixel size and seeing of the images (typically a full-width at half-maximum, FWHM, of $\sim 0.8\,{\rm arcsec}$).
Foreground objects were avoided using the {\sc segmentation} map produced by {\sc SExtractor}.
Local depths across the images were calculated using the `median absolute deviation' estimator from the closest 200 apertures to each point (where $\sigma = 1.48\times MAD$).
The global depths for each image or image region (e.g. ultra-deep/deep in UltraVISTA) were then derived by taking the median of the calculated local depths.
The UltraVISTA `ultra-deep' stripes and the UDS sub-field provide the deepest tiers of our search, while the VIDEO imaging provides a significantly wider area but at a shallower depth (Table~\ref{table:areas}).
The combination of this range of data in our search enables us to probe a greater dynamic range in apparent and hence absolute magnitude than previous studies.

\begin{table}\caption{The $5\sigma$ limiting magnitudes for the imaging data used in this study.
Depths were calculated using randomly placed empty circular apertures over the data, using an aperture diameter of $1.8$ arcsec.
The depths in the COSMOS field are shown on the left, where the range in depths for the $YJHK_{s}$ bands corresponds to the different `stripes' visible in Fig.~\ref{fig:cosmos}.
The XMM-LSS values are shown on the right, split by tile as shown in Fig.~\ref{fig:xmm}.
The~\emph{Spitzer}/IRAC depths were calculated in a $2.8$ arcsec diameter aperture to account for the poorer resolution of these data.
}\label{table:depths}
\begin{tabular}{l c c c c l}

\hline

& \multicolumn{2}{l}{COSMOS} & \multicolumn{2}{l}{XMM-LSS} & \\
Filter & ultra-deep & deep & UDS & VIDEO & Source \\
\hline

$u^*$ & $ 27.3$ &-- & --  &$ 27.3$& CFHT\\
$g$ & $ 27.5$ &--  & --  &$ 27.6$& CFHT\\
$r$ & $ 27.1$ & --  & --  &$ 27.1$& CFHT\\
$i$ & $ 26.8$ &--   & --  &$ 26.7$& CFHT\\
$z$ & $ 25.7$ &--  & --  &$ 25.6$& CFHT\\
$G$ & $ 27.3$ &-- &$ 27.2$  &$ 26.7$& HSC\\
$R$ & $ 26.9$ &-- &$ 26.7$ &$ 26.3$& HSC\\
$I$ & $ 26.8$ &-- &$ 26.5$  &$ 25.7$& HSC\\
$z'$ & $ 26.7$ &-- & --  & --   & Suprime-Cam\\
$Z$ & $ 26.1$ &-- &$ 25.9$ &$ 24.9$& HSC\\
$y$ & $ 25.6$ &-- &$ 25.2$ &$ 24.3$& HSC\\
$Y$ & 26.2-26.3 &25.1-25.2 &$ 25.3$ &$ 25.4$& VISTA\\
$J$ & 26.0-26.1 &24.8-25.0 &$ 25.8$ &$ 24.9$& VISTA/UKIRT\\
$H$ & 25.6-25.7 &24.5-24.6 &$ 25.2$ &$ 24.4$& VISTA/UKIRT\\
$K_s$ & 25.2-25.4 &24.9-25.0 &$ 25.5$ &$ 24.0$& VISTA/UKIRT\\
$3.6$ & $ 25.5$ &24.9 &$ 25.5$ &$ 24.3$ & \emph{Spitzer}/IRAC\\
$4.5$ & $ 25.5$ &24.8 &$ 25.4$ &$24.0$  & \emph{Spitzer}/IRAC\\

\hline

\end{tabular}
\end{table}

\section{Galaxy Selection}\label{sect:selection}

We searched for bright $z \gtrsim 7.5$ Lyman-break galaxy candidates in the XMM-LSS and COSMOS fields using a photometric redshift fitting analysis.
Such an approach allows up to 17 bands of broad-band photometric data from the optical to mid-infrared to be utilized in the selection process.
The initial candidates were extracted from the near-infrared selected catalogues described in Section~\ref{sect:cats} by requiring the object to be detected at $> 5\sigma$ significance according to the global depth in either the $J$ or $H$-bands ($z \simeq 8$ search) or $H$ or $K_{s}$-bands ($z \simeq 9$ search; $K$-band in the UDS sub-field).
We then required the object to be undetected at the $2\sigma$ level (according to the local depth) in all filters blue-ward of the expected Lyman-break for each redshift selection.
For the $z \simeq 8$ search the reddest band where we required a non-detection was the $z'$-band, and for the $z \simeq 9$--$10$ search we also required a non-detection up to and including the VISTA $Y$-band.
These catalogues were visually inspected in the detection band to remove obvious artefacts such as diffraction spikes or halos around bright stars.
The UKIRT $JHK$ imaging contains a strong `cross-talk' artefact that results in repeating ghost images at 128 pixels from bright stars.
Therefore in the case of UDS-detected objects, we also require a detection in the corresponding VISTA $JHK_{s}$ band at the $2\sigma$ level.
During follow-up of $z \simeq 7$ LBG candidates with~\emph{HST} we also identified a cross-talk artefact in the VISTA $YJHK_{s}$ data~\citep{Bowler2016}.
The cross-talk in the VISTA VIRCAM data is significantly fainter than that in the UDS for a given bright source, however it can mimic $z > 7$ LBGs close to the detection limit of the data.
To account for this possibility in both the UltraVISTA and VIDEO data we created a cross-talk mask by simulating the position of the cross-talk from all of the bright ($J < 14$) stars in the image.
This process was verified with visual inspection of the flagged cross-talk artefacts, where faint cross-talk has an extended, diffuse, appearance~\citep{Bowler2016} and the object in question is not detected in any other bands. 

\subsection{Photometric redshift analysis}
The resulting catalogues, with errors derived from our local depth analysis, were then fitted with a range of galaxy and brown-dwarf templates to form the final high-redshift sample.
We used the photometric redshift fitting code {\sc LePhare}~\citep{Arnouts1999, Ilbert2006} with a wide-range of galaxy templates from the~\citet{Bruzual2003} model library.
A declining star-formation history was assumed, with characteristic time-scales of $\tau = 0.05,0.1,0.2,0.5,1,2,5,10\,{\rm Gyrs}$ to approximate a burst and constant star formation at the extremes.
Dust attenuation was applied assuming the~\citet{Calzetti2000} dust law, with $A_{\rm V} = 0.0$--$6.0$ (steps of $0.2\,{\rm mag}$) to account for very dusty low-redshift interlopers~\citep{Dunlop2007}.
Metallicities of $1/5$ \Zsun and \Zsun were considered~\citep{Steidel2016}.
We performed the fitting with and without strong emission lines, which were included within {\sc LePhare} following the prescription presented in~\citet{Ilbert2009}.
We do not present the resulting best-fit galaxy physical properties such as the stellar mass, star-formation rate, the rest-frame UV slope, $A_{\rm V}$, $\tau$ or $Z$ as the low number of photometric detections at these redshifts does not warrant such an analysis.
Typically when over-fitting to data in this way, the resulting errors completely span the parameter space.
This effect can be observed in the fitting presented in S19 (e.g. see their fig. 10).

The high-redshift candidates were first required to have a best fitting photometric redshift in the range $7.0 < z < 9.0$ for the preliminary $z \simeq 8$ sample, and $8.0 < z < 11.0$ for the preliminary $z \simeq 9$--$10$ sample.
The best-fit solution had to be formally acceptable given the number of available bands for that object in that sub-field.
The lower redshift solution was required to be worse than the high-redshift fit at the $2\sigma$ level, corresponding to a $\Delta \chi^2 > 4.0$.
The initial selection was undertaken with the optical and near-infrared bands only (i.e. excluding the~\emph{Spitzer}/IRAC data) and with the results of the photometric redshift fitting without emission lines.
We then calculated the photometric redshifts including the deconfused~\chone~and~\chtwo~data points, which formed the final photometric redshifts in this work.
In this work we retained objects with best-fitting photometric redshifts $z > 7.4$ without lines.
The derived redshifts with nebular emission lines included were consistent with the line-free fits, with a small shift to higher redshifts up to $\delta z \simeq 0.1$ due to the Lyman-$\alpha$ and {\sc OIII + H}$\beta$ emission lines occupying the near-infrared bands and \chtwo~respectively (see Table~\ref{table:muv}).
The final candidates were then all carefully visually inspected in all filters to remove subtle artefacts and single-band detections.
In particular visual checks in the deep Subaru $z'$-band imaging and in a stack of the optical bands was successful at removing low-redshift contaminants that appeared to be acceptable high-redshift LBG candidates.
Prior to this final visual selection, the catalogues of sources that passed the automatic cuts on S/N and photometric redshift contained 495 objects.

\subsection{Brown dwarf interlopers}\label{sect:bd}

\begin{figure}
\includegraphics[width = 0.49\textwidth]{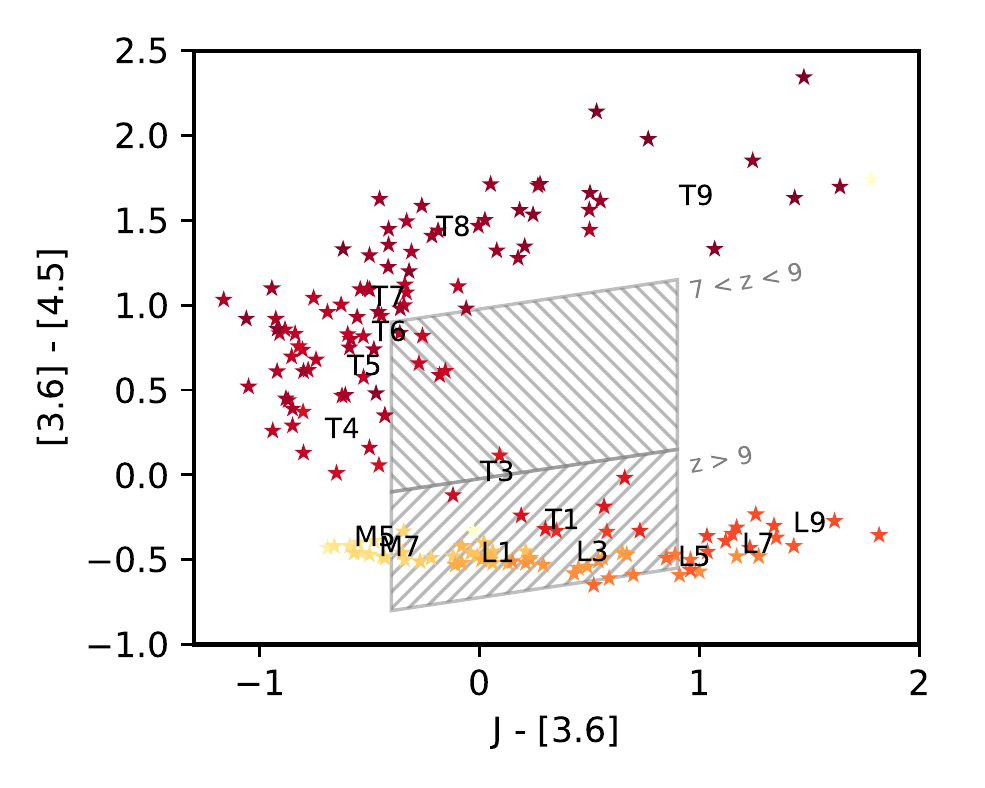}\\
\caption{The observed colours of brown dwarfs in the $J-[3.6]$ vs. $[3.6]-[4.5]$ colour space derived from the data in~\citet{DavyKirkpatrick2011} and~\citet{Patten2006}.
The sub-type corresponding to each region of colour space is labelled at the median colour of brown-dwarfs of that type.
The grey hatched areas show the expected colours of LBGs in the range $7 < z < 9$ (upper region) and $z > 9$ (lower region) due to the impact of nebular line contamination (as shown in Fig.~\ref{fig:neb}).
The spread in $J-[3.6]$ colour (which measures the underlying stellar continuum) was derived from~\citet{Bruzual2003} SED models with a constant SF history, $Z = 0.2 Z_{\odot}$, $A_{\rm V} = 0.0$--$0.5$ and ages in the range $50$--$500\,{\rm Myr}$.
 }\label{fig:bd}
\end{figure}

Cool Galactic brown dwarfs can mimic the colours of $z \gtrsim 7$ LBGs, as they are typically undetected in the optical bands but peak in the near-infrared.
To identify brown dwarfs in our initial LBG sample we fit the photometry with reference stellar spectra from the SpeX prism library\footnote{\url{http://pono.ucsd.edu/~adam/browndwarfs/spexprism/index.html}}.
These templates extend from $\lambda = 0.8$--$2.5\,\mu{\rm m}$.
While brown-dwarf templates that extend into the mid-infrared do exist, these are derived from models (e.g.~\citealp{Burrows2006}) and we found that these were unable to reproduce the observed colours of L and T-dwarfs beyond the $K$-band (e.g. see fig 4 in~\citealp{Leggett2007};~\citealp{Leggett2019}).
We instead use empirical brown-dwarf colours in the~\emph{Spitzer}/IRAC~\chone~and~\chtwo~bands to inform our selection.
We compiled observed~\emph{Spitzer}/IRAC photometry from~\citet{Patten2006} and~\citet{DavyKirkpatrick2011}.
These studies also had sub-typing of the brown dwarfs through spectroscopy.
In Fig.~\ref{fig:bd} we show the $J-[3.6]$ and $[3.6] -[4.5]$ colours of brown-dwarfs from these studies, coloured and labelled by sub-type.
We see that M- and L-dwarfs occupy a tight locus with $[3.6] -[4.5] \sim -0.5$ and a range of $J-[3.6]$ colours.
T-dwarfs instead occupy much redder $[3.6] -[4.5]$ colours.
As shown in Fig.~\ref{fig:neb}, the expected~\emph{Spitzer}/IRAC colour of $7 < z < 9$ sources is $[3.6] -[4.5] \sim 0.0$--$1.0$.
Hence, T-dwarfs, which also show a break in the $Y$-band, can mimic the colours of the high-redshift galaxies we are searching for and must be carefully considered and removed.
At $z > 9$ the expected IRAC colour of LBGs due to nebular line emission in these bands changes sign.
The blue $[3.6] -[4.5] \sim -0.7$--$0.0$ can then be reproduced by both early L- and T-dwarfs.
Our primary method for removal of this contaminant was through SED fitting of the red optical and near-infrared photometry using the SpeX templates.
Using the best-fitting sub-type we were then able to predict the \chone~and \chtwo~magnitudes from the colours shown in Fig.~\ref{fig:bd} and the observed $J$-band for each candidate.
We show the predicted brown-dwarf magnitudes in the \chone~and \chtwo~bands in the SED figures presented in Appendix~\ref{ap:stamps}.
The predicted average colour (or colours, if multiple sub-types were acceptable brown dwarf fits) were then compared to the observed photometry and used to discriminate between the brown-dwarf and galaxy fits.
M- and L-dwarf contaminants are expected to have detections in the red-optical or $Y$-band, and hence even if the IRAC colour is the same as that expected from a genuine high-redshift LBG, this type of contaminant can be clearly excluded from the sample according to the optical/NIR fitting.
The removal of T-dwarf contaminants is more challenging, as they can reproduce the optical/NIR SED (i.e. a spectral break) and in some cases reproduce the $J-[3.6]$ and $[3.6] -[4.5]$ colours as shown in Fig.~\ref{fig:bd}.
While the average colour of the T-dwarfs sub-types are outside the region occupied by high-redshift sources, the T-dwarfs show a large intrinsic scatter in colour that means they can occasionally reproduce the expected colours of $z > 7$ LBGs.
Fig.~\ref{fig:bd} shows that this scatter is predominantly a problem at $J-[3.6] < 0.2$, while the majority of our sample show redder colours than this (see Table~\ref{table:photz8} and~\ref{table:photz9}.
Thus in combination with our optical/NIR fitting, we are confident that brown dwarf contamination is not significant in our sample.

\begin{table*}
\caption{The coordinates and observed photometry for the LBG candidates in the $z \simeq 8$ sample presented in this work.
The XMM-LSS candidates are displayed in the upper part of the table, with the COSMOS objects shown in the lower part.
Each section is ordered by $J$-band magnitude.
The ID number is shown in the first column, with the first part denoting which sub-field each candidate is found in.
For example XMM2-4314 is found within the second tile of XMM-LSS as shown in Fig.~\ref{fig:xmm}.
The second and third columns display the Right Ascension (R.A.) and Declination of the sources.
The following columns show the total magnitudes in the $z$- and $y$-bands from HSC or SC, the near-infrared data from VISTA or UKIRT (if available) and finally the~\emph{Spitzer}/IRAC~\chone~and~\chtwo~bands.
In the case of a non-detection at the $2\sigma$ level (derived from the local depth), the measurement is shown as an upper limit.
The measured aperture photometry (in a $1.8$ arcsec diameter circular aperture) has been corrected to total assuming a point-source correction.
}\label{table:photz8}
\begin{tabular}{lcccccccccc}
\hline
ID & R.A. & Dec. & $z'$ & $y$ & $Y$ & $J$ & $H$ & $K_s$ & $[3.6]$ & $[4.5]$ \\
\hline
XMM3--5645 & 02:25:52.26 & -05:02:46.07 & $ > 25.16 $ & $ > 25.09 $ & $ 25.15_{-0.18}^{+0.21}$ & $ 23.86_{-0.11}^{+0.12}$ & $ 23.83_{-0.18}^{+0.21}$ & $ 23.76_{-0.22}^{+0.28}$ & $ 23.02_{-0.20}^{+0.2
4}$ & $ 22.77_{-0.20}^{+0.24}$ \\[1ex]
XMM2--3904 & 02:21:54.15 & -04:24:12.29 & $ > 25.68 $ & $ > 24.99 $ & $ 25.37_{-0.27}^{+0.36}$ & $ 24.03_{-0.14}^{+0.16}$ & $ 24.25_{-0.23}^{+0.29}$ & $ 24.02_{-0.28}^{+0.37}$ & $ > 24.67 $ & $ > 24.
31 $ \\[1ex]
XMM2--4314 & 02:20:09.28 & -04:11:43.12 & $ > 25.19 $ & $ > 24.74 $ & $ > 26.00 $ & $ 24.16_{-0.16}^{+0.18}$ & $ 23.61_{-0.13}^{+0.15}$ & $ 23.42_{-0.14}^{+0.17}$ & $ 22.57_{-0.20}^{+0.24}$ & $ 22.64
_{-0.20}^{+0.24}$ \\[1ex]
XMM1--994 & 02:16:33.48 & -04:30:07.91 & $ > 26.46 $ & $ > 25.82 $ & $ > 25.88 $ & $ 24.20_{-0.17}^{+0.20}$ & $ 23.94_{-0.21}^{+0.26}$ & $ 23.98_{-0.35}^{+0.52}$ & $ 22.70_{-0.20}^{+0.24}$ & $ 22.54_
{-0.20}^{+0.24}$ \\[1ex]
XMM3--6787 & 02:26:16.52 & -04:07:04.07 & $ > 25.66 $ & $ > 24.93 $ & $ > 25.98 $ & $ 24.42_{-0.21}^{+0.25}$ & $ 24.19_{-0.21}^{+0.27}$ & $ 24.16_{-0.28}^{+0.37}$ & $ 23.61_{-0.20}^{+0.24}$ & $ 23.82
_{-0.22}^{+0.27}$ \\[1ex]
UDS--254 & 02:16:12.56 & -04:59:28.99 & $ > 27.70 $ & $ > 25.61 $ & $ > 25.99 $ & $ 24.80_{-0.12}^{+0.14}$ & $ 24.69_{-0.19}^{+0.23}$ & $ 25.32_{-0.23}^{+0.30}$ & $ 24.55_{-0.20}^{+0.25}$ & $ 24.77_{
-0.30}^{+0.41}$ \\[1ex]
UDS--299 & 02:17:18.55 & -04:54:58.50 & $ > 27.51 $ & $ > 25.89 $ & $ > 26.13 $ & $ 25.39_{-0.20}^{+0.25}$ & $ 25.52_{-0.34}^{+0.49}$ & $ 25.91_{-0.34}^{+0.50}$ & $ 25.25_{-0.22}^{+0.28}$ & $ 24.61_{
-0.26}^{+0.33}$ \\[1ex]
UDS--74 & 02:16:45.78 & -05:23:33.32 & $ > 27.58 $ & $ > 25.78 $ & $ > 26.05 $ & $ 25.47_{-0.20}^{+0.25}$ & $ 25.48_{-0.32}^{+0.47}$ & $ 26.01_{-0.34}^{+0.51}$ & $ 25.43_{-0.43}^{+0.72}$ & $ 24.46_{-
0.20}^{+0.25}$ \\
\hline
UVISTA--914 & 10:02:12.55 & +02:30:45.74 & $ > 27.17 $ & $ > 26.20 $ & $ > 26.81 $ & $ 24.84_{-0.10}^{+0.11}$ & $ 24.98_{-0.16}^{+0.18}$ & $ 25.18_{-0.25}^{+0.32}$ & $ 25.00_{-0.28}^{+0.38}$ & $ 24.3
6_{-0.20}^{+0.24}$ \\[1ex]
UVISTA--762 & 09:57:47.90 & +02:20:43.55 & $ > 26.81 $ & $ > 26.35 $ & $ > 26.86 $ & $ 24.89_{-0.12}^{+0.13}$ & $ 24.69_{-0.10}^{+0.11}$ & $ 24.56_{-0.13}^{+0.15}$ & $ 24.27_{-0.21}^{+0.26}$ & $ 24.0
7_{-0.20}^{+0.24}$ \\[1ex]
UVISTA--301 & 10:00:58.48 & +01:49:56.00 & $ > 27.35 $ & $ > 26.38 $ & $ 25.76_{-0.28}^{+0.37}$ & $ 24.89_{-0.15}^{+0.17}$ & $ 25.05_{-0.26}^{+0.35}$ & $ 24.98_{-0.21}^{+0.26}$ & $ 24.92_{-0.31}^{+0.
44}$ & $ 24.72_{-0.28}^{+0.37}$ \\[1ex]
UVISTA--1043 & 09:58:38.95 & +02:42:32.05 & $ > 27.08 $ & $ > 26.27 $ & $ > 26.02 $ & $ 25.18_{-0.23}^{+0.29}$ & $ 25.65_{-0.44}^{+0.74}$ & $ > 25.69 $ & $ 25.56_{-0.40}^{+0.63}$ & $ > 25.56 $ \\[1ex]
UVISTA--879 & 09:57:54.69 & +02:27:54.90 & $ > 26.67 $ & $ > 26.03 $ & $ 26.57_{-0.36}^{+0.54}$ & $ 25.19_{-0.14}^{+0.17}$ & $ 25.55_{-0.23}^{+0.30}$ & $ 25.54_{-0.36}^{+0.55}$ & $ 24.68_{-0.27}^{+0.
37}$ & $ 24.36_{-0.20}^{+0.24}$ \\[1ex]
UVISTA--839 & 09:57:54.26 & +02:25:08.41 & $ > 26.56 $ & $ > 26.08 $ & $ > 26.86 $ & $ 25.41_{-0.20}^{+0.24}$ & $ 25.72_{-0.29}^{+0.39}$ & $ 25.66_{-0.35}^{+0.52}$ & $ 24.92_{-0.28}^{+0.39}$ & $ 24.5
3_{-0.23}^{+0.30}$ \\[1ex]
UVISTA--1032 & 10:00:30.67 & +02:42:09.23 & $ > 26.80 $ & $ > 26.11 $ & $ > 26.45 $ & $ 25.44_{-0.25}^{+0.33}$ & $ 25.47_{-0.32}^{+0.45}$ & $ > 25.73 $ & $ > 25.13 $ & $ > 25.06 $ \\[1ex]
UVISTA--598 & 10:01:47.49 & +02:10:15.39 & $ > 27.25 $ & $ > 26.28 $ & $ > 26.90 $ & $ 25.54_{-0.19}^{+0.22}$ & $ 25.87_{-0.33}^{+0.48}$ & $ 25.76_{-0.35}^{+0.52}$ & $ 25.44_{-0.29}^{+0.39}$ & $ 25.1
0_{-0.26}^{+0.34}$ \\[1ex]
UVISTA--213 & 10:00:32.32 & +01:44:31.21 & $ > 27.23 $ & $ > 26.39 $ & $ 26.46_{-0.30}^{+0.42}$ & $ 25.56_{-0.15}^{+0.18}$ & $ 25.02_{-0.12}^{+0.14}$ & $ 25.42_{-0.26}^{+0.34}$ & $ 24.54_{-0.20}^{+0.
24}$ & $ 24.30_{-0.20}^{+0.24}$ \\[1ex]
UVISTA--953 & 10:01:56.33 & +02:34:16.21 & $ > 27.11 $ & $ > 26.24 $ & $ > 26.65 $ & $ 25.57_{-0.19}^{+0.23}$ & $ 25.83_{-0.35}^{+0.52}$ & $ 25.79_{-0.40}^{+0.64}$ & $ > 25.79 $ & $ 25.39_{-0.33}^{+0
.48}$ \\[1ex]
UVISTA--356 & 10:00:17.89 & +01:53:14.35 & $ > 27.48 $ & $ > 26.41 $ & $ > 26.97 $ & $ 25.59_{-0.15}^{+0.18}$ & $ 26.03_{-0.36}^{+0.55}$ & $ > 26.03 $ & $ > 25.60 $ & $ > 25.99 $ \\[1ex]
UVISTA--919 & 10:00:22.93 & +02:31:24.36 & $ > 27.20 $ & $ > 26.21 $ & $ > 26.87 $ & $ 25.61_{-0.17}^{+0.20}$ & $ 25.59_{-0.22}^{+0.28}$ & $ 25.49_{-0.27}^{+0.36}$ & $ 24.88_{-0.25}^{+0.32}$ & $ 24.8
6_{-0.20}^{+0.24}$ \\[1ex]
UVISTA--266 & 10:01:45.05 & +01:48:28.53 & $ > 27.49 $ & $ > 26.27 $ & $ > 26.69 $ & $ 25.68_{-0.23}^{+0.29}$ & $ 25.74_{-0.29}^{+0.40}$ & $ 25.37_{-0.25}^{+0.33}$ & $ > 25.44 $ & $ 25.08_{-0.33}^{+0
.48}$ \\[1ex]
UVISTA--634 & 10:00:41.18 & +02:12:23.95 & $ > 27.13 $ & $ > 26.53 $ & $ > 26.83 $ & $ 25.73_{-0.19}^{+0.23}$ & $ 26.30_{-0.42}^{+0.69}$ & $ > 26.12 $ & $ 24.59_{-0.20}^{+0.24}$ & $ 24.59_{-0.20}^{+0
.24}$ \\

\hline
\end{tabular}
\end{table*}

\begin{table*}
\caption{The coordinates and observed photometry for the LBG candidates at $z > 8.5$ found in this study.
The columns are as in Table~\ref{table:photz8}.
The first row shows the photometry for the $z = 10.9$ candidate found in the XMM-LSS field.
Following this we show the $z \simeq 9$ sample with the XMM-LSS candidates followed by the COSMOS sources.
Object UVISTA-1212 is shown with an asterisk to denote that it was found within the `deep' part of the UltraVISTA data, not the `ultra-deep' stripes.
}
\begin{tabular}{lcccccccccc}
\hline
ID & R.A. & Dec. & $z'$ & $y$ & $Y$ & $J$ & $H$ & $K_s$ & $[3.6]$ & $[4.5]$ \\

 \hline
 XMM3--3085 & 02:26:59.08 & -05:12:17.49 & $ > 25.03 $ & $ > 24.31 $ & $ > 25.70 $ & $ > 25.43 $ & $ 23.87_{-0.17}^{+0.21}$ & $ 23.96_{-0.25}^{+0.32}$ & $ 23.69_{-0.26}^{+0.35}$ & $ 23.52_{-0.25}^{+0.33}$ \\
\hline
UDS--355 & 02:17:42.47 & -04:58:57.80 & $ > 27.51 $ & $ > 25.95 $ & $ > 26.04 $ & $ 25.18_{-0.15}^{+0.17}$ & $ 24.80_{-0.17}^{+0.20}$ & $ 25.11_{-0.18}^{+0.21}$ & $ 24.49_{-0.20}^{+0.24}$ & $ 24.02_{-0.20}^{+0.24}$ \\[1ex]
UDS--787 & 02:16:27.92 & -04:42:29.21 & $ > 26.97 $ & $ > 25.40 $ & $ > 25.49 $ & $ 24.87_{-0.16}^{+0.19}$ & $ 24.85_{-0.20}^{+0.25}$ & $ 25.30_{-0.25}^{+0.32}$ & $ 25.33_{-0.42}^{+0.68}$ & $ 24.15_{-0.21}^{+0.26}$ \\[1ex]
UDS--320 & 02:18:38.44 & -04:52:59.16 & $ > 26.90 $ & $ > 24.97 $ & $ > 25.78 $ & $ 25.38_{-0.20}^{+0.24}$ & $ 25.39_{-0.34}^{+0.49}$ & $ 25.60_{-0.29}^{+0.39}$ & $ > 24.52 $ & $ > 24.79 $ \\
\hline
UVISTA--1212* & 10:02:31.81 & +02:31:17.10 & $ > 27.42 $ & $ > 25.97 $ & $ > 25.69 $ & $ 24.72_{-0.22}^{+0.28}$ & $ 24.39_{-0.22}^{+0.28}$ & $ 24.42_{-0.16}^{+0.18}$ & $ 25.05_{-0.33}^{+0.49}$ & $ 24
.06_{-0.20}^{+0.24}$ \\[1ex]
UVISTA--237 & 10:00:31.88 & +01:57:50.04 & $ > 27.50 $ & $ > 26.33 $ & $ > 27.07 $ & $ 25.78_{-0.19}^{+0.24}$ & $ 25.33_{-0.18}^{+0.22}$ & $ 25.77_{-0.37}^{+0.57}$ & $ > 25.94 $ & $ 24.80_{-0.20}^{+0
.24}$ \\

\hline
 
\end{tabular}\label{table:photz9}
\end{table*}

\section{The Sample}\label{sect:sample}

The result of our photometric redshift selection procedure was a sample of 28 candidate LBGs at $z \gtrsim 7.5$ from $\sim 6\,$\ds~ of optical/near-infrared imaging in the XMM-LSS and COSMOS fields.
Of this full sample, five have photometric redshifts in the range $8.5 < z < 9.5$ and one object has a best-fit photometric redshift of $z \simeq 10.9$.
We present the observed photometry of the sample in Tables~\ref{table:photz8} and~\ref{table:photz9}.
Postage-stamp images of the candidates and the best-fitting galaxy and brown-dwarf templates are presented in Appendix~\ref{ap:stamps}.

\subsection{The XMM-LSS sample}
The XMM-LSS field covers $4.5$\ds~to shallower depths than COSMOS (except in the UDS sub-field), and hence the candidates found in these data contribute to the very bright-end of our sample.
Within the XMM-LSS field we find eight candidate LBGs at $z \simeq 8$, three at $z\simeq 9$ and one with a best-fit photometric redshift of $z \simeq 10.9 \pm 1.0$.
As shown in Table~\ref{table:photz8}, these objects have typical near-infrared magnitudes of $J \simeq 24$--$24.5$.
We find fewer $z \simeq 8$ objects over the UDS sub-field in comparison to the `ultra-deep' part of COSMOS.
This is to be expected as the UDS has shallower $Y$-band data and hence the selection of $Y$-dropout sources is less efficient here.
The $z \simeq 9$ candidates were found within the deeper near-infrared data in the UKIRT UDS field.
The shallower optical data over the wide XMM-LSS field makes selecting clean samples of high-redshift galaxies more challenging.
We therefore view these six objects in the wide XMM-LSS field as the least secure high-redshift LBG candidates in our sample.

\begin{figure}
\centering
\includegraphics[width = 0.4\textwidth]{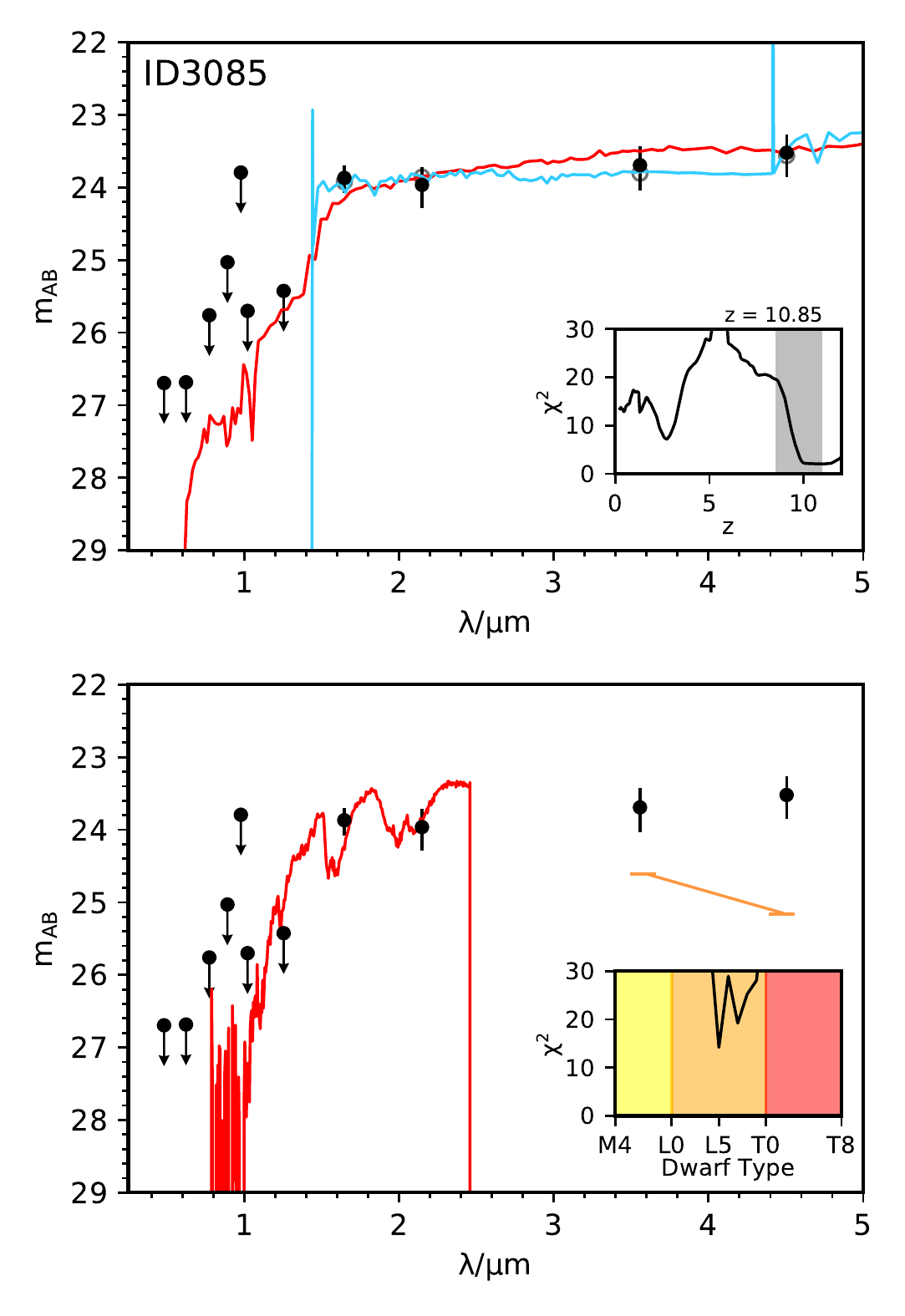}
\caption{The observed photometry and best-fitting SED models from our photometric redshift analysis for the highest redshift source in our sample.
Selected within the XMM-LSS field, the object XMM3-3085 has a best-fitting photometric redshift of $z = 10.9 \pm 1.0$.
The black points show the observed photometry for this galaxy.
In the upper plot we show the best-fitting galaxy templates, with the blue line showing the preferred high-redshift solution and the red line showing the low-redshift best-fit.
The inset shows the model $\chi^2$ as a function of redshift.
In the lower plot we show the results of fitting the photometry with brown-dwarf models.
The expected \chone--\chtwo~colour for the best-fitting brown-dwarf sub-type is shown as the orange line.
The inset here shows the $\chi^2$ for each sub-type of brown dwarf.
In this case the brown dwarf fit is significantly worse than the galaxy fit.
}\label{fig:z10}
\end{figure}

We find one $z \simeq 11$ candidate that passes all of our selection criterion in the wide part of our survey.
As shown in Fig.~\ref{fig:z10}, for XMM3-3085 the low-redshift fit is unable to reproduce the near-infrared photometry.
The $\chi^2$ distribution shows that a $z \simeq 3$ solution is the next most probable, and this solution is formally acceptable in the SED fitting analysis.
The object is excluded as a brown dwarf based on the poor $\chi^2$ in our brown dwarf fitting.
If we take the best-fitting brown-dwarf sub-type, which is L5, we would expect to measure an IRAC colour of $[3.6] -[4.5] = -0.5$ according to the measured brown dwarf colours shown in Fig.~\ref{fig:bd}.
This predicted IRAC colour is inconsistent with the observed photometry as shown in Fig.~\ref{fig:z10}, further strengthening our conclusion that it is not a brown dwarf.
This source, if confirmed, would be the brightest object known at $z > 8$.
Deeper imaging or spectroscopy of this source will be required to determine robustly the redshift.
However, as we show in Section~\ref{sect:lf}, the existence of such a luminous source at this redshift is fully consistent with our derived evolution of the LF at the very bright-end.

\subsection{The COSMOS sample}

Within the COSMOS field we select 14 candidate LBGs at $z \simeq 8$, and two at $z \simeq 9$.
The brightest object, UVISTA-1212, was found in the `deep' part of the UltraVISTA data (see Fig.~\ref{fig:cosmos}) whereas the remaining candidates were found within the `ultra-deep' region.
As the `ultra-deep' stripes are the deepest near-infrared region analysed, the COSMOS candidates selected here form the faint end of our sample with $J \simeq 24.8$--$25.7$.
It is to be expected that fewer candidates will be selected in the `deep' component of UltraVISTA, as this imaging constitutes the shallowest $Y$-band data analysed in this study.
The majority of the LBGs are detected in the~\emph{Spitzer}/IRAC data, with some objects visually showing the red colours expected from strong rest-frame optical emission lines (e.g. UVISTA-953).
Three of the candidates were selected outside the region of very deep $z'$-band data from Subaru.
UVISTA-1212 is the brightest object we find within COSMOS, and with a best-fitting photometric redshift of $z = 9.12^{+0.20}_{-0.26}$, it is one of the most luminous $z \simeq 9$ LBG candidates known with $H = 24.4^{+0.3}_{-0.2}\,{\rm mag}$.
The brown-dwarf fit can be strongly excluded based on the~\emph{Spitzer}/IRAC colour.

\subsection{Comparison to~\citet{Stefanon2017, Stefanon2019}}\label{sect:S19}

A search for $z > 7$ LBGs in the COSMOS field was performed by~\citet{Stefanon2017, Stefanon2019}, who utilized a $Y$ and $J$-drop colour-colour cut methodology to find 16 galaxy candidates with photometric redshifts in the range $7.4 < z < 9.2$. 
In our COSMOS sample we recover seven of this sample, predominantly the brightest objects, which are labelled in Table~\ref{table:muv} and in the SED plots in the Appendix.
Of the nine S19 LBG candidates that were not reselected in this study, all but one (Y7) were present in our initial catalogues but later excluded for being likely low-redshift galaxies in our analysis.
The candidate Y7 was not detected as a distinct object in any of our initial catalogues due to being in the wings of a bright source.
The sources we do not reselect are also the least secure objects in S19, where they determine that Y6, Y9, Y13, Y14, Y15 all have $\gtrsim 20$\% likelihood of being at $z < 7$.
We note that objects Y6 and Y11 are both detected at $\simeq 2 \sigma$ in the deep $z'$-band imaging from~\citet{Furusawa2016} that was not utilized by S19, supporting the conclusion that they are at low-redshift.
We find that the remaining objects that we fail to reselect are particularly faint in our catalogues, leading to poorly-constrained photometric redshifts.
We compare the $H$-band magnitudes from our work and S19, who used the UltraVISTA DR3 data, in Appendix~\ref{app:S19}.
We find evidence that the magnitudes of the fainter galaxies in the S19 sample, which were derived from the shallower UltraVISTA DR3 imaging, are too bright by $\simeq 0.5$ mag.
This could explain the down-weighting of the low-redshift solution in their analysis, as it would lead to an overestimated S/N in the near-infrared bands.

In~\citet{Stefanon2017}, two additional candidate $z \gtrsim 8.5$ were presented that appeared as $J$-dropout sources in the ground-based photometry.
When followed-up with~\emph{HST}/WFC3, both objects (J1 and J2) were subsequently found to be likely low-redshift interlopers when they included this new photometry in an SED fitting analysis.
\citet{Stefanon2017} claim that the~\emph{HST}/WFC3 data was essential to determine an accurate photometric redshift, however we find that these objects are best-fit as low-redshift ($z \simeq 2$) galaxies using photometry derived from the same ground-based imaging utilized in their study (and also when using both the DR3 and DR4 releases of UltraVISTA).
Furthermore, we find J2 to be detected at $2\sigma$ in the deeper $z'$-band imaging, which excludes it being a $z > 7$ source.
Our analysis suggests that in this case the particularly deep ($m_{{\rm AB}, 5\sigma} \simeq 27$) $z'$-band imaging from~\citet{Furusawa2016} is more valuable as a discriminator between competing low and high-redshift models than the shallower ($m_{{\rm AB}, 5\sigma} \simeq 24.5$--$25.8$) near-infrared follow-up with WFC3.

\subsection{Rest-frame optical colours}\label{sect:neb}

\begin{figure}
\includegraphics[width = 0.47\textwidth]{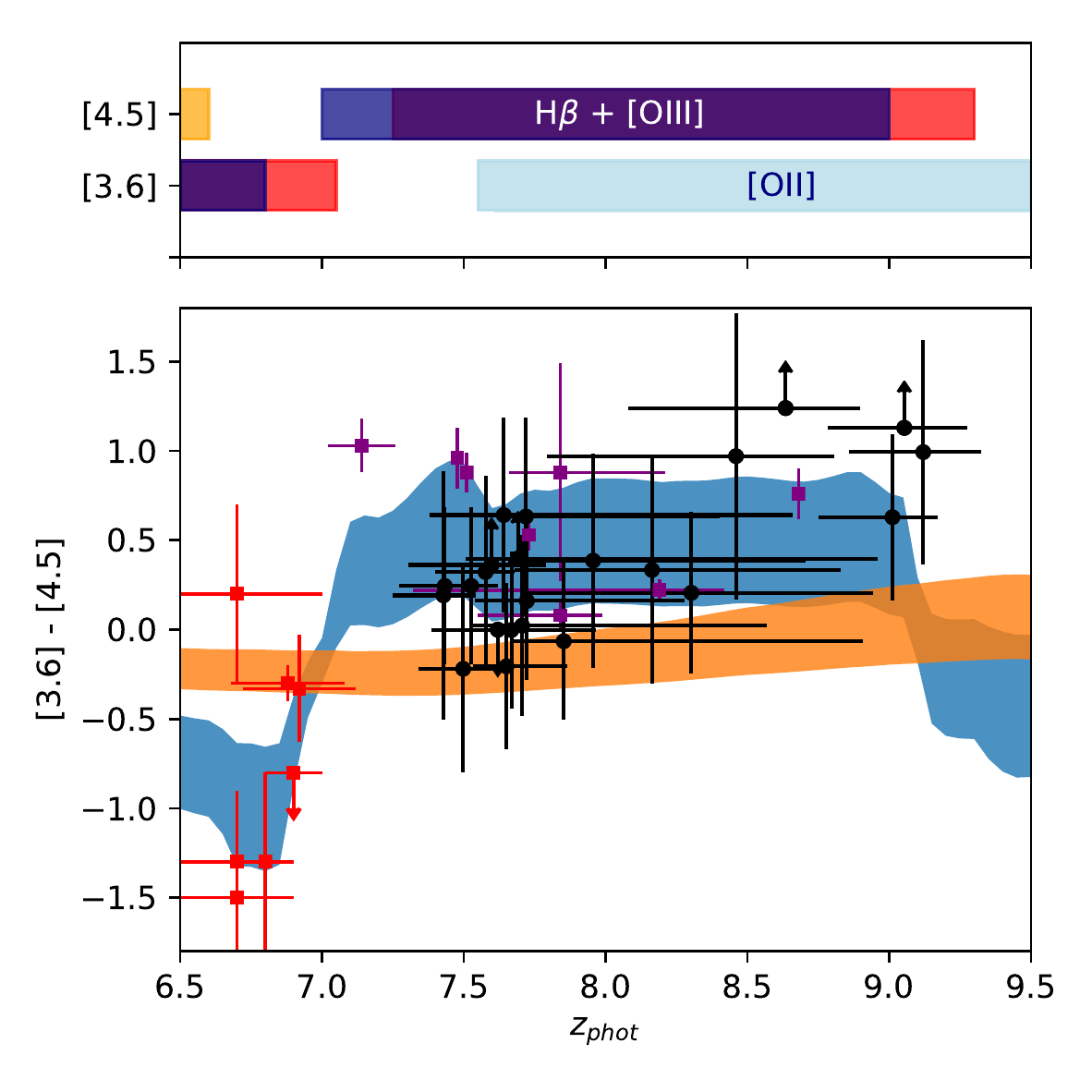}\\
\caption{The observed $[3.6]-[4.5]$ colours of our sample of bright $z \simeq 8$ and $z \simeq 9$ LBGs.
The upper plot shows which strong rest-frame optical emission lines populate the $[3.6]$ and $[4.5]$ bands at a given redshift.
In the lower plot our results are shown as the back circles.  
The purples squares show the results of~\citet{RobertsBorsani2015} and~\citet{Oesch2016}.
The red squares show the results of~\citet{Smit2014}.
The horizontal orange shaded region shows the expected colour in these bands without emission lines.
The blue shaded region shows the expected colour with emission lines of $EW_{0}({\rm [OIII]} + {\rm H}\beta) = 500$--$2000$\AA.
}\label{fig:neb}
\end{figure}

At $7 < z < 9 $ the {\sc H}$\beta$ and {\sc [OIII]} $\lambda 4959, 5007$ rest-frame optical emission lines occupy the~\emph{Spitzer}/IRAC $[4.5]$ band, with the $[3.6]$ band only containing the weaker {\sc [OII]} $\lambda 3727$ line.
The result is that a red $[3.6]-[4.5]$ colour is expected if these emission lines are strong, as has been observed in samples of similarly luminous galaxies at $z \simeq 7$~\citep{Bowler2016}.
We show the measured $[3.6]-[4.5]$ colours of our sample plotted against photometric redshift in Fig.~\ref{fig:neb}, in comparison to other results derived from fainter sources in~\citet{Smit2014} and a compilation of $z > 7$ objects presented in~\citet{RobertsBorsani2015}.
We show the expected colours from a range of~\citet{Bruzual2003} models with emission lines added according to a simple redshift dependent emission line strength model described in~\citet{Bowler2014}.
We assume rest-frame equivalent widths in the range $EW_{0}(${\sc [OIII]}$ + {\rm H}\beta) = 500-2000$\AA~(fixed at $z = 6.8$ and allowed to evolve according to $(1+z)^{1.8}$;~\citealp{Fumagalli2012}).
While the error bars are large, we find a preference for red colours in our sample, which is in agreement with some line contamination.
We find a wide distribution of colours within the sample spanning $[3.6]-[4.5] \simeq 0.0$--$1.0$.
As shown in Fig.~\ref{fig:neb} these colours span the expected range from continuum only emission (orange shaded region) to significant contamination by extreme emission lines (blue shaded region).
Our results therefore indicate that there is a distribution in rest-frame optical emission line strengths in bright ($M_{\rm UV} \lesssim -21$) star-forming galaxies at $z > 6$, from no discernible emission to rest-frame equivalent widths exceeding $EW_{0}(${\sc [OIII]}$ + {\rm H}\beta) > 1500$\AA.
This is consistent with previous measurements of the IRAC colours of bright $z \simeq 7$ LBGs, which also show a large spread in $[3.6]-[4.5]$ under a deconfusion analysis using~\emph{HST}/WFC3 imaging~\citep{Bowler2016}.
\citet{Hashimoto2018} find evidence for a strong Balmer break indicative of an evolved stellar population in a lensed $z \sim 9$ LBG.
The IRAC colour in this galaxy, MACS1149-JD1 was measured to be $[3.6]-[4.5] = 0.9$, which if interpreted as a Balmer break, suggests that this galaxy was forming stars only 250 Myrs after the Big Bang.
The majority of our sample do not show such red colours, which suggests that either MACS1149-JD1 is not representative of the galaxy population at these redshifts or that the IRAC colour in this object is due to contamination by the {\sc [OIII]} emission line.

\section{Determination of the LF}\label{sect:lf}

In this section, we use the results of our search for $z =  8$--$10$ star-forming galaxies to determine the bright-end of the rest-frame UV luminosity function.
The best-fitting photometric redshifts and absolute UV magnitudes for our sample of 28 LBGs are presented in Table~\ref{table:muv}.
We calculate the absolute UV magnitude from the best-fitting SED model using a top-hat filter of width 100\AA~centred on 1500\AA.
Due to the wide-area imaging utilized by our study, we are able to select extremely luminous LBG candidates with absolute magnitudes as bright as $M_{\rm UV} \simeq -23$.
While the majority of the most luminous sources we detect were found in the XMM-LSS field, we also detect one particularly luminous $z \simeq 9$ source in the COSMOS field, as well as recovering the bright sub-sample of LBG candidates found by S19.
At the faint-end of our sample the sources exclusively come from the COSMOS/UltraVISTA `ultra-deep' tier of data, which crucially provides very deep $Y$-band imaging that allows the selection of $z > 7$ sources down to an $M_{\rm UV} \simeq -21.5$.
Due to the lack of photometric filters around the Lyman-break region (e.g. as compared to $z \simeq 7$ where the break is bracket by multiple close $Z$- and $Y$-band filters) the photometric redshifts at $z =8$--$9$ have a broader probability distribution and hence have larger errors.
We nevertheless split our sample into $z = 8$ and $z = 9$ bins for the LF calculation with the expectation that once spectroscopically confirmed, the sources will span the expected range in photometric redshifts as derived from the fitting.
We find one ultra-luminous candidate LBG with a photometric redshift of $z = 10.85_{-1.02}^{+1.00}$, leading to a derived absolute magnitude of $M_{\rm UV} \simeq -23.68$.
Given the large errors on the photometric redshift of this source, due to the putative Lyman-break from this candidate occupying the space between the $J$ and $H$-bands, we estimate the number density associated with this candidate and compare it to previous results at $z \simeq 10$.
In Table~\ref{table:muv} we present the photometric redshifts with and without nebular emission lines included in the fitting.
For the LF analysis we utilize the photometric redshifts derived when fitting with nebular emission lines, as there is evidence from the~\emph{Spitzer}/IRAC colours that line emission is important in bright LBGs at high-redshift (Section~\ref{sect:neb}).
The line-free results are typically $\delta z = 0.0$--$ 0.1$ lower, and hence our LF results are unchanged if we use these values.

\begin{table}\caption{The photometric redshifts and absolute UV magnitudes for our sample.
The first column gives the object ID, where the first part of the name describes which field/sub-field the object was selected in.
We show the best-fit photometric redshift from continuum only fitting in column 2, followed by the best-fit when emission lines are included in column 3.
The next column shows the rest-frame UV absolute magnitude for each object.
The objects have been ordered by $M_{\rm UV}$.
The final column shows the name of the source in the S19 sample.
}
\begin{tabular}{lcccc}
\hline
ID & $z_{\rm phot}$ & $z_{\rm phot}$ & $M_{\rm UV}$ & S19 \\
& (line free) & (with lines) & $/{\rm mag}$ &  \\
\hline
XMM3--3085 & $10.76_{-0.95}^{+0.93}$ & $10.85_{-1.02}^{+1.00}$ & $-23.68_{-0.15}^{+0.18}$ &  \\[1ex]
XMM3--5645 & $7.48_{-0.09}^{+0.07}$ & $7.53_{-0.12}^{+0.09}$ & $-23.20_{-0.09}^{+0.10}$ &  \\[1ex]
XMM2--4314 & $7.77_{-0.13}^{+0.62}$ & $7.85_{-0.19}^{+1.05}$ & $-23.06_{-0.10}^{+0.11}$ &  \\[1ex]
XMM2--3904 & $7.56_{-0.07}^{+0.06}$ & $7.58_{-0.08}^{+0.08}$ & $-23.05_{-0.14}^{+0.16}$ &  \\[1ex]
UVISTA--1212 & $9.07_{-0.23}^{+0.21}$ & $9.12_{-0.26}^{+0.20}$ & $-23.01_{-0.27}^{+0.37}$ &  \\[1ex]
XMM1--994 & $7.73_{-0.20}^{+0.51}$ & $7.72_{-0.18}^{+0.56}$ & $-22.92_{-0.13}^{+0.15}$ &  \\[1ex]
XMM3--6787 & $7.64_{-0.12}^{+0.16}$ & $7.65_{-0.12}^{+0.22}$ & $-22.68_{-0.20}^{+0.24}$ &  \\[1ex]
UDS--355 & $8.95_{-0.25}^{+0.16}$ & $9.01_{-0.26}^{+0.16}$ & $-22.48_{-0.18}^{+0.21}$ &  \\[1ex]
UDS--787 & $8.58_{-0.42}^{+0.26}$ & $8.63_{-0.55}^{+0.26}$ & $-22.37_{-0.17}^{+0.20}$ &  \\[1ex]
UVISTA--762 & $8.19_{-0.49}^{+0.67}$ & $8.30_{-0.59}^{+0.64}$ & $-22.36_{-0.08}^{+0.09}$ & Y1 \\[1ex] 
UVISTA--914 & $7.67_{-0.08}^{+0.66}$ & $7.72_{-0.10}^{+0.68}$ & $-22.20_{-0.09}^{+0.10}$ & Y2 \\[1ex]
UDS--254 & $7.46_{-0.14}^{+0.14}$ & $7.50_{-0.16}^{+0.16}$ & $-22.17_{-0.11}^{+0.13}$ &  \\[1ex]
UVISTA--301 & $7.36_{-0.12}^{+0.12}$ & $7.43_{-0.18}^{+0.11}$ & $-22.14_{-0.13}^{+0.14}$ & Y4 \\[1ex]
UVISTA--237 & $9.01_{-0.26}^{+0.21}$ & $9.05_{-0.27}^{+0.22}$ & $-21.92_{-0.22}^{+0.27}$ & Y5 \\[1ex]
UDS--320 & $8.54_{-0.99}^{+0.45}$ & $8.62_{-1.04}^{+0.44}$ & $-21.87_{-0.42}^{+0.68}$ &  \\[1ex]
UVISTA--879 & $7.49_{-0.12}^{+0.11}$ & $7.58_{-0.18}^{+0.10}$ & $-21.78_{-0.13}^{+0.14}$ &  \\[1ex]
UVISTA--1043 & $7.55_{-0.20}^{+0.24}$ & $7.62_{-0.23}^{+0.20}$ & $-21.76_{-0.19}^{+0.22}$ &  \\[1ex]
UVISTA--1032 & $7.84_{-0.21}^{+1.16}$ & $7.87_{-0.21}^{+1.16}$ & $-21.67_{-0.19}^{+0.22}$ &  \\[1ex]
UDS--74 & $8.46_{-0.68}^{+0.26}$ & $8.46_{-0.67}^{+0.35}$ & $-21.66_{-0.15}^{+0.18}$ &  \\[1ex]
UDS--299 & $7.56_{-0.21}^{+1.06}$ & $7.64_{-0.26}^{+1.02}$ & $-21.63_{-0.16}^{+0.19}$ &  \\[1ex]
UVISTA--213 & $7.39_{-0.14}^{+0.12}$ & $7.43_{-0.16}^{+0.13}$ & $-21.61_{-0.11}^{+0.12}$ & Y3 \\[1ex]
UVISTA--839 & $8.12_{-0.48}^{+0.52}$ & $7.96_{-0.30}^{+0.75}$ & $-21.61_{-0.16}^{+0.19}$ &  \\[1ex]
UVISTA--598 & $8.19_{-0.52}^{+0.58}$ & $8.16_{-0.48}^{+0.67}$ & $-21.53_{-0.18}^{+0.21}$ & Y10 \\[1ex]
UVISTA--953 & $7.64_{-0.16}^{+1.27}$ & $7.69_{-0.18}^{+1.27}$ & $-21.47_{-0.16}^{+0.19}$ & Y16 \\[1ex]
UVISTA--919 & $7.68_{-0.18}^{+0.79}$ & $7.71_{-0.18}^{+0.86}$ & $-21.45_{-0.14}^{+0.16}$ &  \\[1ex]
UVISTA--356 & $7.67_{-0.11}^{+0.21}$ & $7.68_{-0.11}^{+0.22}$ & $-21.39_{-0.15}^{+0.17}$ &  \\[1ex]
UVISTA--266 & $7.54_{-0.24}^{+0.21}$ & $7.60_{-0.29}^{+0.19}$ & $-21.37_{-0.23}^{+0.28}$ &  \\[1ex]
UVISTA--634 & $7.67_{-0.28}^{+0.25}$ & $7.67_{-0.28}^{+0.30}$ & $-21.18_{-0.17}^{+0.20}$ &  \\

\hline
\end{tabular}\label{table:muv}
\end{table}

\subsection{Completeness simulations}
We perform a full simulation of our selection process by injecting and recovering fake high-redshift sources into the imaging data.
This process allows an estimate of the incompleteness of our selection methodology.
In this study we use predominantly optical and near-infrared imaging over the XMM-LSS and COSMOS extragalactic fields as shown in Figs.~\ref{fig:xmm} and~\ref{fig:cosmos}.
To accurately derive the comoving number density of galaxies from our data it is necessary to take into account the differing image depths across the two fields.
To do this we simulate six separate regions of the fields which are detailed in Table~\ref{table:areas}. 
Fake sources were created with a realistic rest-frame UV slope (mean $\beta = -2.0$), with a Gaussian scatter of $\Delta \beta = 0.2$~\citep{Rogers2014}.
We injected these sources as point sources~\citep{Bowler2014} into all of the available ground-based images within each sub-field and recovered them following the steps undertaken for real sources. 
The proportion of fake galaxies that passed this selection procedure were then used to estimate the completeness as a function of $M_{\rm UV}$ and $z$, which was then folded into the LF calculation. 
The absolute magnitude distribution of injected fake sources was calculated according to an assumed underlying LF.
We ran simulations assuming both a DPL function form derived in~\citet{Bowler2015}, and a Schechter function form with the parameters and redshift evolution of~\citet{Bouwens2015}.
As expected, the completeness derived from these two assumed LF functional forms were comparable for the faintest sources in our sample, where the DPL and Schechter forms converge.
For the brightest sources (at $M_{\rm UV} < -21$) the results differ due to the steep exponential slope of the Schechter function, which results in a larger contribution from up-scattered sources and the derived ``completeness'' typically exceeding one.
If the underlying functional form was Schechter, then applying this correction would bring the observed excess down, and the data points would reflect the steep exponential decline.
When we do this however, we still find an excess of sources bright-ward of the knee, suggestive of a deviation from this function form.  
Hence in our final LF, we calculate the binned points assuming the completeness derived from an underlying DPL function.

\subsection{Binned results}

We calculate the binned LF data points, $\Phi(M_{\rm UV})$, from our sample using the classic $1/V_{\rm max}$ estimator~\citep{Schmidt1968}.
Here the fiducial volume that each galaxy could occupy is given by the shell between the limits of the redshift bin (e.g. $8.5 < z < 9.5$).
The upper redshift limit is then modulated according to the point at which that object would be undetected in our selection, taken as when the redshifting of the galaxy SED causes it to drop below the $5\sigma$ limit of the detection band.
Hence for the faintest objects in a given dataset the $V_{\rm max}$ is lower than that for brighter objects, which typically could be recovered in the full redshift range of the selection.
The incompleteness is taken into account by effectively reducing this volume by $1/C(M_{i}, z_{i})$ where $C$ is the completeness as a function of the absolute magnitude and redshift of each galaxy (and is determined from the simulations described previously).

\begin{table}
\centering
\caption{The rest-frame UV LF data points derived in this work at $z = 8$ and $z  = 9$.  The first column gives the central redshift, where we take the bin width to be $\delta z = 1.0$.  The second and third columns show the absolute UV magnitude of the bin and the bin width.  The final column shows the derived comoving number density of galaxies.}\label{table:binned}
\begin{tabular}{cccc}
\hline
Redshift & $M_{\rm UV}$ &$\Delta\,M$ & $\phi$\\
& /mag & /mag & $/10^{-6} /{\rm mag}/{\rm Mpc}^{3}$\\
\hline
8 & $-21.65$ & $0.5$ & $2.95 \pm 0.98$\\
8 & $-22.15$ & $0.5$ & $0.58 \pm 0.33$\\
8 & $-22.90$ & $1.0$ & $0.14 \pm 0.06$\\
\hline
9 & $-21.9$ & $1.0$ & $0.84 \pm 0.49$\\
9 & $-22.9$ & $1.0$ & $0.16 \pm 0.11$\\

\hline
\end{tabular}
\end{table}

\begin{figure}
\includegraphics[width = 0.49\textwidth]{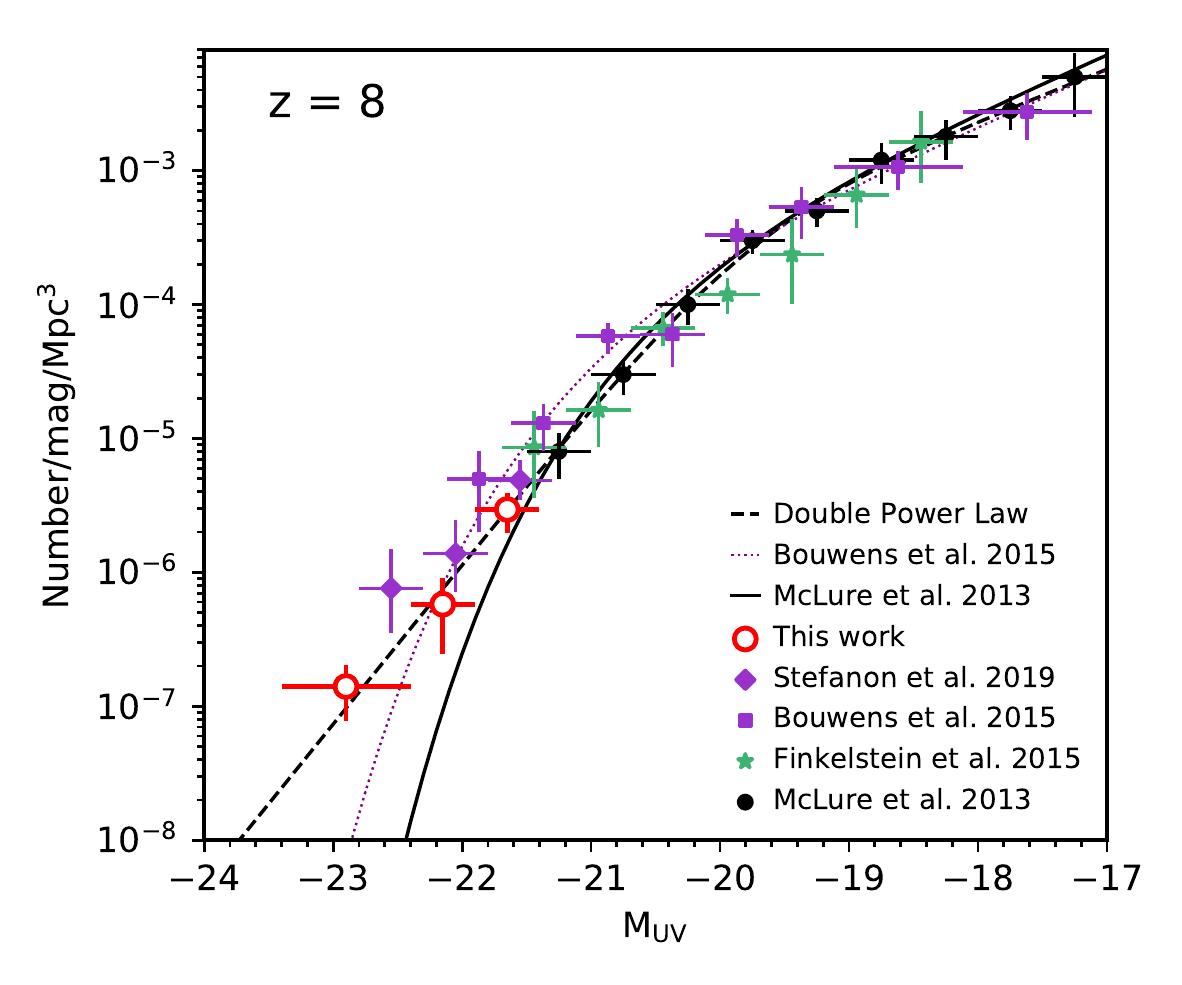}\\
\includegraphics[width = 0.49\textwidth]{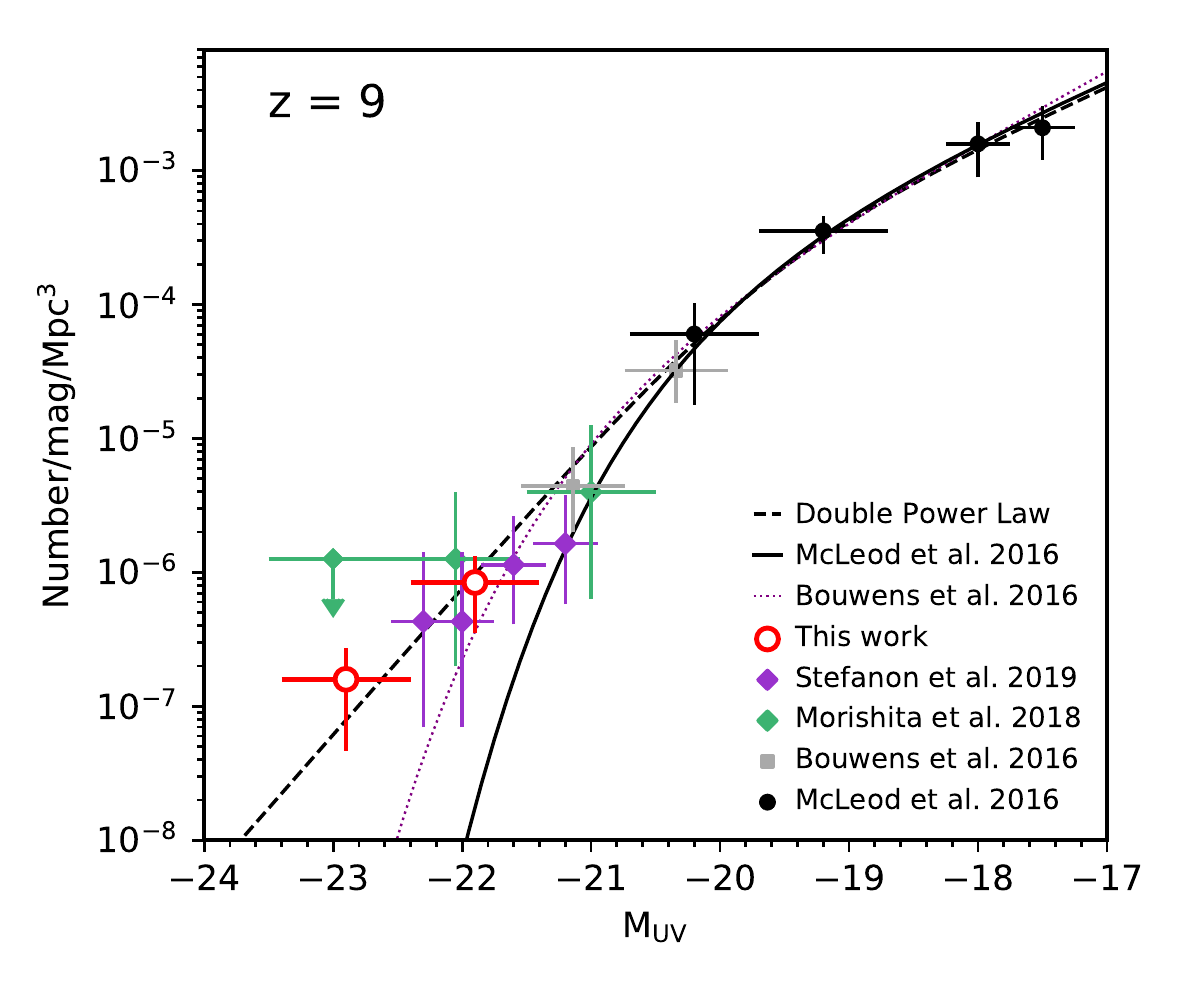}
\caption{The rest-frame UV LF at $z = 8$ and $z = 9$.
The results of this work are shown as the open red points, with data points from previous studies shown as detailed on each plot.
We extend to brighter absolute magnitudes than previous studies as a consequence of the larger area of near-infrared data.
The black dashed line is the best-fitting DPL function fitted to our data combined with the fainter results of~\citet{McLure2013}~\citep{McLeod2016} at $z = 8$ ($z = 9$).
We show the best-fitting Schechter functions from~\citet{McLure2013} and~\citet{McLeod2016} as the black solid lines, and the fits from~\citet{Bouwens2015, Bouwens2016} as the purple dotted lines,
}\label{fig:lf}
\end{figure}

For our final rest-frame UV LF results we combine the XMM-LSS and COSMOS samples to span the full range in absolute UV magnitude probed by the different depths of data in the two fields.
We calculate the $z = 8$ and $z = 9$ LF bright-ward of $M_{\rm UV} = -21.4$, as this is the magnitude at which our simulations demonstrate that we become more than 50 percent incomplete.
At $z = 8$ we compute the results in bins of width $0.5\,{\rm mag}$ near to the faint cut-off.
For the brightest bin at $z = 8$ and for the two bins at $z = 9$, we use larger bin widths of $1.0\,{\rm mag}$ to account for the smaller number of objects in these magnitude and redshift ranges.
The resulting binned points are shown in Fig.~\ref{fig:lf} and are tabulated in Table~\ref{table:binned}.
The wide area we probe using the ground-based XMM-LSS field in combination with COSMOS enables us to determine the number density of LBGs as bright as $M_{\rm UV} \simeq -23$ for the first time at these redshifts.
At $z \simeq 8$, where our three bins span $\simeq 1.5\,{\rm mag}$ in absolute UV magnitude, we see a clear decline in the number density of bright galaxies by more than a factor of ten.
However, the decline we see is not as rapid as expected from the Schechter function fits of previous studies, extrapolated to brighter magnitudes.
Our results at $M_{\rm UV } \lesssim -22$ are significantly in excess of the Schechter fits from~\citet{McLure2013} and~\citet{McLeod2016} at $z = 8$ and $z = 9$ respectively. 
In comparison to the fits from~\citet{Bouwens2015} and~\citet{Bouwens2016}, which find a brighter characteristic magnitude than the~\citet{McLure2013} and~\citet{McLeod2016} studies, we still find an excess of sources around $M_{\rm UV } \simeq -23$.
We also fit a DPL function to our results (combined with fainter studies as described in Section~\ref{sect:fit}), and show this as the dashed line.
This function appears to better reproduce the decline we see at the bright-end for these redshift bins.
When comparing the binned results at $z \simeq 8$ and $z \simeq 9$, we do not see strong evolution in the number densities at the absolute magnitudes probed by this study.
We find fewer sources at $z \simeq 9$ than at $z \simeq 8$ but because the volume for selection is smaller at $z \simeq 9$ (due to the requirement for deep $H$-band imaging which is only satisfied in the deeper COSMOS and UDS regions) the derived LF is similar between the two bins.
We discuss the inferred number density of $z \simeq 10$ LBGs from our single candidate in this bin in Section~\ref{sect:z10}.

\subsubsection{Comparison to Previous Studies}
Our derived LF points are consistent (within the errors) with previous results in the magnitude regime where they overlap.
The most comparable work to this study was undertaken by S19, who searched for $z > 7$ LBGs in the COSMOS field using the shallower UltraVISTA DR3 data.
As described in Section~\ref{sect:S19} we reselect $\sim 50$ percent of their sample as high-redshift objects.
When comparing our LF points at $z \simeq 8$ we find a number density that is approximately a factor of two lower, in agreement with the direct sample comparison.
Note that when calculating the $z \sim 9$ LF S19 use a subset of their full sample with $z_{\rm phot} > 8.6$.
Hence the five objects which they assign to the $z \simeq 9$ bin are also included in the $z \simeq 8$ results.
If instead their sample was split by best-fitting photometric redshift as we do in this study, the evolution they find in the number counts between $z = 8$--$9$ would be reduced. 
The~\emph{HST} pure-parallel program BoRG has provided wide-area NIR imaging for the selection of high-redshift galaxies over multiple sight-lines.
With filter-sets especially designed for selection $z \simeq 8$ and $z \simeq 9$ sources, the BoRG and BoRG[z9] surveys have covered an area of approximately $\simeq 350\,{\rm arcmin}^2$ each, allowing constraints on the bright-end of the LF.
At $z = 9$,~\citet{Morishita2018} used the full BoRG[z9] survey to search for galaxy candidates, finding two sources consistent with being at this redshift.
As shown in Fig.~\ref{fig:lf} their derived number density is consistent with our results, although the small number of objects (two sources in three magnitude bins) makes the errors on the binned points particularly large.

\subsection{Schechter and DPL function fitting}\label{sect:fit}

\begin{table}\caption{The DPL and Schechter function best-fit parameters derived in this study.
In the fitting we combined our results at bright magnitudes with the~\citet{McLure2013}~\citep{McLeod2016} results at $M_{\rm UV} > -21$ at $z = 8$ ($z = 9$).
The first column denotes the redshift in question.
This is followed by the best-fitting $\phi^*$, $M^*$ and $\alpha$ parameters.
The final column shows the best-fitting bright-end slope ($\beta$) for the DPL parameterisation, which is shown in the upper row for each redshift.
The faint-end slope at $z = 9$ for the DPL fit is denoted by an asterisk to signal that it was fixed in our fitting analysis.
}\label{tab:lfparam}
\centering
\begin{tabular}{c c c c c}
\hline
$z$&$\phi^*$ & $M^*$ & $\alpha $ & $\beta$ \\
&$/10^{-4} /{\rm mag}/{\rm Mpc}^{3}$ & $/{\rm mag}$ & & \\
\hline
8 & $4.83\pm2.25$ & $-19.80\pm0.26$ & $-1.96\pm0.15$ & $-3.98\pm0.14$\\
8 &$1.92\pm1.07$ & $-20.48\pm0.23$ & $-2.18\pm0.16$ & --\\
\hline

$9$ & $2.85\pm1.39$ & $-19.67\pm0.33$ & $-2.10^*$ & $-3.75\pm0.22$\\
9& $0.53\pm0.56$ & $-20.80\pm0.43$ & $-2.31\pm0.24$ & --\\

\hline

\end{tabular}
\end{table}

\begin{figure}
\includegraphics[width = 0.49\textwidth]{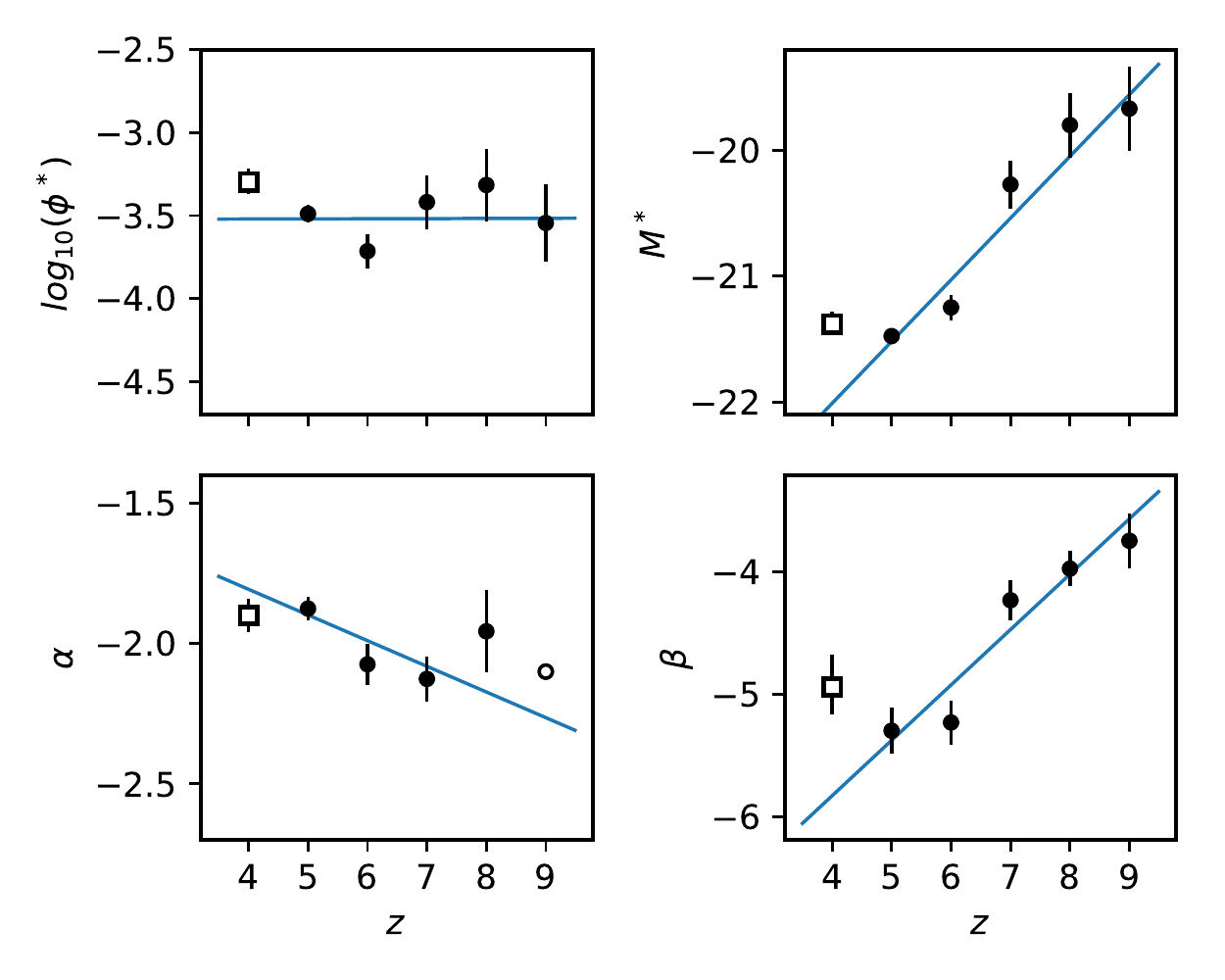}\\

\caption{The evolution of the four DPL parameters from $z = 5$--$9$ derived in our fitting analysis.
The best-fit value and error were derived at each redshift by fitting to a compilation of LF measurements as described in the text.
The blue lines in each plot show the linear fit to these results.
We fit to all points at $z \ge 5$ except the faint-end slope at $z = 9$, which cannot be constrained by our data and is shown as an open circle.
The open squares at $z \simeq 4$ show the results of an independent analysis by~\citet{Adams2019} and is shown for comparison.
The equivalent plot for a Schechter function fit is presented in Fig.~\ref{fig:paramsSH}.
}\label{fig:params}
\end{figure}

To determine the best-fit parameters for the DPL and Schechter functional forms we combine our results with data points at fainter magnitudes derived in previous studies.
We include the results of~\citet{McLure2013} and~\citet{McLeod2016} for the fitting at $z = 8$ and $z = 9$ respectively.
These studies used a similar photometric-redshift fitting methodology as that utilized in this work.
We checked that our conclusions are unchanged if we include the other available studies over the same range in absolute magnitude.
We determined the best-fitting parameters using least squares fitting, with the one dimensional errors on the parameters derived by minimising the $\chi^2$ over the other free parameters.
We present the best-fitting DPL and Schechter function parameters and errors in Table~\ref{tab:lfparam}, and show the best-fitting DPL fit as the dashed lines in Fig.~\ref{fig:lf}.
We also fit Schechter and DPL functions to a selection of data at $z = 5$--$7$ to derive the evolution of the fit parameters.
Here we include the results of~\citet{Ono2018} that probe predominantly bright-ward of the LF knee.
At $z = 6$ and $z = 7$ we add our previous results from~\citet{Bowler2015, Bowler2014}.
For the faint-end of the LF we include the results of~\citet{Bouwens2015} at $z = 5$ and $z = 6$, and~\citet{McLure2013} at $z = 7$.
We stress that our results are not sensitively dependent on which studies we choose to fit.
All parameters are allowed to be free in the fitting process, except for the faint-end slope of the DPL at $z = 9$, which we cannot constrain from our data.
Here we fix the slope to the best-fitting value found for a DPL fit at $z = 6$ and $z = 7$ ($\alpha = -2.1$;~\citealp{Bowler2015}).
Once the best-fit parameters and errors have been derived for each redshift bin and functional form, we combine these results to derive a simple linear evolution model.
We fit the parameters derived for each redshift between $z = 5$ to $z = 9$, and measure the gradient and intercept of $\phi$, $M^*$, $\alpha$ and $\beta$ (for DPL) over this range.
The results and the linear fits are shown in Fig.~\ref{fig:params} for the DPL form.
The evolution can be described by the following equations, with reference to $z = 6$:

\begin{equation}
\begin{gathered}
\phantom{{\rm log}_{10}(0)}M^* = (-21.03 \pm 0.49) + (0.49 \pm 0.09) \,(z - 6)\\
{\rm log}_{10}(\phi^*) = (-3.52 \pm 0.32) + (0.00 \pm 0.06)\,(z - 6)\\
\phantom{{\rm log}_{10}(0)}\alpha = (-1.99 \pm 0.29) - (0.09 \pm 0.05)\,(z - 6)\\
\phantom{{\rm log}_{10}()}\beta = (-4.92 \pm 0.60)+ (0.45 \pm 0.08)\,(z - 6)
\end{gathered}
\end{equation}
The results of this simple analysis show that with a DPL form the LF evolution is dominated by changes in $M^*$ and $\beta$, with $\alpha$ and $\phi^*$ showing little change.
In Fig.~\ref{fig:params} we also show the best-fitting parameters at $z \simeq 4$ derived in an independent analysis by~\citet{Adams2019}.
Their analysis show a similarly bright $M^*$ and steep bright-end slope ($\beta \simeq -5.0$) to our fitting results at $z \simeq 5$.
As we discuss in Section~\ref{dis:ev}, at $z \simeq 4$ the impact of AGN on the bright-end of the LF becomes significant.
However, even with this added complication, the fits of~\citet{Adams2019} show best-fitting $M^*$ and $\beta$ values that are lower than our $z \gtrsim 7$ results, while the $\phi^*$ and $\alpha$ values are comparable in agreement with our proposed evolution.
We perform an identical analysis assuming a Schechter function.
The evolution of the parameters are shown in Appendix~\ref{app:SH}.
The assumption of a Schechter function dramatically changes the derived form of the evolution, with the best-fitting absolute magnitude becoming nearly constant at $M^* \simeq -21$ while the $\phi^*$ and $\alpha$ parameters strongly evolve.
We discuss the implications of these different evolutionary scenarios in the next section.

\section{Discussion}\label{sect:discussion}
In this study we searched for $z \gtrsim 7.5$ LBGs within $\sim 6\,$\ds~of optical, near and mid-infrared imaging in the XMM-LSS and COSMOS fields.
The result was a sample of 27 candidate galaxies with best-fitting photometric redshifts in the range $7.4 \lesssim z \lesssim 9.1$, and one extremely bright candidate $z \sim 10$ LBG in the XMM-LSS field.
With this sample we computed the rest-frame UV LF and used this, in combination with data at $z = 5$--$7$, to derive the shape and evolution of the best-fitting function parameters in the range $z \simeq 5$--$10$.

\subsection{Shape of the rest-frame UV LF at \boldmath$z = 8$--$9$}\label{sect:shape}
With the advent of sufficiently deep~\emph{and} wide-area surveys to select samples of luminous ($L > L^*$) galaxies at high redshifts, there has been an increased discussion on the functional form of the rest-frame UV LF.
Prior to surveys from VISTA, UKIRT and HSC, the observed LFs at $z > 6$ were derived almost exclusively from~\emph{HST} data covering at most $0.2\,$\ds~(e.g from CANDELS).
These results were well described by a Schechter functional form, as the lack of galaxies at bright magnitudes, in addition to the larger errors in these bins, permitted an exponential decline bright-ward of the knee in the number counts.
The UltraVISTA survey has revolutionised the study of the very bright-end of the $z \simeq 7$ LF, as it crucially provided deep $Y$-band data that probes just red-ward the Lyman-break at these redshifts, on a degree-scale for the first time.
In~\citet{Bowler2012, Bowler2014} we presented a sample of very bright $z \simeq 7$ LBGs selected predominantly from the UltraVISTA data, which provided the first evidence for an excess in the number density above that expected from the previously assumed Schechter function.
In subsequent deeper data releases, the high-redshift nature of these sources has been confirmed with significantly deeper optical to near-infrared photometry~\citep{Bowler2018}.
Furthermore, several of these extremely luminous $z \simeq 7$ sources have now been spectroscopically confirmed with ALMA (e.g.~\citealp{Hashimoto2019}, Schouws et al. in prep.), thus strengthening the conclusion that the rest-frame UV LF at this redshift deviates from a Schechter from.
In this study we find evidence that a shallower functional form continues out to $z \simeq 8$ and $z \simeq 9$ (and potentially even $z \simeq 10$, see Section~\ref{sect:z10}).
A number of the very bright LBG candidates we present in this work were selected from regions of shallower optical data, as a consequence of the wide-areas needed to find them.
This could lead to a higher rate of contamination in these sub-samples.
Even in a pessimistic case of high contamination, leading to the confirmation of only one or two sources in these bins, this would still significantly challenge a Schechter function decline.
At lower redshifts there is now additional evidence for a deviation from a Schechter function.
Using the HSC SSP data,~\citet{Ono2018} found that the $z \simeq 4$--$7$ LFs show an excess of very bright-galaxies and are preferentially fit with a DPL or lensed Schechter function.
The excess of sources at $M_{\rm UV} \simeq -24$ from this study can be seen in Fig.~\ref{fig:lf}, in comparison to our previous work at $z \simeq 6$ and $z \simeq 7$.

In light of this evidence from previous studies, and our new results at $z > 7$, it is pertinent to discuss what functional form is to be expected for the rest-frame UV LF at very high redshift.
In the local Universe, the mass and rest-frame optical luminosity functions of galaxies can be well described by single or double Schechter functions (e.g.~\citealp{Peng2010, Baldry2012, Loveday2012}).
However, when measurements have been made using a waveband that probes recent SF rather than mass, several studies have found a shallower decline than expected from a Schechter function at the high-luminosity end.
For example, measurements of the far-IR LF (e.g.~\citealp{Soifer1987}) and~\emph{dust-corrected} near-UV~\citep{Jurek2013} and H$\alpha$~\citep{Gunawardhana2013} LFs, have all shown deviations from a Schechter form.
Scatter in the mass-to-light ratio of galaxies, for example due to stochastic star formation, can naturally explain this observed shallower decline for SFR-based LFs as opposed to mass functions (MFs) in these studies.
\citet{Salim2012} and more recently~\citet{Ren2019} have theoretically demonstrated this effect, showing that scatter in SFR as a function of galaxy or halo mass causes a shallower decline in LF measurements that trace the galaxy SFR.
Given this theoretical prediction, why is it that rest-frame UV LFs at intermediate redshifts ($2 \lesssim z \lesssim 5$;~\citealp{Parsa2016, Shapley2011, VanderBurg2010}) show an apparent Schechter function form, despite tracing the recent SFR of the galaxies in question?
An answer to this question may be found by inspecting the results of galaxy formation simulations and models.
The majority of these models initially over-predict the number of luminous galaxies in the rest-frame UV LF, potentially due to the effect of scatter between SFR and mass (e.g.~\citealp{Cai2014, Genel2014, Henriques2014, Paardekooper2013}).
The models are then brought into agreement with the observed number densities with the addition of significant dust attenuation (see discussion in~\citealp{Bowler2015}).
Therefore from both theoretical arguments and the results of simulations, it is expected that without the effects of dust, the rest-frame UV LF should have a shallower decline at the bright-end, inconsistent with a Schechter function form.
While the presence of dust in intermediate redshift LBGs is expected and has been comprehensively measured (e.g.~\citealp{Fudamoto2017, McLure2017}), the same is not true at the very high-redshifts considered in this study.
Depending on the dust formation mechanism, it is argued that early galaxies have limited dust (e.g.~\citealp{Michalowski2015}).
Indeed, low dust attenuation is often assumed for high-redshift galaxies and is what is expected from the evolution of the colour-magnitude relation~\citep{Rogers2014, Bouwens2015}.
We therefore would expect the observed rest-frame UV LF to approach a power-law like form at the bright-end as the effects of dust become less significant.
While there have been direct observations of dust continuum emission from $z \gtrsim 7$ LBGs (e.g.~\citealp{Tamura2019, Bowler2018, Laporte2017}), the derived dust masses in these sources are reduced compared to low-redshift observations, because of the higher assumed dust temperature (e.g.~\citealp{Hashimoto2019}). 
In addition to the effect of reduced dust at the highest redshifts, there is reason to believe that the underlying MF of galaxies during this epoch is shallower than observed in the local Universe.
In the phenomenological model presented in~\citet{Peng2010}, the exponential decline in the number of massive galaxies at low-redshift is a result of a characteristic quenching stellar mass ($M_{\star} = 10^{10.2}\,{\rm M}_{\odot}$) above which SF, and hence mass-growth, is halted.
The expected stellar masses of the galaxies we find are significantly lower than this quenching mass (e.g.~\citealp{Bowler2014}), and hence it is reasonable to assume that the stellar mass function at this time has a different form.
The detection of the very bright star-forming galaxies in this work suggests that we may be observing this transition into an era before mass quenching and significant dust attenuation.

One other potentially important effect on the observed shape of the bright-end of the LF is the role of magnification bias.
In the case of a steeply declining galaxy LF, gravitational lensing can have a significant effect on the number of very luminous sources detected.
For example, if an underlying Schechter function is assumed for high-redshift galaxies, all sources detected bright-ward of $M_{\rm UV} \simeq -23$ are strongly lensed objects~\citep{Mason2015, Barone-Nugent2015a}.
As in our previous works at $z = 6$ and $z = 7$, we directly measured the gravitational lensing of our sources using a simple model of the magnification from foreground galaxies in our images~\citep{Bowler2014, Bowler2015}.
We find no evidence that the brightest sources are preferentially lensed compared to a random sky position.
The typical magnification due to foreground galaxies was a brightening of $0.1$--$0.4{\rm mag}$, and this was found to be uncorrelated with the observed magnitude of the source.
We therefore exclude strong lensing as a cause of the observed shape of the LF.
Even when demagnifying the sources at $z \simeq 7$, we still find an excess in the number of the brightest galaxies compared to the Schechter function prediction~\citep{Bowler2014}.
The importance of the magnification bias on the observed LF depends on the steepness of the underlying function.
We have argued that the rest-frame UV LF at very high-redshifts is expected to be shallower than the typically assumed Schechter function.
In the case of a DPL or power-law form, the effects of lensing will be significantly reduced, particularly in the magnitude ranges probed by this study (see figure 13 in~\citealp{Mason2015}).
The direct measurement of a low magnification for the brightest objects in our sample thus further supports a more gentle decline in the bright-end of the LF than expected from a Schechter function.

\begin{figure}
\includegraphics[width = 0.49\textwidth]{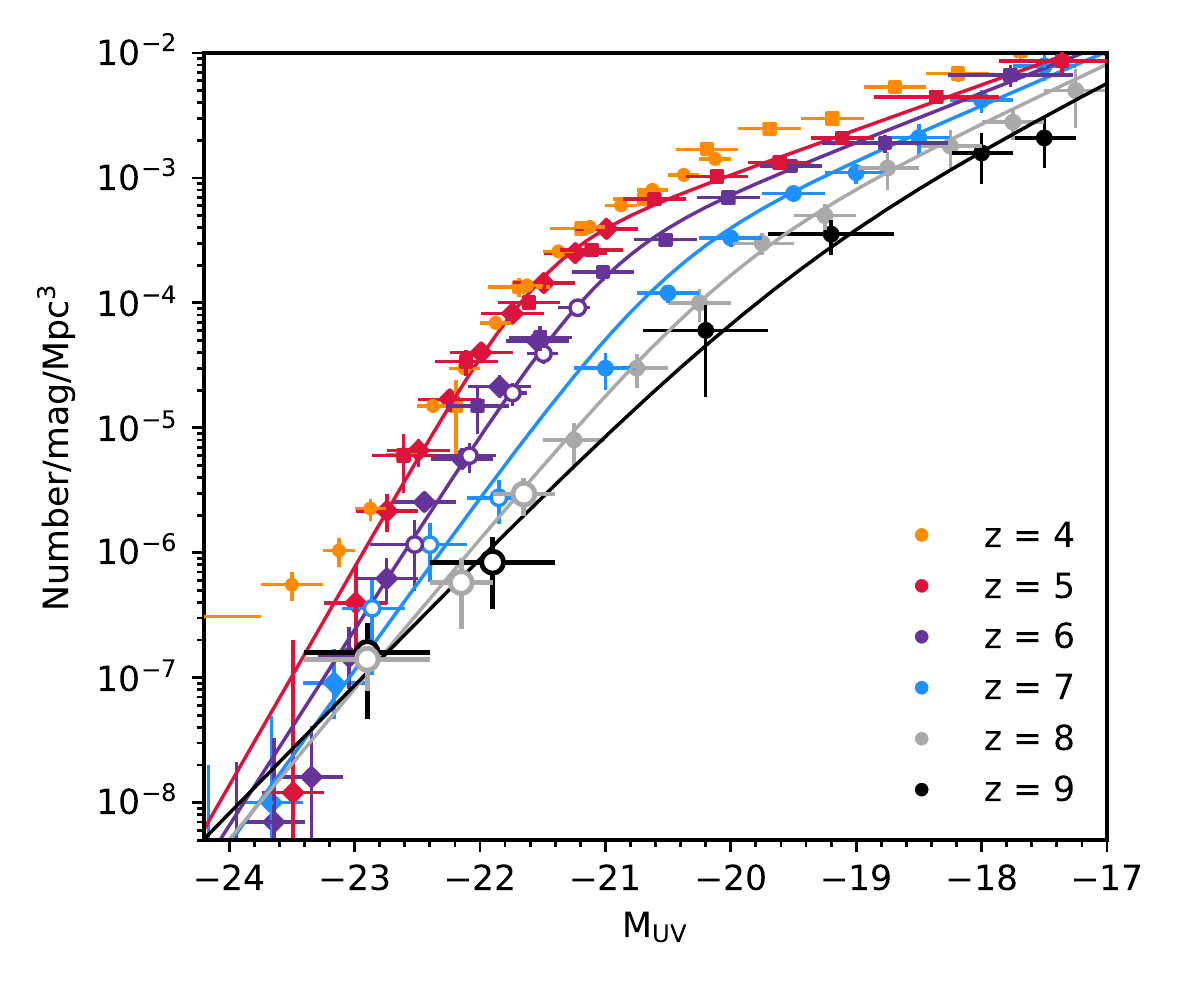}
\caption{The evolution of the rest-frame UV LF from $z = 4$ to $z = 9$ as described by our evolving DPL model.
The results of this study at $z = 8$ and $z = 9$ are shown as the open grey and black points respectively.
The lines show the derived DPL parameterisation from fitting to the data shown as described in the text.
Our previous work at $z = 6$ and $z = 7$ is shown with the open purple and blue circles respectively~\citep{Bowler2015, Bowler2014}.
A lack of evolution is seen at the very bright-end from our studies and also from~\citet{Ono2018} (diamonds).
At $z = 4$, we show the data points from~\citet{Adams2019} and~\citet{Bouwens2015}.
The excess observed at the bright-end at this redshift is due to the presence of AGN.
In the higher redshift bins AGN are sufficiently rare that they make a negligible impact according to recent evidence for accelerated evolution~\citep{Jiang2016}.
}\label{fig:ev}
\end{figure}

\subsection{Form of the LF evolution from \boldmath$z \simeq 5$--$10$}\label{dis:ev}
In Fig.~\ref{fig:ev} we show a comparison between the observed rest-frame UV LF data points from $z = 4$--$9$ and our evolving DPL model.
Remarkably, the LF is now measured over six magnitudes even at $z \simeq 9$.
From the data alone, it is clear that there is a rapid change in the number density of star-forming galaxies over this epoch ($z \simeq 5$--$9$; $\sim 800\,{\rm Myr}$), and that this evolution predominantly happens around the knee of the function at $M_{\rm UV} \simeq -21$.
In this study we have focused on determining the bright-end of the very high-redshift LF.
Between $z \simeq 8$ and $z \simeq 9$ we find no evidence for a change in the number density of the brightest galaxies ($M_{\rm UV} \simeq -23$).
It is clear from Fig.~\ref{fig:ev} however, that this lack of evolution is also observed down to $z \simeq 5$, as seen in our previous studies~\citep{Bowler2014, Bowler2015} and at brighter magnitudes in the study of~\citet{Ono2018}.
The results of our DPL function fitting in Section~\ref{sect:fit} demonstrate that this observed lack of evolution at the very bright-end is a result of a change in shape in the rest-frame UV LF over this redshift interval.
As we have argued in the previous sub-section, this conclusion is theoretically motivated by a change in the presence of dust in galaxies over this timescale, such that between $z \simeq 9$ and $z \simeq 5$ the rest-frame UV LF transitions from being a DPL-like function to being better described by a Schechter function.
One added complication when considering the shape of the $z \simeq 2$--$4$ LF is the presence of high-redshift AGN that have comparable number densities to LBGs at $M_{\rm UV} \simeq -23$.
These faint AGN show similar broad-band colours to LBGs and `contaminate' the measurement of the galaxy UV LF, leading to a boost in the bright-end of the function that must be accounted for~\citep{Adams2019,Ono2018,Stevans2018, Bian2013}).
At $z > 5$ the quasar LF is observed to rapidly drop ($\phi \propto 10^{\,k(1+z)}$, $k = -0.72$;~\citealp{Jiang2016}), making this effect insignificant at the magnitudes probes by this study (see~\citealp{Bowler2014} for further discussion).
The presence of AGN at $z \simeq 4$ is clear visible in Fig.~\ref{fig:ev} at $M_{\rm UV} < -23$.

The lack of evolution we observe at the bright-end of the $z > 7$ LF is a consequence of the general evolution we see at $z = 5$--$7$ in which the bright-end slope steepens and $M^*$ brightens with time.
The additional freedom in the bright-end slope that is granted in the DPL formalism, and the addition of our bright LF points, results in a different evolutionary scenario than previous studies that have typically assumed a Schechter function form.
In the last decade, there has been continued discussion on which parameters drive the LF evolution at high redshift, for example whether changes in $M^*$ (e.g.~\citealp{Bouwens2011, McLure2009}) or $\phi^*$ (e.g.~\citealp{VanderBurg2010}) are dominant.
Recent analyses by~\citet{Bouwens2015} and~\citet{Finkelstein2015}, who used a compilation of~\emph{HST} data, concluded instead that the LF evolves predominantly by $\phi^*$-evolution between $z \simeq 4$--$8$ with the absolute UV magnitude appearing to remain  constant at $M^* \simeq -21$.
In contrast, by allowing the functional form to change over the range $z \simeq 5$--$7$, we found evidence for changes in $M^*$ over this epoch~\citep{Bowler2015}.
Here we have extended this analysis and have shown that an evolving DPL formalism holds up to $z \simeq 9$.
We also exploited our function fitting framework to explore what evolution we would derive if we assumed a Schechter function when fitting our compilation of data.
The results of this analysis are presented in Appendix~\ref{app:SH}.
Interestingly, the resulting evolution from this procedure is a predominantly $\phi^*$-evolution, consistent with the previous results from~\citet{Bouwens2015,Finkelstein2015}.
This comes about due to the approximately constant number density of $M_{\rm UV} \simeq -23$ sources, coupled with the hard exponential cut-off imposed in the Schechter function formalism.
The results of this analysis do not adequately reproduce the observed LF points at $M_{\rm UV} \lesssim -22$ at $z > 7$ however, as can be seen in Fig.~\ref{fig:evSH}.
In light of our new observations, and the compilation of other studies at bright-magnitudes at $z = 5$--$7$, we argue that the previously derived $\phi^*$-evolution was caused by the fitting of a different and possibly not correct function to the data.
While the relatively steep bright-end slope at $z \simeq 5$--$6$ can be equally well reproduced by a DPL or Schechter function~\citep{Bowler2015}, at $z > 6$ the DPL better describes the drop-off as demonstrated in Fig.~\ref{fig:ev}.
As we have discussed in Section~\ref{sect:shape}, the observed change in the shape of the rest-frame UV LF can be explained as the combined effect of a lack of mass quenching and a lack of dust obscuration in $z > 7$ star-forming galaxies.

\begin{figure}
\includegraphics[width = 0.49\textwidth]{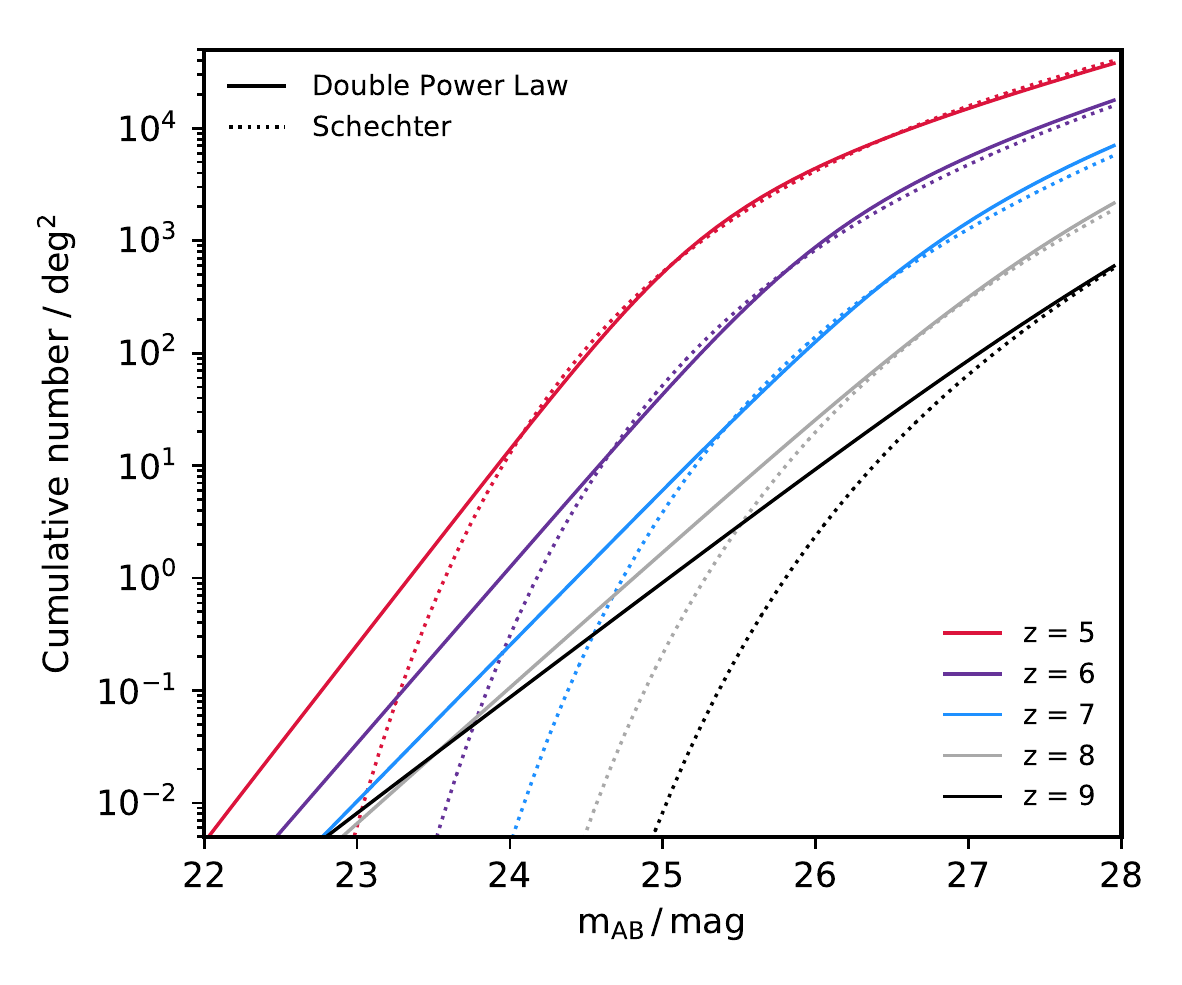}
\caption{The cumulative number density of LBGs from $z \simeq 5$ to $z \simeq 9$ derived from our fitting analysis.
At each redshift we assume a bin of width $\delta z = 1.0$.
The DPL and Schechter function results are shown as the solid and dotted lines respectively.
Note that this calculation assumes 100 percent completeness in the selection, and hence represents a maximal yield per \ds~at each $m_{\rm AB}$ value.
}\label{fig:euclid}
\end{figure}

Our results have strong implications for the yield of upcoming wide-area near-infrared surveys from~\emph{Euclid} and~\emph{WFIRST}.
In Fig.~\ref{fig:euclid} we compute the cumulative number density of galaxies as a function of apparent magnitude with our evolving DPL and Schechter function parameterization.
The DPL formalism dramatically increases the predicted number of very bright $z > 6$ LBGs over the Schechter function predictions.
For reference the~\emph{Euclid} satellite will provide $\sim 40\,$\ds~of $YJH$ data to a $5\sigma$ depth of $m_{\rm AB} = 26.0$ as part of the deep survey component, and $\sim 15,000\,$\ds~to a depth of $m_{\rm AB} = 24.0$ in the wide component.
If our derived DPL formalism is an accurate representation of the galaxy number counts at $z > 7$, then we expect numerous detections (thousands) of very bright LBGs at these redshifts even in the wide survey from~\emph{Euclid}.

\subsection{The existence of very bright \boldmath$z > 9$ LBGs}\label{sect:z10}

\begin{figure}
\includegraphics[width = 0.49\textwidth]{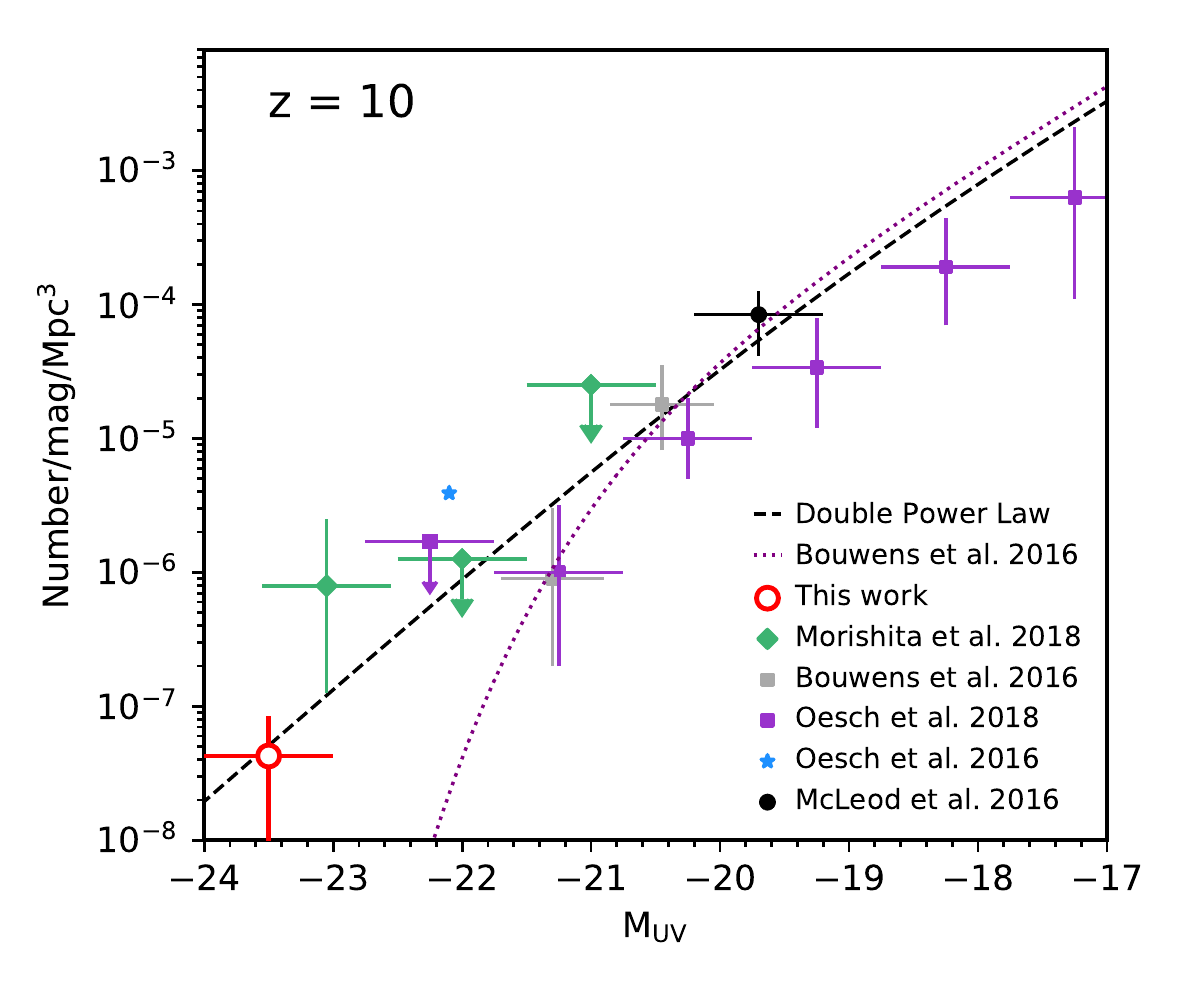}
\caption{A compilation of derived rest-frame UV LF results at $z \simeq 10$.
The open red circle shows the estimated number density of very bright LBGs at this redshift derived from our single LBG candidate with a best-fitting photometric redshift of $z = 10.9 \pm 1.0$.
We also show the estimated number density of the $z = 11.1$ source from~\citet{Oesch2016}.
Note that the~\citet{Morishita2018} results were calculated with a non-standard $\Delta z = 1.2$, hence the volumes here should be a considered a lower limit in this comparison.
The dashed line shows our DPL model extrapolated to $z = 10$, and the dotted purple line shows the Schecter function fit from~\citet{Bouwens2016}.
}\label{fig:lfz10}
\end{figure}

Using our formalism for the evolution of the rest-frame UV LF at $z \ge 5$ we can speculate on the prevalence of very bright $z > 9$ LBGs.
The predicted DPL function at $z = 10$ from our evolution parameterization is shown in Fig.~\ref{fig:lfz10}, along with the results of previous studies.
The extrapolated $z = 10$ parameters are $M^* = -19.07$, $\phi^* = 3.05 \times 10^{-4}\,/{\rm mag}/{\rm Mpc}^3$, $\alpha = -2.36$ and $\beta = -3.12$.
We estimate the number density of our single $z \simeq 10$ source using a LF bin at $M_{\rm UV} = -23.5$.
The candidate LBG, XMM3-3085, has a photometric redshift of $z_{\rm phot} = 10.85^{+1.00}_{-1.02}$.
If confirmed as a high-redshift source, this object would be the most luminous LBG known at $z > 7$ with a $M_{\rm UV} \simeq -23.7$.
The detection of one such source over our survey area however, is not unexpected from our evolving DPL model.
The extrapolated DPL almost exactly matches the derived number density of this single source.
The expected number density of AGN at this $M_{\rm UV}$ and redshift is a factor of $1000$ times lower than of this source assuming the evolving DPL model of~\citet{Jiang2016}.
\citet{Morishita2018} also presented one surprisingly bright $z \simeq 10$ candidate found within the BoRG[z9] survey.
The source, $2140+0241-303$, has an apparent magnitude of $m_{\rm AB} = 24.5$ and was found in only $350\,{\rm arcmin}^2$.
The derived number density of $z \simeq 10$ LBGs from~\citet{Morishita2018} is an order of magnitude higher than our extrapolated DPL prediction. 
In Fig.~\ref{fig:lfz10} we also plot the estimated number density derived from the $z = 11.1$ source GN-z11 presented in~\citet{Oesch2016}.
This object is fainter than the BoRG[z9] source, with $m_{\rm AB} = 26.0$, however the derived number density is also significantly in excess of our extrapolated DPL prediction. 
At the faint-end of the LF at $z \simeq 10$, our extrapolated DPL LF and the Schechter function parameterization of~\citet{Bouwens2016} are in excess of the binned LF points at this redshift derived by~\citet{Oesch2018}.
\citet{Oesch2018} argue that there is an accelerated decline in the number of sources from $z = 8$--$10$.
If confirmed this would make the detection of extremely bright sources at these ultra-high redshifts even more unusual (see also~\citealp{McLeod2016}).
The detection of these ultra-high-redshift sources will continue to be extremely challenging prior to the launch of~\emph{JWST}, with contamination being more likely given the lack of detections redward of the Lyman-break.
If any of these extremely bright sources are spectroscopically confirmed, it would lend significantly weight to our proposed model of the evolving LF, as current Schechter function parameterisations essentially do not predict any sources to exist brightward of $M_{\rm UV} \simeq -22$ at $z \gtrsim 10$.

\section{Conclusions}\label{sect:conc}
We have undertaken a search for bright $z \gtrsim 7.5$ Lyman-break galaxies over $6\,$\ds~of ground-based data in the COSMOS and XMM-LSS fields.
Using a full photometric redshift fitting method to the UltraVISTA DR4, VIDEO and UKIRT UDS near-infrared imaging combined with deep optical and~\emph{Spitzer}/IRAC data, we find 27 candidate LBGs in the redshift range $7.4 < z < 9.1$.
The galaxies are some of the most luminous galaxies known at ultra-high redshift, with absolute UV magnitudes in the range $-23.2 < M_{\rm UV} < -21.3$.
We also find one candidate LBG with a best-fitting photometric redshift of $z = 10.9 \pm 1.0$ in the XMM-LSS field.
We carefully exclude brown-dwarf contaminants that can mimic the colours of high-redshift galaxies, by incorporating the expected~\emph{Spitzer}/IRAC colours using an empirical relation between the $J$-,~\chone-~and~\chtwo-band magnitudes and sub-type.
We compute the rest-frame UV LF from our sample at $z = 8$ and $z = 9$, extending the measurements to $M_{\rm UV} \simeq -23$ for the first time at these redshifts.
When compared to the Schechter function predictions from previous studies based on fainter samples, we find an excess in the number density of very bright galaxies in our samples.
We find instead that a double power law provides a good fit to the data.
When comparing the derived number density of very bright LBGs from this study and previous works, we find a lack of evolution between $z \simeq 5$ and $z \simeq 9$ at $M_{\rm UV} \lesssim -23$.
By fitting a simple linear evolution model to the data at $z \ge 5$, we find that a DPL model with a brightening characteristic magnitude ($\Delta M^*/\Delta z \simeq -0.5$) and a steepening bright-end slope ($\Delta \beta/\Delta z \simeq -0.5$) can reproduce the observed evolution in the rest-frame UV LF in the range $5 < z < 10$.
We argue that a shallower decline in the number density of the most luminous sources is to be expected at very high redshifts, due to the reduction in the dust obscuration that has been shown in both simulations and observations to shape the bright-end of the rest-frame UV LF at $z \simeq 5$.
The lack of mass quenching for galaxies at these very high redshifts further acts to soften the bright-end decline of the observed LF.
Further insights into the interplay between these important astrophysical effects will be obtained from new larger samples derived from upcoming surveys (e.g.~\emph{Euclid} and~\emph{WFIRST}), robust measurements of the mass function at very high-redshifts from including deep~\emph{Spitzer} and~\emph{JWST} data, and detailed dust continuum measurements from ALMA. 

\section*{Acknowledgements}
This work was supported by the Glasstone Foundation and the Oxford Hintze Centre for Astrophysical Surveys which is funded through generous support from the Hintze Family Charitable Foundation.
The Cosmic Dawn Center is funded by the Danish National Research Foundation.
BMJ is supported in part by Independent Research Fund Denmark grant DFF - 7014-00017.
This work is based on data products from observations made with ESO Telescopes at the La Silla Paranal Observatory under ESO programme ID 179.A-2005 and ID 179.A-2006 and on data products produced by CALET and the Cambridge Astronomy Survey Unit on behalf of the UltraVISTA and VIDEO consortia.
The Hyper Suprime-Cam (HSC) collaboration includes the astronomical communities of Japan and Taiwan, and Princeton University. The HSC instrumentation and software were developed by the National Astronomical Observatory of Japan (NAOJ), the Kavli Institute for the Physics and Mathematics of the Universe (Kavli IPMU), the University of Tokyo, the High Energy Accelerator Research Organization (KEK), the Academia Sinica Institute for Astronomy and Astrophysics in Taiwan (ASIAA), and Princeton University. Funding was contributed by the FIRST program from Japanese Cabinet Office, the Ministry of Education, Culture, Sports, Science and Technology (MEXT), the Japan Society for the Promotion of Science (JSPS), Japan Science and Technology Agency (JST), the Toray Science Foundation, NAOJ, Kavli IPMU, KEK, ASIAA, and Princeton University. 
This paper makes use of software developed for the Large Synoptic Survey Telescope. We thank the LSST Project for making their code available as free software at \url{http://dm.lsst.org}.
Based in part on data collected at the Subaru Telescope and retrieved from the HSC data archive system, which is operated by Subaru Telescope and Astronomy Data Center at National Astronomical Observatory of Japan.
This study was based in part on observations obtained with MegaPrime/MegaCam, a joint project of CFHT and CEA/DAPNIA, at the Canada-France-Hawaii Telescope (CFHT) which is operated by the National Research Council (NRC) of Canada, the Institut National des Science de l'Univers of the Centre National de la Recherche Scientifique (CNRS) of France, and the University of Hawaii. This work is based in part on data products produced at TERAPIX and the Canadian Astronomy Data Centre as part of the CFHTLS, a collaborative project of NRC and CNRS. 
This work is based in part on observations made with the Spitzer Space Telescope, which is operated by the Jet Propulsion Laboratory, California Institute of Technology under NASA contract 1407.
This research made use of {\sc Astropy}, a community-developed core Python package for Astronomy (Astropy Collaboration, 2013).
This research has benefitted from the SpeX Prism Spectral Libraries, maintained by Adam Burgasser at~\url{http://pono.ucsd.edu/~adam/browndwarfs/spexprism}.




\bibliographystyle{mnras}
\bibliography{library_abbrv} 



\appendix

\section{Comparison to photometry of S19}\label{app:S19}
\begin{figure}
\centering
\includegraphics[width=0.4\textwidth]{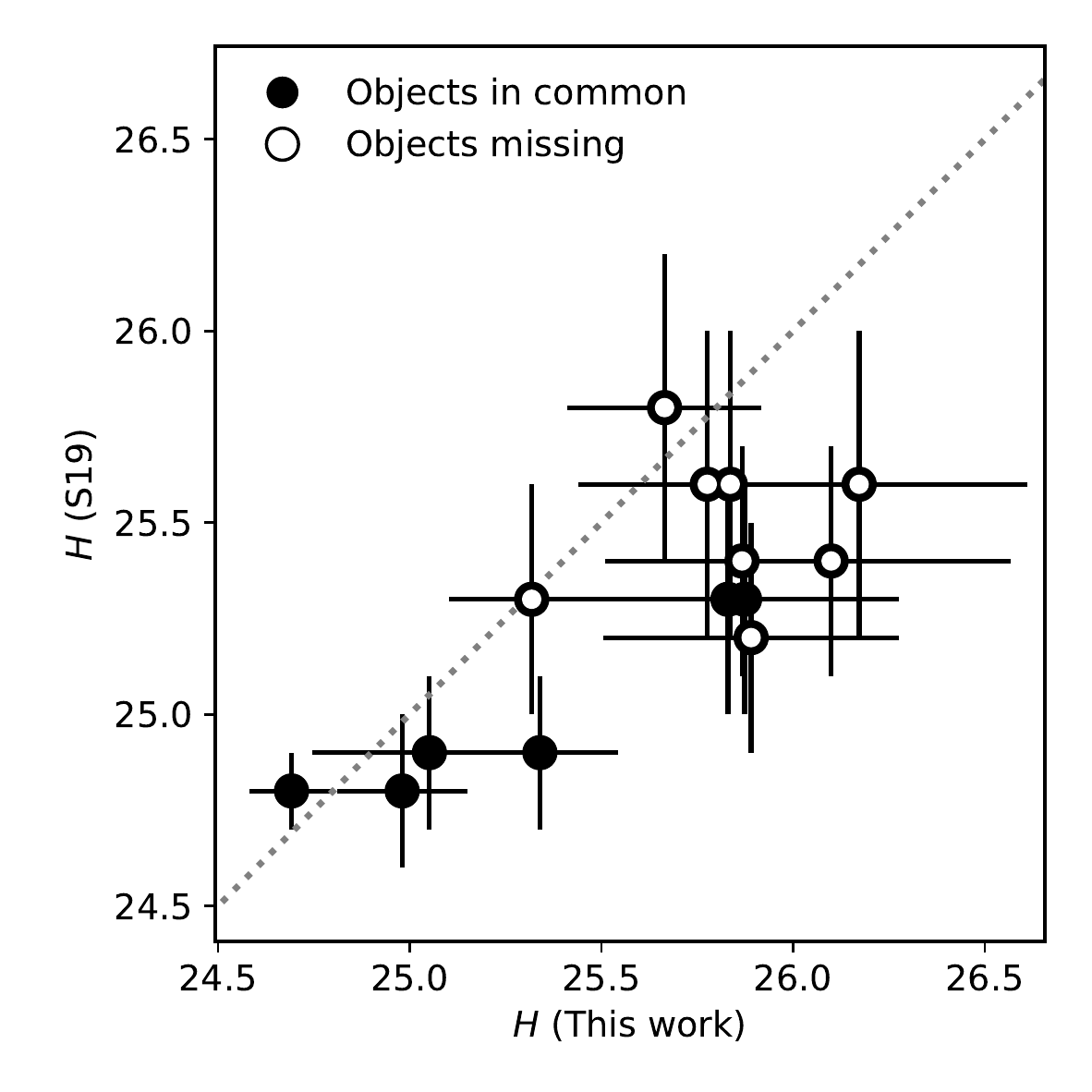}
\caption{A comparison of the $H$-band magnitudes for the $z \simeq 8$--$9$ sample of S19, who used the shallower UltraVISTA DR3 data, and the photometry derived in this work.
Filled points show the overlapping LBG candidates between this study and S19 sample, while open points show the objects that were not recovered as high-redshift sources in our analysis.
The majority of the sources that we do not reselect are at the faint end.
Comparing to the one-to-one line (dotted) we find that S19 derive brighter magnitudes by $\gtrsim 0.5\,$mag that we measure.
}\label{fig:S19}
\end{figure}

Here we present a comparison between the photometry in S19 and our catalogues, in an effort to understand why we do not recover their full sample of LBG candidates.
At the bright-end of our samples we find good agreement within the errors ($\delta {\rm m} < 0.2$), however we find that for the fainter  S19 candidates their photometry is systematically fainter than that measured in our catalogues as shown in Fig.~\ref{fig:S19}.
For objects Y5, Y8-Y11, Y14 and Y16 we find offsets exceeding $0.4 {\rm mag}$, with offsets of $0.7\,{\rm mag}$ for Y9 and Y11.
S19 used the UltraVISTA DR3 photometry, whereas we use the more recent DR4 release.
If we instead measure our photometry on the DR3 images, the offset is significantly reduced, demonstrating that the observed difference in magnitude is mainly due to a difference between the DR3 and DR4 data.
There is no zeropoint offset between the two data releases.
As the objects with the biggest discrepancy between DR3 and DR4 are close to the magnitude limit of the survey, it is likely that they were up-scattered by noise into the S19 sample (while other high-redshift candidates were down-scattered).
S19 require a $5\sigma$ detection in a stack of five bands for selection, whereas we impose a more conservative cut of $5\sigma$ significance in a single band.
The result is that our selection is less effected by noise, because we do not select as close to the limit of the data.
To test this hypothesis we compared the photometry for a sample of $z \simeq 7$ sources that were first identified in the UltraVISTA DR1 to the resulting photometry from DR2, using our method of cutting at $5\sigma$ in a single band~\citep{Bowler2014}.
We find no systematic offset between these measurements, providing reassurance that our photometry is robust for the sample presented in this work.
\citet{Stefanon2017} note that they find offsets between their~\emph{HST}/WFC3 data and the UltraVISTA imaging. 
For objects Y5, J1 and J2 they find that the UltraVISTA $H$-band measurement was $1{\rm mag}$ brighter than the WFC3 $H_{160}$-band magnitude.
Our photometry of these objects suggest that the UltraVISTA $H$-band magnitudes should be $\sim 0.5\,{\rm mag}$ fainter than those presented in~\citet{Stefanon2017}, somewhat reducing the observed discrepancy.

\section{Schechter function evolution}\label{app:SH}

\begin{figure}
\includegraphics[width = 0.49\textwidth]{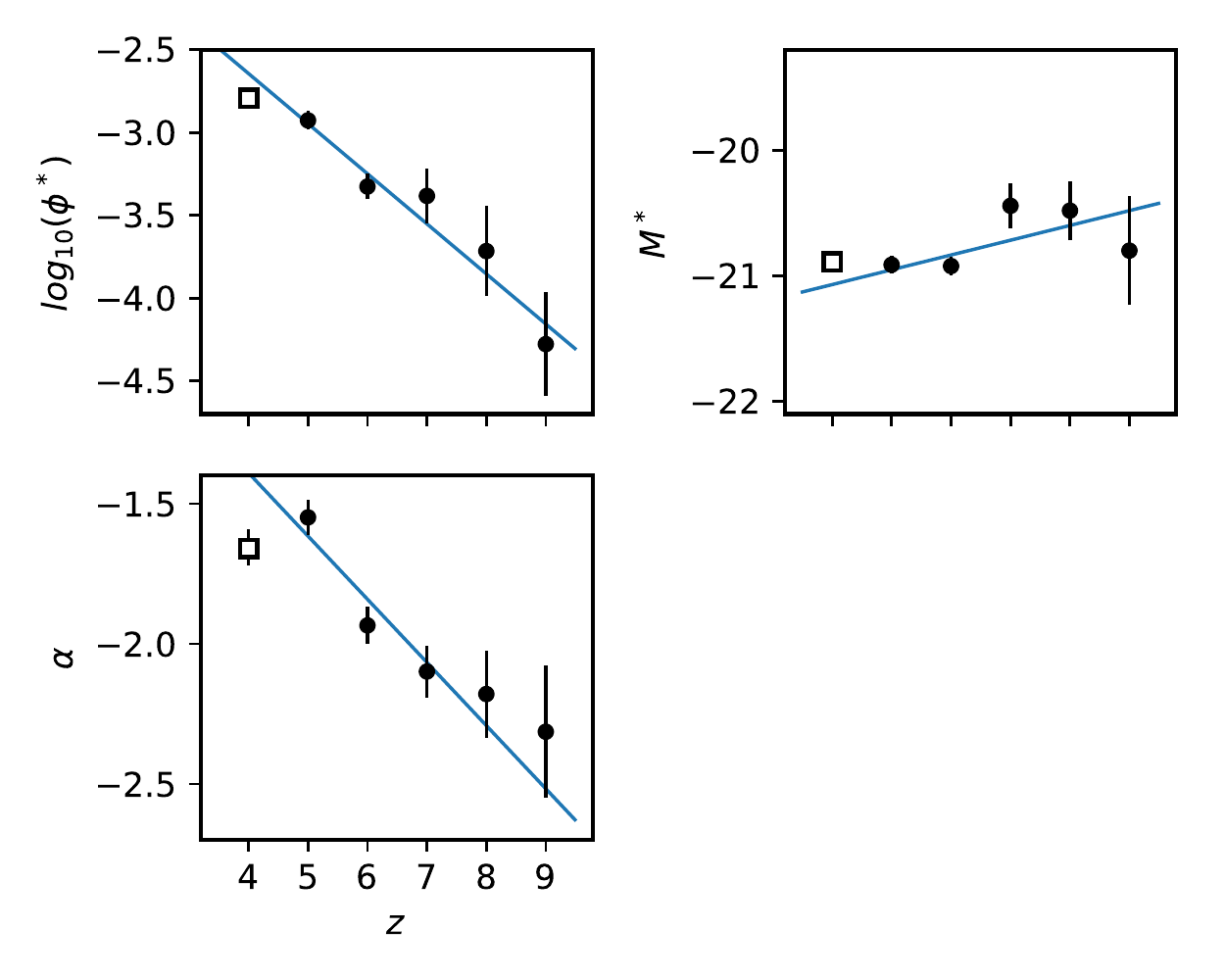}\\

\caption{The evolution of the three Schechter function parameters from $z = 5$--$9$ derived from fitting to a compilation of LF measurements as described in the text.
The blue lines in each plot show the linear fit to the results.
The open square show the results of an independent fitting analysis by~\citet{Adams2019} for comparison.
}\label{fig:paramsSH}
\end{figure}

\begin{figure}
\includegraphics[width = 0.49\textwidth]{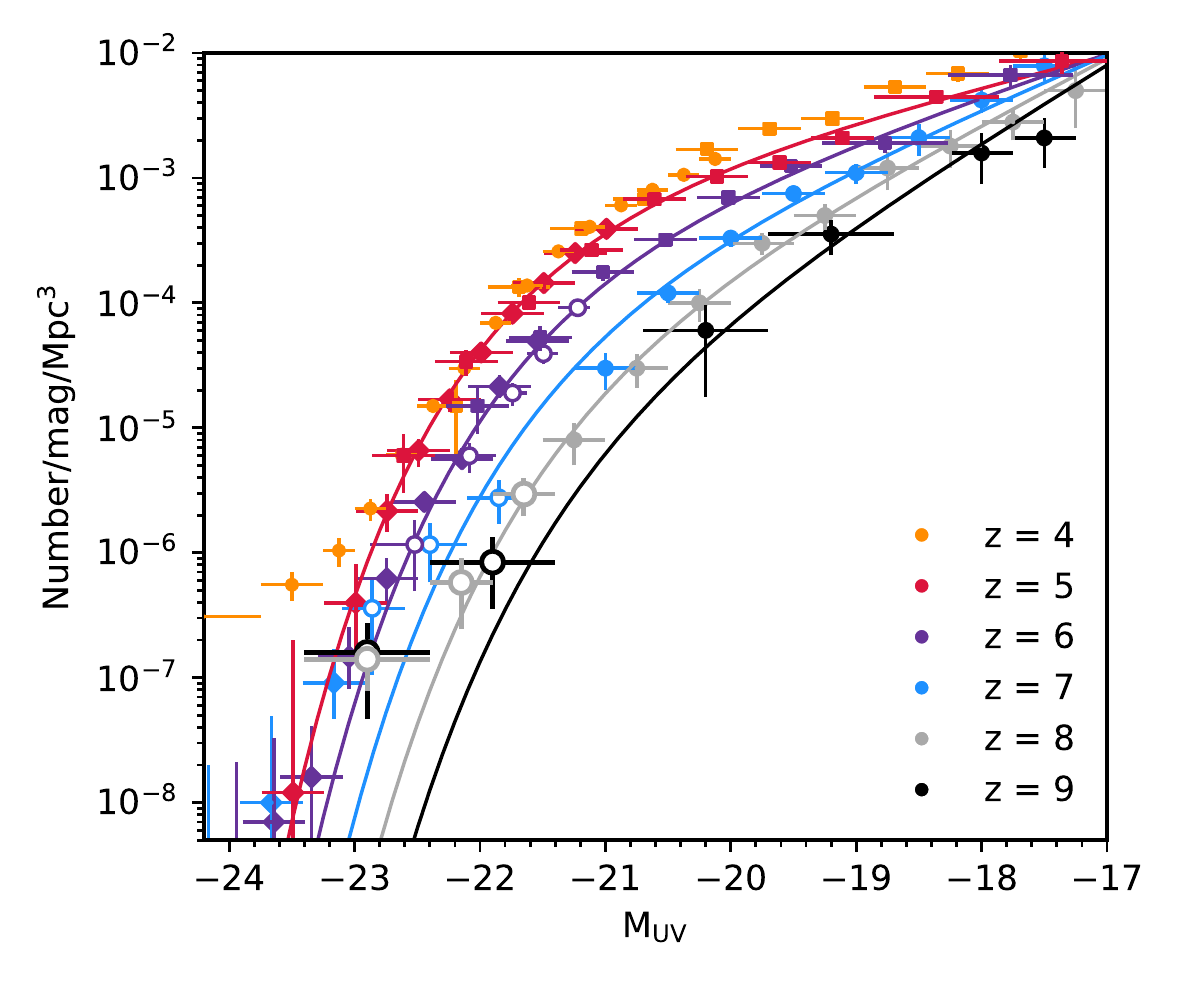}

\caption{The observed rest-frame UV LF results from $z = 5$ to $z = 9$ compared to the results of our evolving Schechter function model. 
The data points shown are as described in the caption to Fig.~\ref{fig:ev}.
The Schechter function parameterization cannot reproduce the observed number of bright ($M_{\rm UV} \lesssim -22$) sources at $z \ge 7$.
}\label{fig:evSH}
\end{figure}

In addition to the DPL fitting, we ran an identical procedure assuming a Schechter function.
The best-fitting Schechter function parameters and the linear fit to these results are shown in Fig~\ref{fig:paramsSH}.
The evolution of the parameters according to the linear fit is given by the following equations for the characteristic magnitude, normalisation and faint-end slope respectively:
\begin{equation}\
\begin{gathered}
M^* = (-20.83 \pm 0.42) + (0.12 \pm 0.07)\,(z - 6)\\
{\rm log}_{10} (\phi^*) = (-3.25 \pm 0.26) - (0.30 \pm 0.05)\,(z - 6)\\
\alpha = (-1.84 \pm 0.28) - (0.23 \pm 0.05)\,(z - 6).
\end{gathered}
\end{equation}
As can be seen in Fig.~\ref{fig:paramsSH} we find that the assumption of a Schechter function changes dramatically the derived evolution, in comparison on our fiducial DPL fits.
We find a strong evolution in $\phi^*$ and $\alpha$, while the characteristic magnitude staying approximately constant at $M^* \simeq -21$. 
These evolving Schechter functions cannot reproduce the measured number density of very bright LBGs at $z \ge 7$.
As shown in Fig.~\ref{fig:evSH}, at $M_{\rm UV} \lesssim -22$ at $z = 7$--$9$, the Schechter function fits dramatically under-predict the number of sources that have been observed, whereas it provides a good description of the decline at $z = 5$--$6$ (see~\citealp{Bowler2015} for further discussion).

\section{Postage-stamps and SED fitting figures}\label{ap:stamps}

\begin{figure*}
\includegraphics[width = \textwidth]{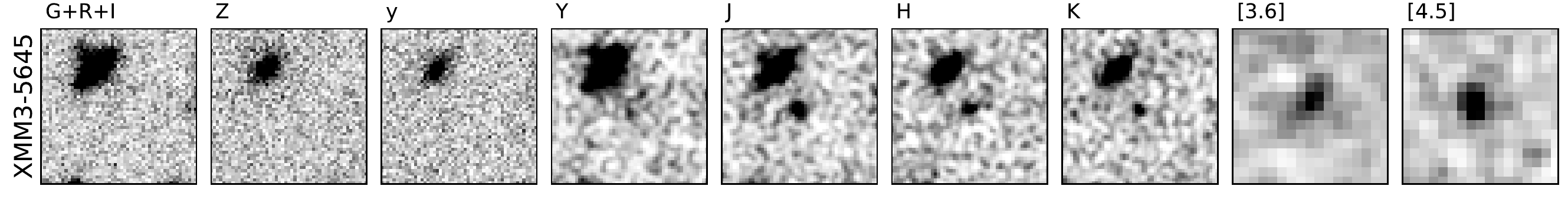}
\includegraphics[width = \textwidth]{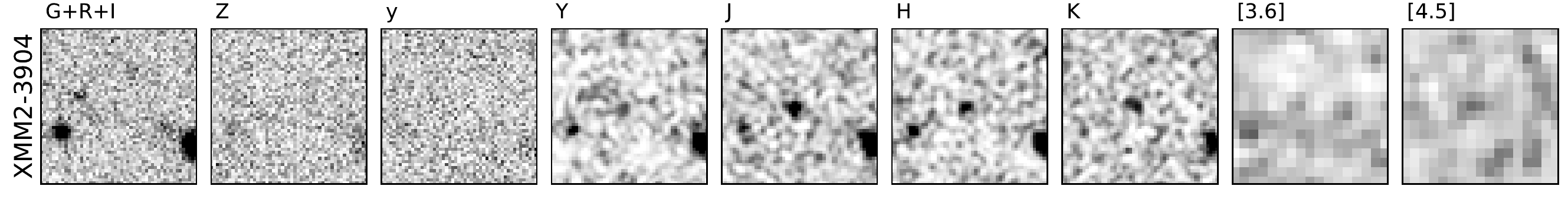}
\includegraphics[width = \textwidth]{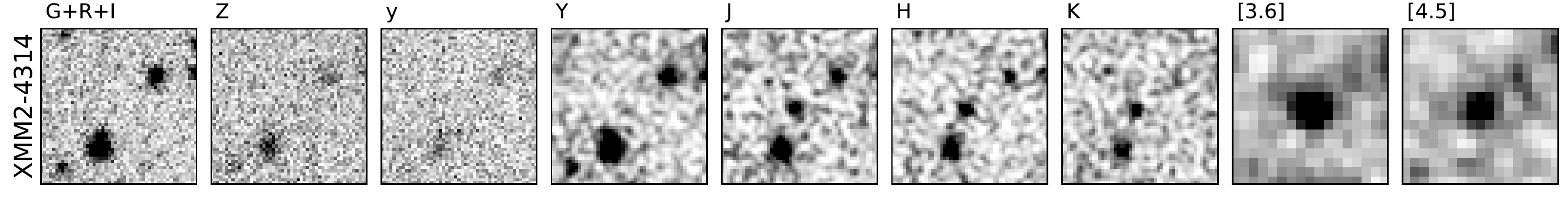}
\includegraphics[width = \textwidth]{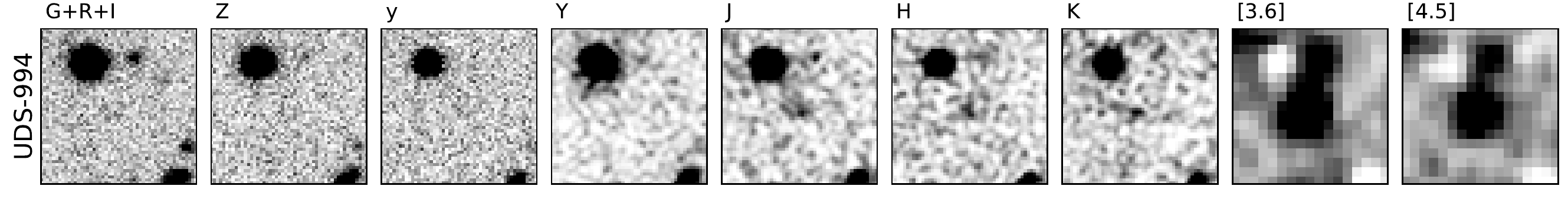}
\includegraphics[width = \textwidth]{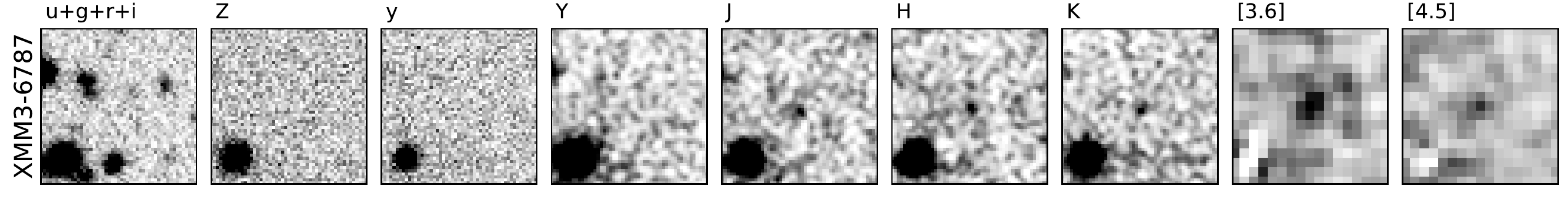}
\includegraphics[width = \textwidth]{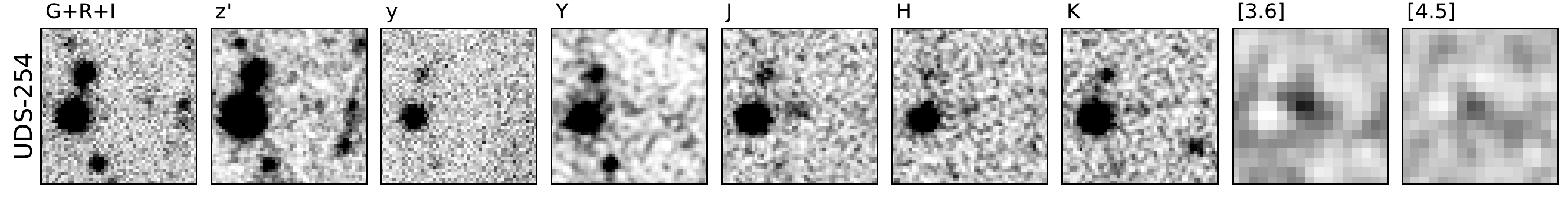}
\includegraphics[width = \textwidth]{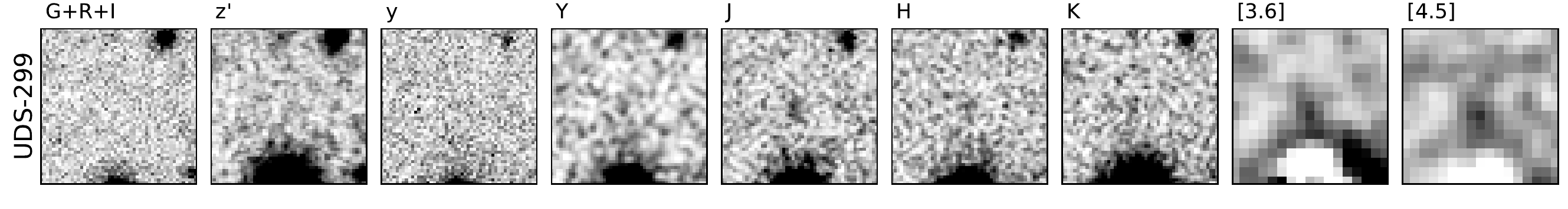}
\includegraphics[width = \textwidth]{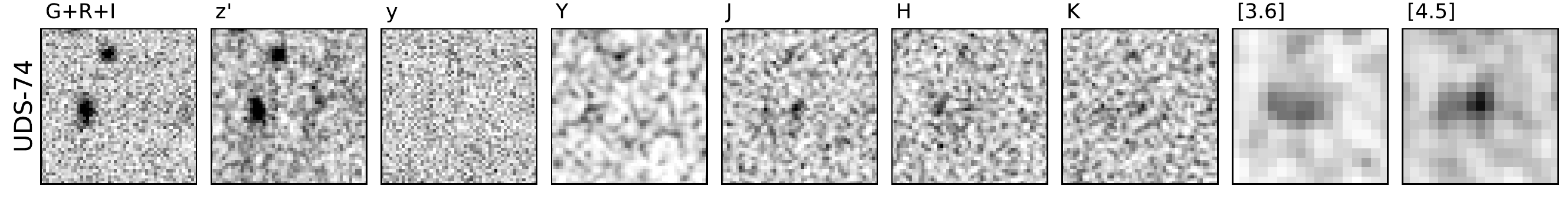}

\caption{Postage-stamp images for the $z \simeq 8$ sample selected in the XMM-LSS field.
Each object corresponds to a single row of stamps, which are ordered from left to right in increasing effective wavelength of the filter.
The stamps are $10\,{\rm arcsec}$ on a side, with North to the top and East to the left.
The objects are ordered by $J$-band magnitude as in Table~\ref{table:photz8}.
The ID of each source is shown on the left, followed by the stacked optical image, the $z'$- or $Z$-band image, the near-infrared bands ($YJHK_{s}$) and finally the deconfused~\emph{Spitzer}/IRAC~\chone~and~\chtwo~bands.
The stamps are saturated beyond the range [-1, 4]$\,\sigma$.}\label{fig:xmmz8}
\end{figure*}

The postage-stamp cutout images and SED fitting results for our sample of 28 high-redshift LBGs.
In Fig.~\ref{fig:xmmz8} and Fig.~\ref{fig:cosz8}--\ref{fig:cosz8ii} we show the postage-stamps for the $z \simeq 8$ sample in the XMM-LSS and COSMOS fields respectively.
In Fig.~\ref{fig:cosz9} we show the postage-stamps for the $z \simeq 9$ sample, and the images for the $z \simeq 10$ source are shown in Fig.~\ref{fig:xmmz10}.
The SED plots for the $z \simeq 8$ sample in the XMM-LSS and COSMOS fields are shown in Fig.~\ref{fig:xmmz8sed}--\ref{fig:xmmz8sedii} and Fig.~\ref{fig:cosz8sed}--\ref{fig:cosz8sedii}, respectively.
The SED fitting results for the $z \simeq 9$ sample are shown in Fig.~\ref{fig:z9sed}.

\begin{figure*}
\includegraphics[width = \textwidth]{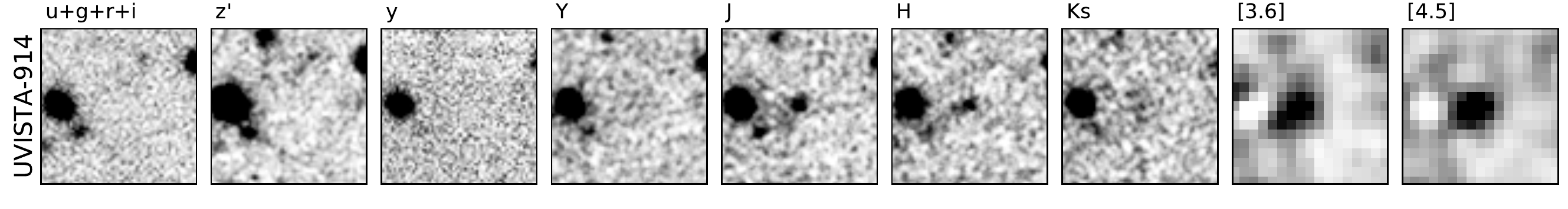}
\includegraphics[width = \textwidth]{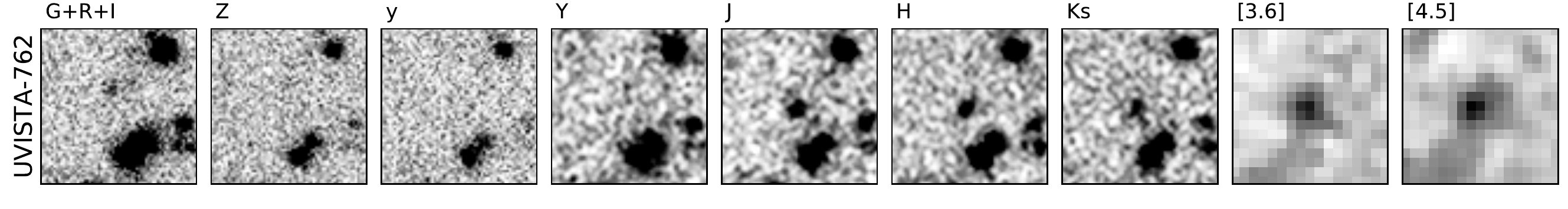}
\includegraphics[width = \textwidth]{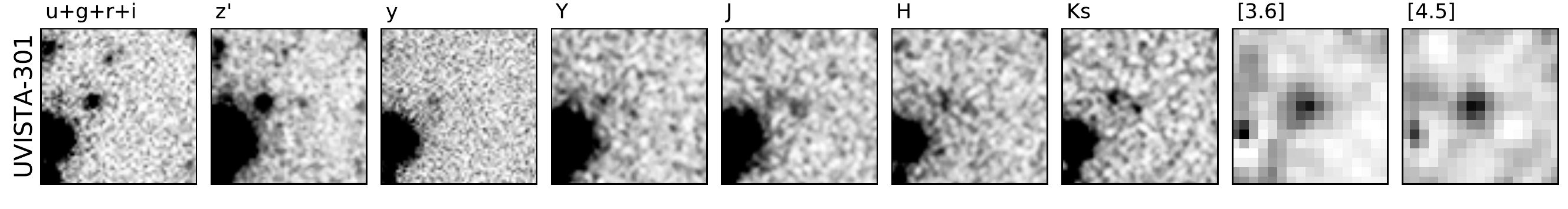}
\includegraphics[width = \textwidth]{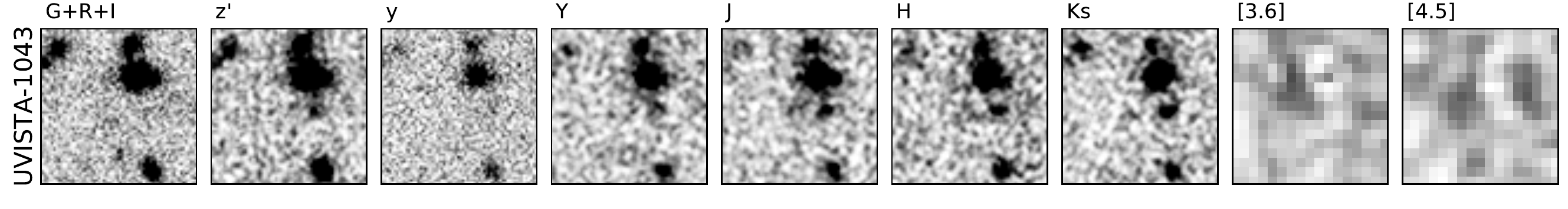}
\includegraphics[width = \textwidth]{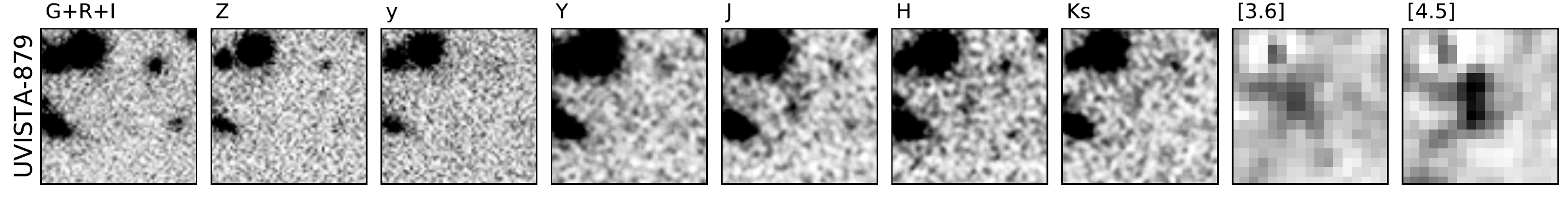}
\includegraphics[width = \textwidth]{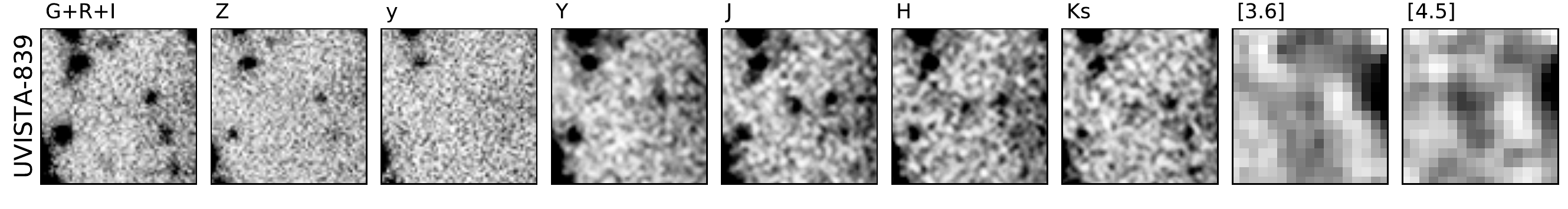}
\includegraphics[width = \textwidth]{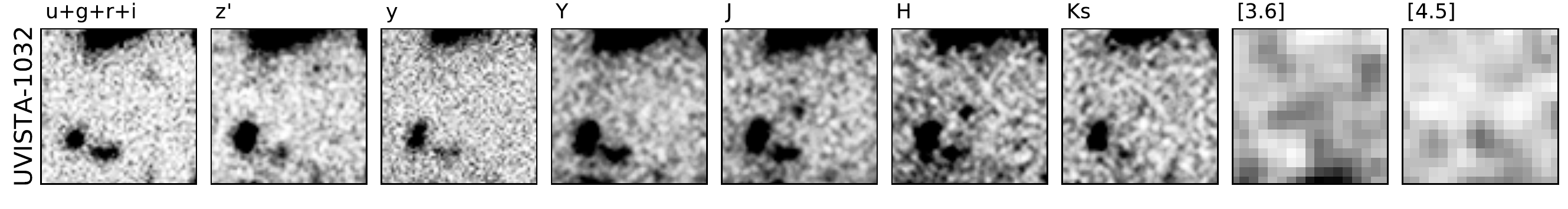}
\includegraphics[width = \textwidth]{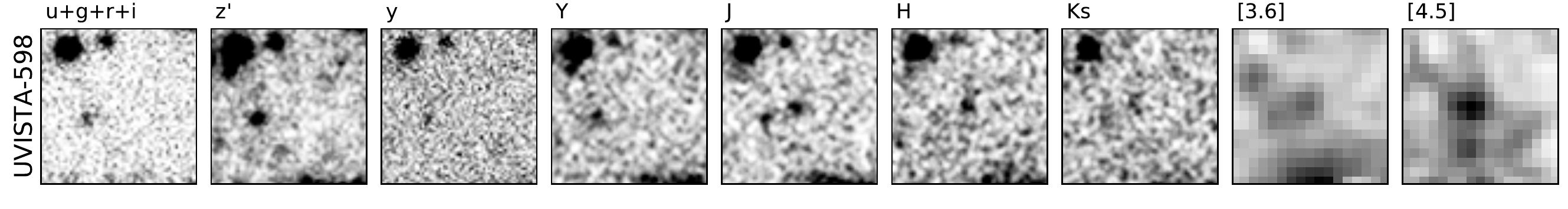}
\includegraphics[width = \textwidth]{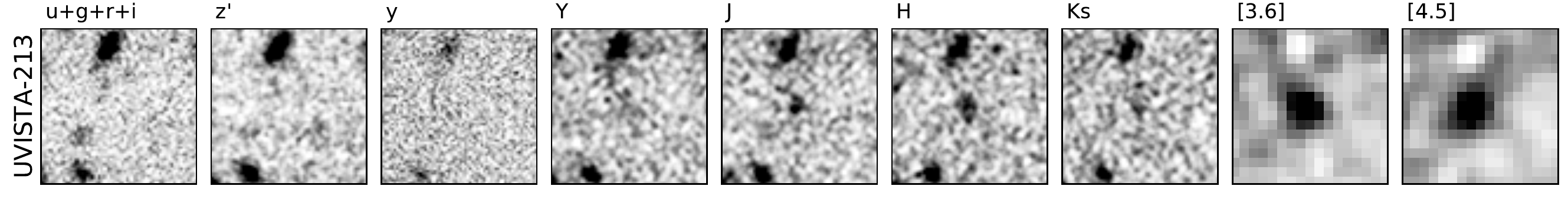}
\caption{Postage-stamp images for the COSMOS sample at $z \simeq 8$.
The figure is in the same format as Fig~\ref{fig:cosz8}.}
\label{fig:cosz8}
\end{figure*}

\begin{figure*}
\includegraphics[width = \textwidth]{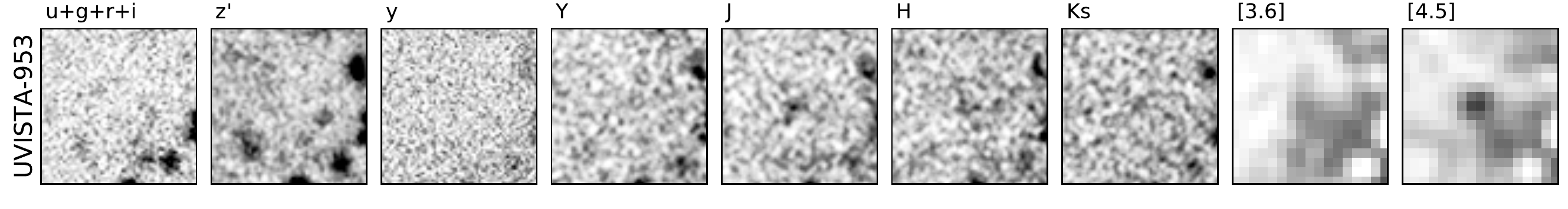}
\includegraphics[width = \textwidth]{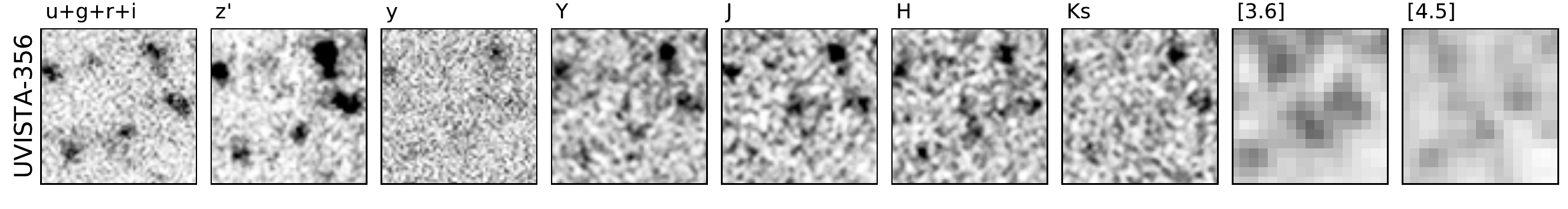}
\includegraphics[width = \textwidth]{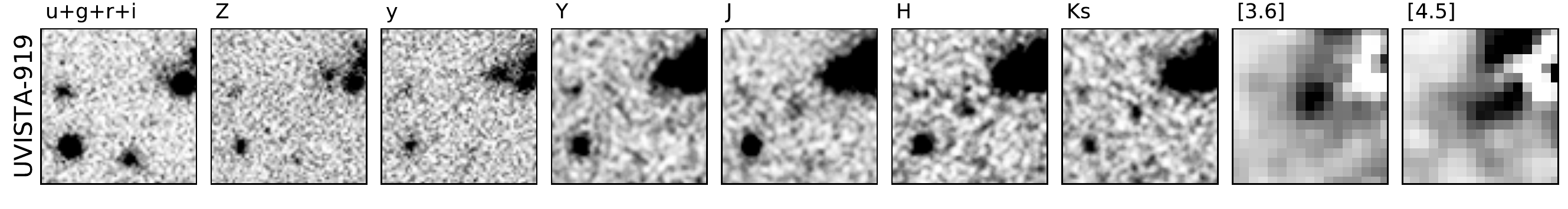}
\includegraphics[width = \textwidth]{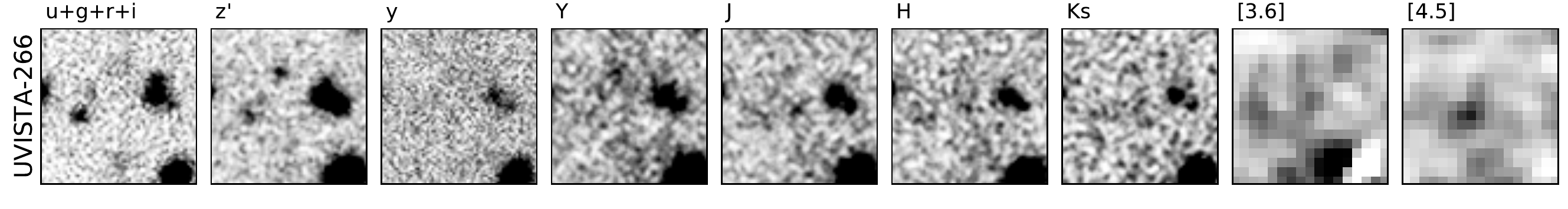}
\includegraphics[width = \textwidth]{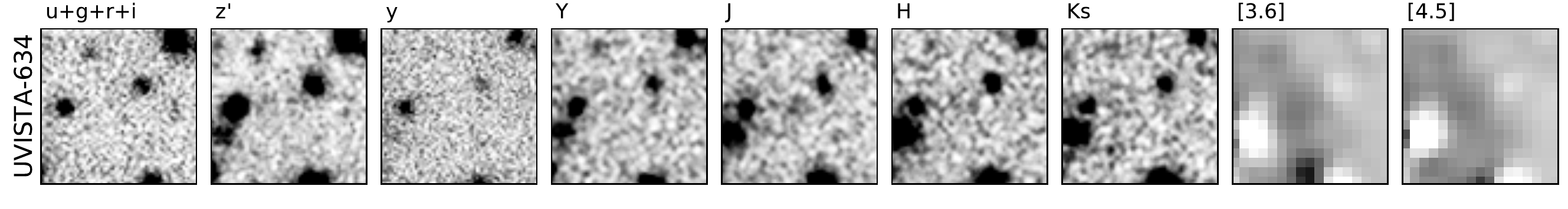}
\caption{Continued.}\label{fig:cosz8ii}
\end{figure*}

\begin{figure*}
\includegraphics[width = \textwidth]{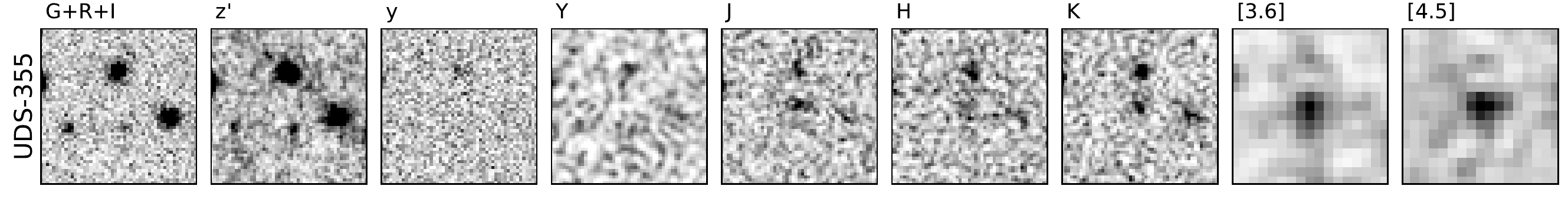}
\includegraphics[width = \textwidth]{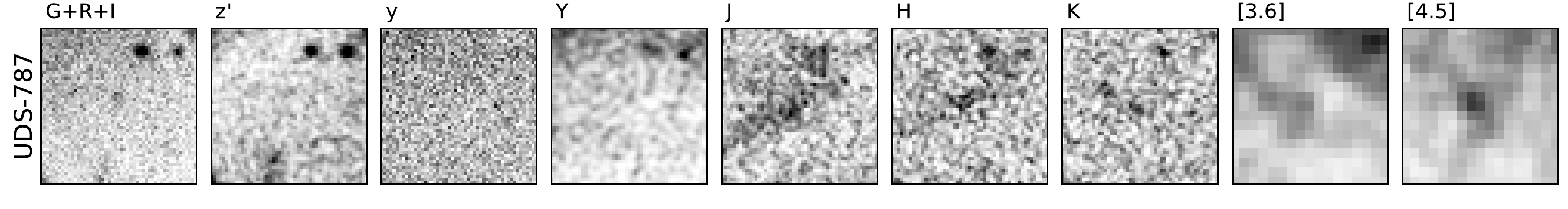}
\includegraphics[width = \textwidth]{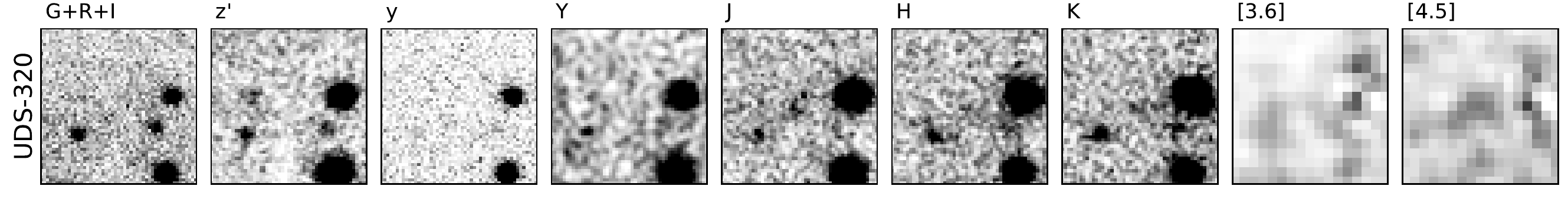}
\includegraphics[width = \textwidth]{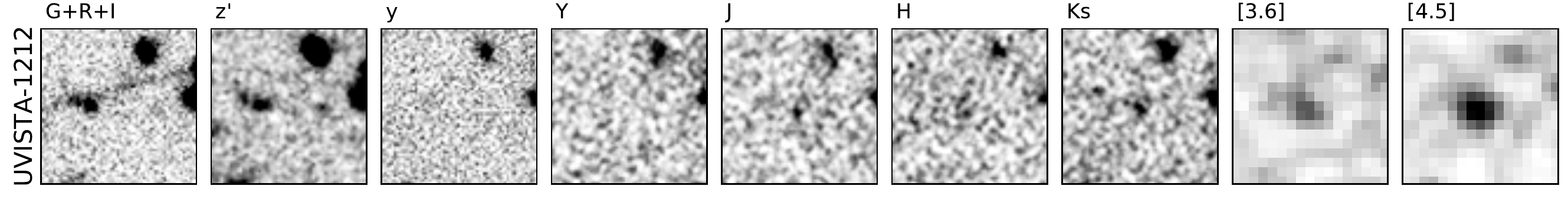}
\includegraphics[width = \textwidth]{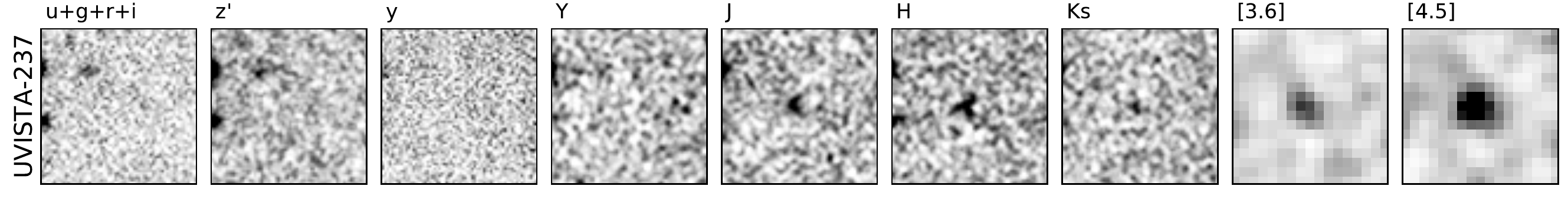}
\caption{Postage-stamp images of the $z \simeq 9$ sample.
The three candidates from XMM-LSS are shown at the top, followed-by the two COSMOS sources.
The objects are ordered by field, and then by $H$-band magnitude as in Table~\ref{table:photz9}.
The scaling of the stamps is the same as described in Fig.~\ref{fig:xmmz8}.
Note that the background for object UDS787 is elevated due to a nearby star.
For this object the NIR detection is confirmed in the $H+K_{s}$ stack (not shown), and the apparent detection in the optical stack results from noise in the $G$-band.}
\label{fig:cosz9}
\end{figure*}

\begin{figure*}
\includegraphics[width = \textwidth]{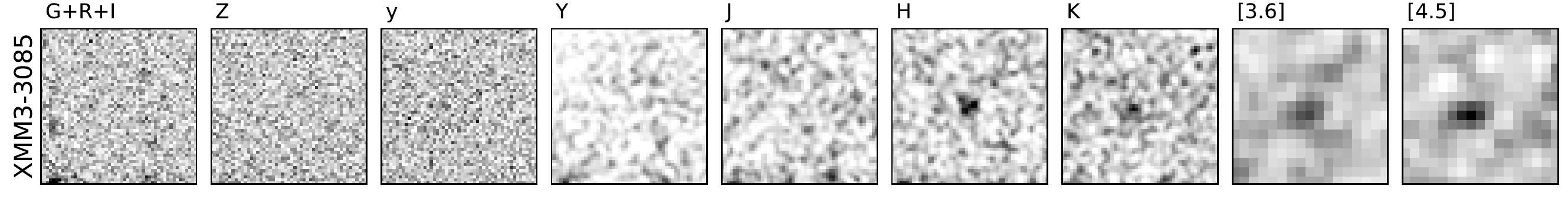}
\caption{Postage-stamp images of the $z \simeq 10$ candidate.
The images are as described in the caption of Fig.~\ref{fig:xmmz8}.}
\label{fig:xmmz10}
\end{figure*}

\begin{figure*}
\includegraphics[width = 0.3\textwidth]{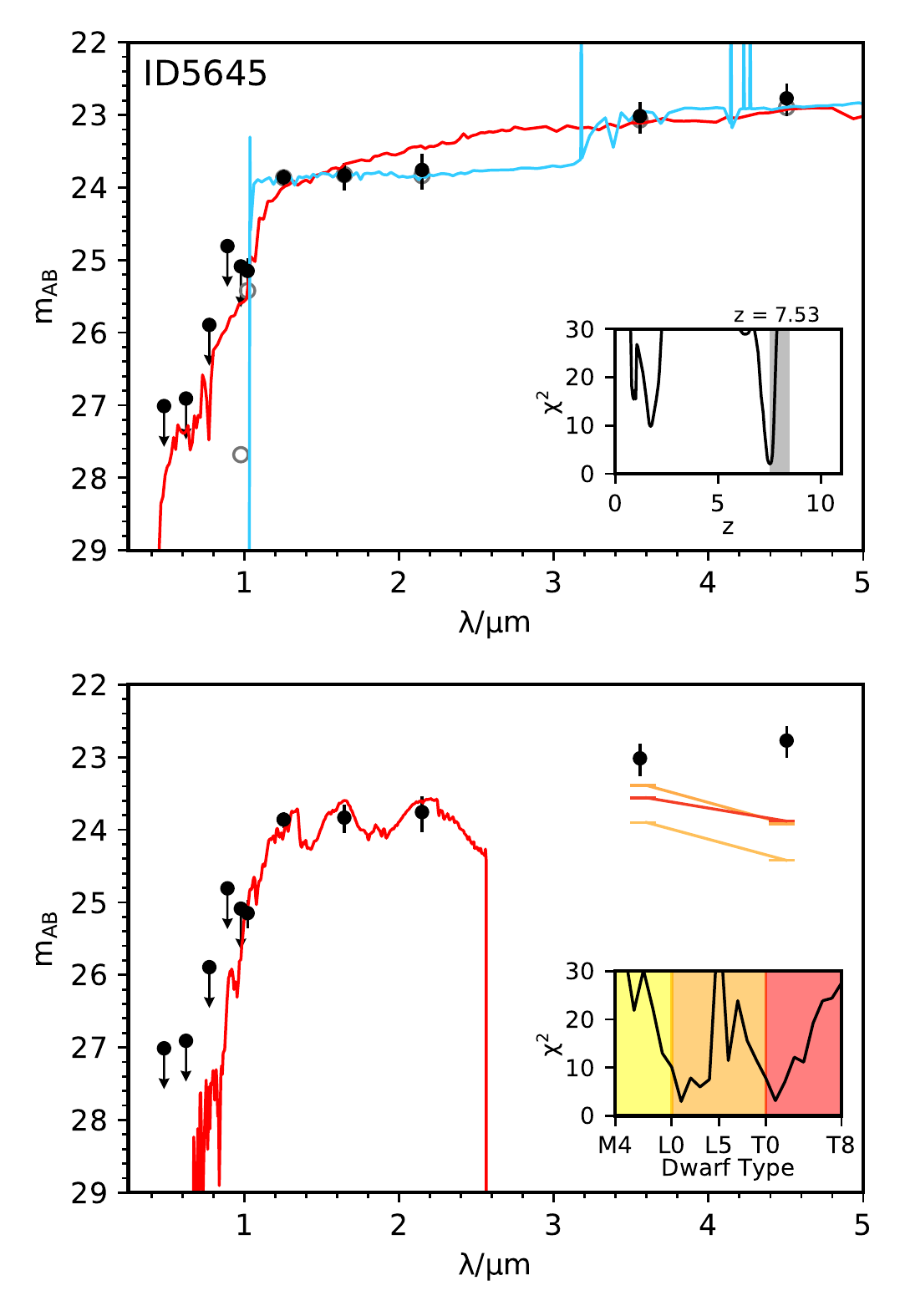}
\includegraphics[width = 0.3\textwidth]{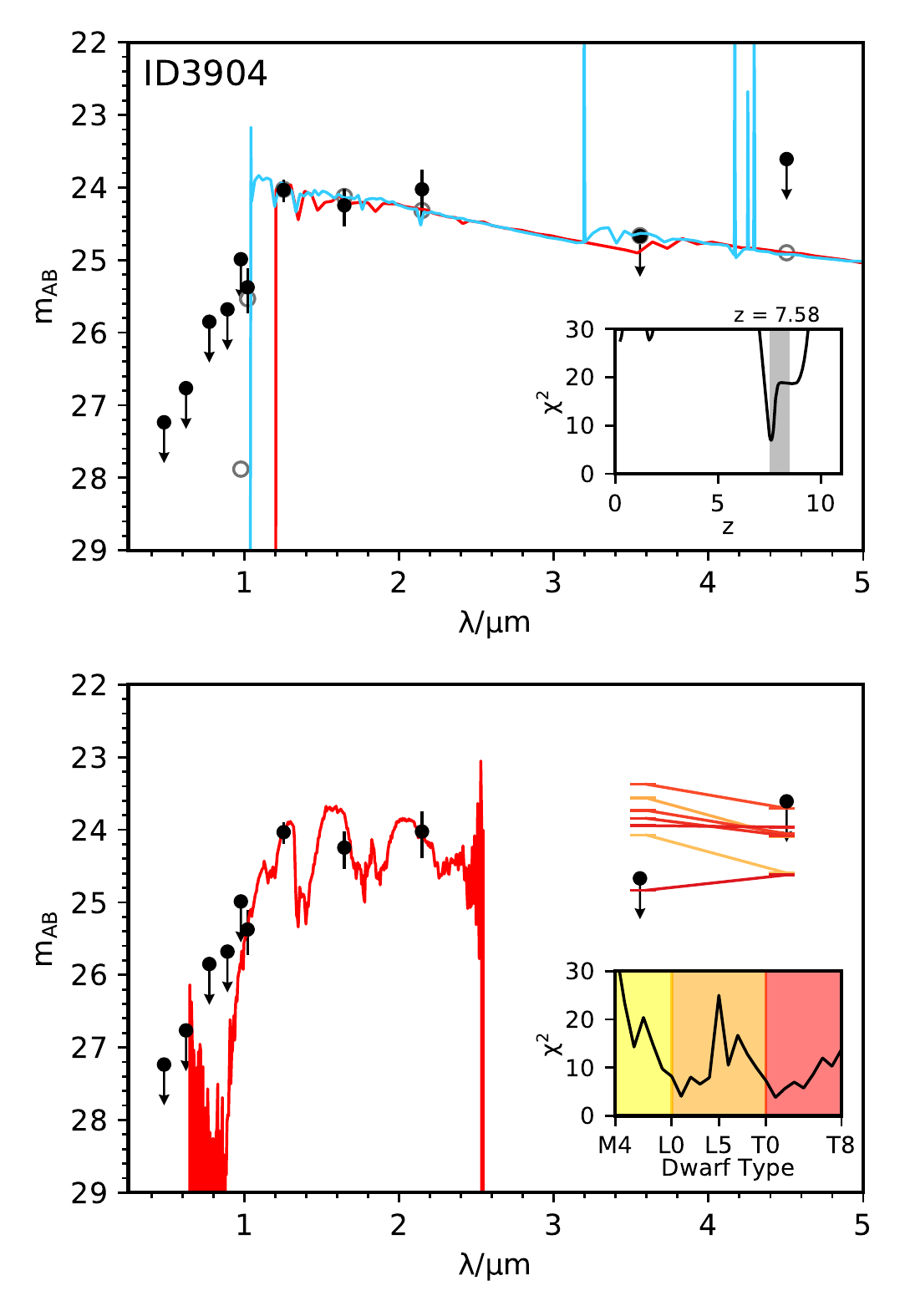}
\includegraphics[width = 0.3\textwidth]{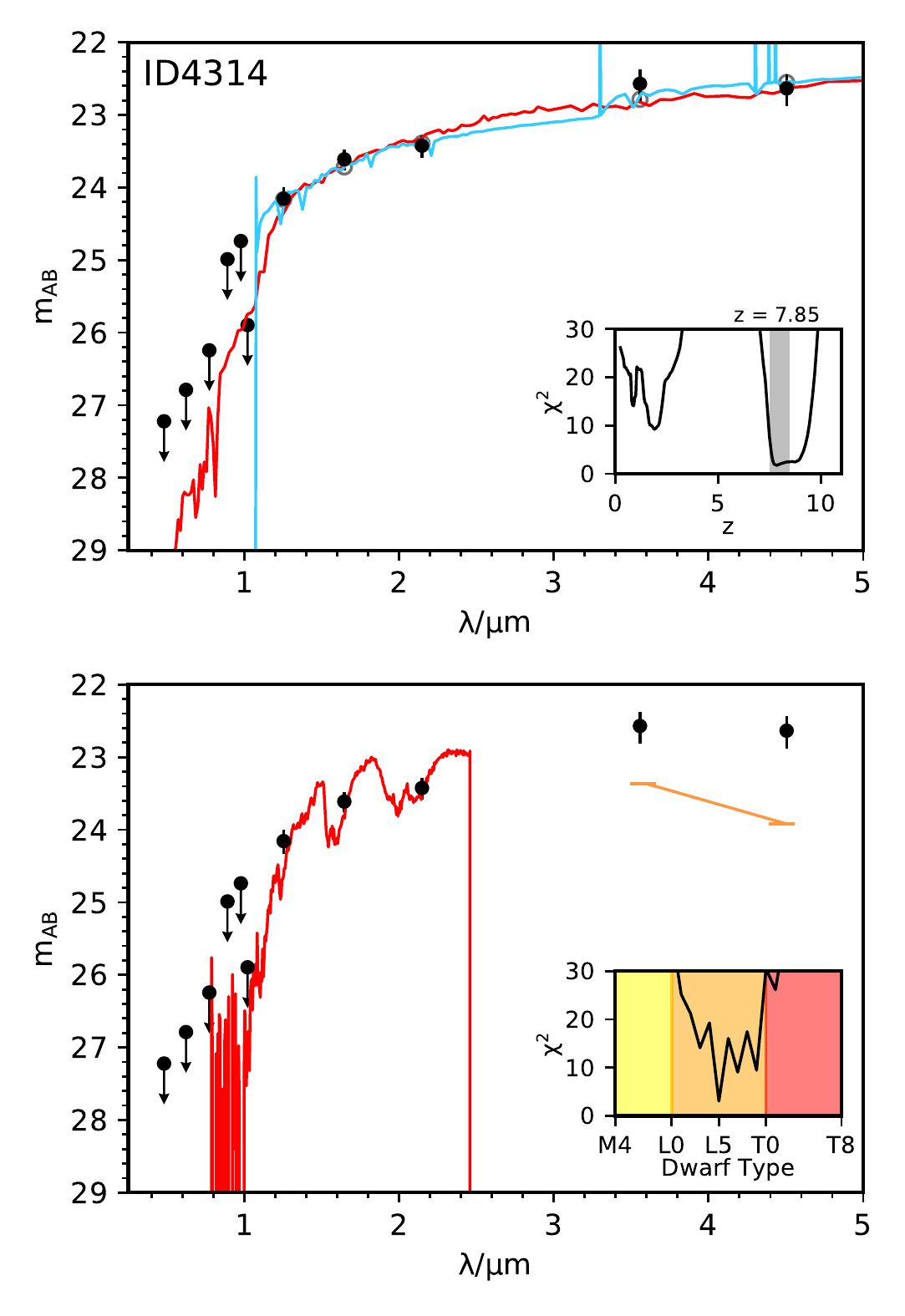}\\
\includegraphics[width = 0.3\textwidth]{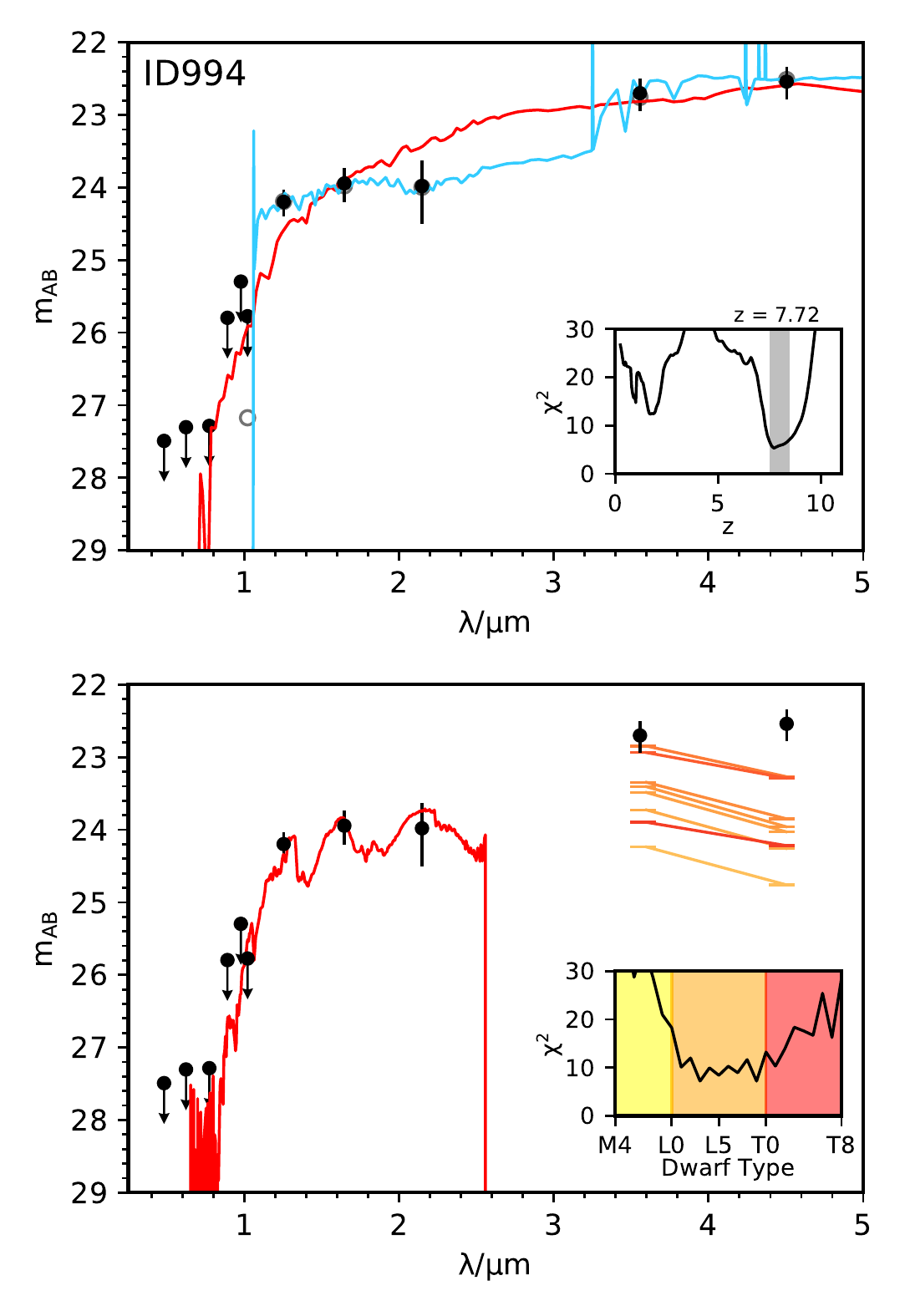}
\includegraphics[width = 0.3\textwidth]{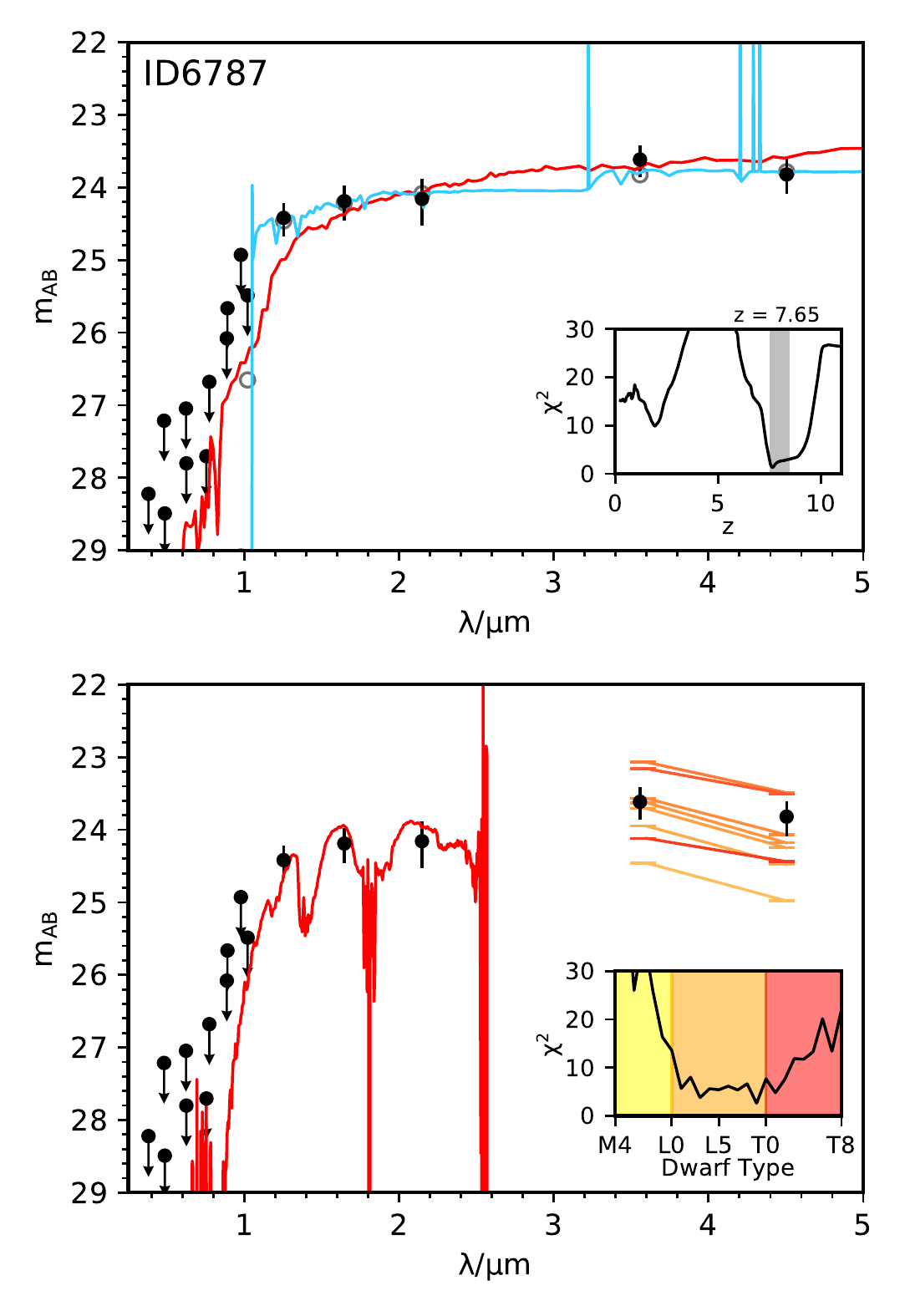}
\includegraphics[width = 0.3\textwidth]{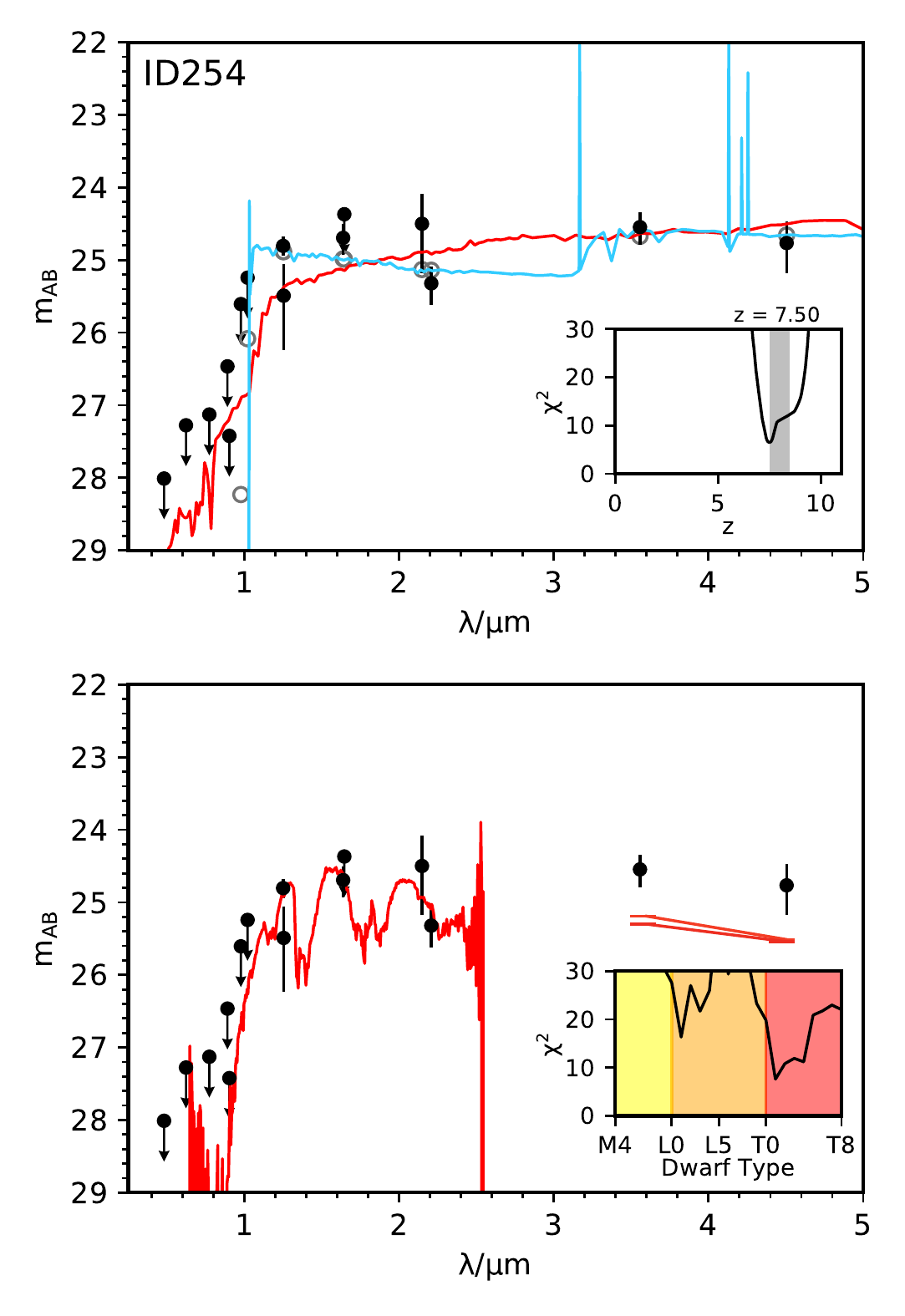}\\
\caption{The observed photometry and best-fitting SED templates for the $z \simeq 8$ candidates in the XMM-LSS field.
The objects are ordered from top-left to bottom-right by $J$-band magnitude as in Table~\ref{table:photz8} and Fig.~\ref{fig:xmmz8}.
For each object we show two plots oriented one above the other, the upper plot shows the galaxy fit results, while the lower plot show the results of fitting brown dwarf templates.
The measured photometry is shown as the black filled points with errors.
In the case of a non-detection at the $2\sigma$ level, the photometry for that band is shown as an upper limit.
In the upper plot the blue line shows the high-redshift best-fit, and the red line shows the second best-fit, which is typically at $z \simeq 2$.
The inset plot displays the $\chi^2$ for each redshift.
In the lower plot at $\lambda > 2.5\,\mu{\rm m}$ we estimate the~\emph{Spitzer}/IRAC photometry by using measured brown dwarf colours as described in Section~\ref{sect:bd}.
For each brown dwarf sub-type that is an acceptable fit to the optical and near-infrared bands we show the expected~\chone ~and~\chtwo~results as a coloured line.
The inset plot displays the $\chi^2$ for each dwarf type.}
\label{fig:xmmz8sed}
\end{figure*}

\begin{figure*}
\includegraphics[width = 0.3\textwidth]{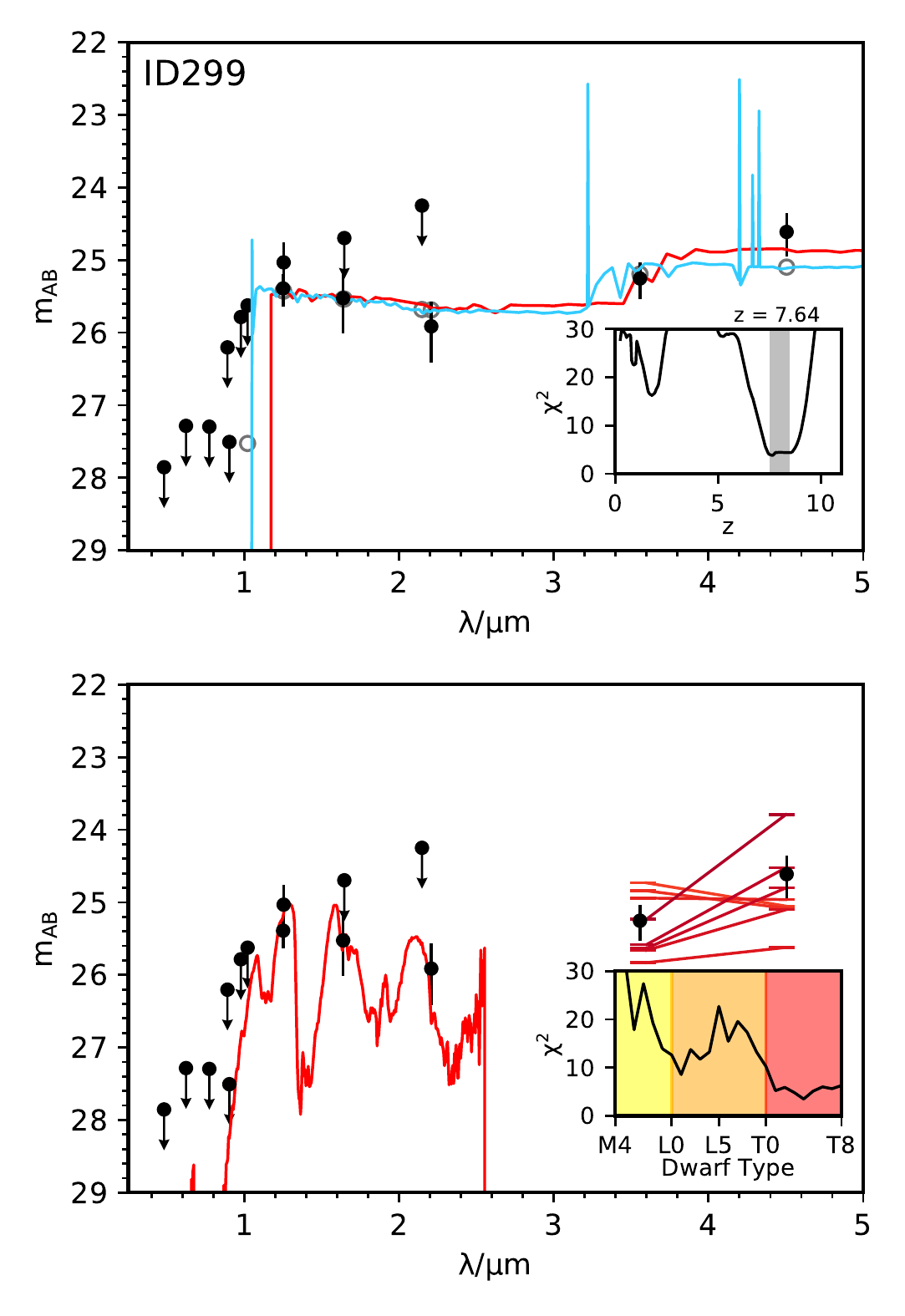}
\includegraphics[width = 0.3\textwidth]{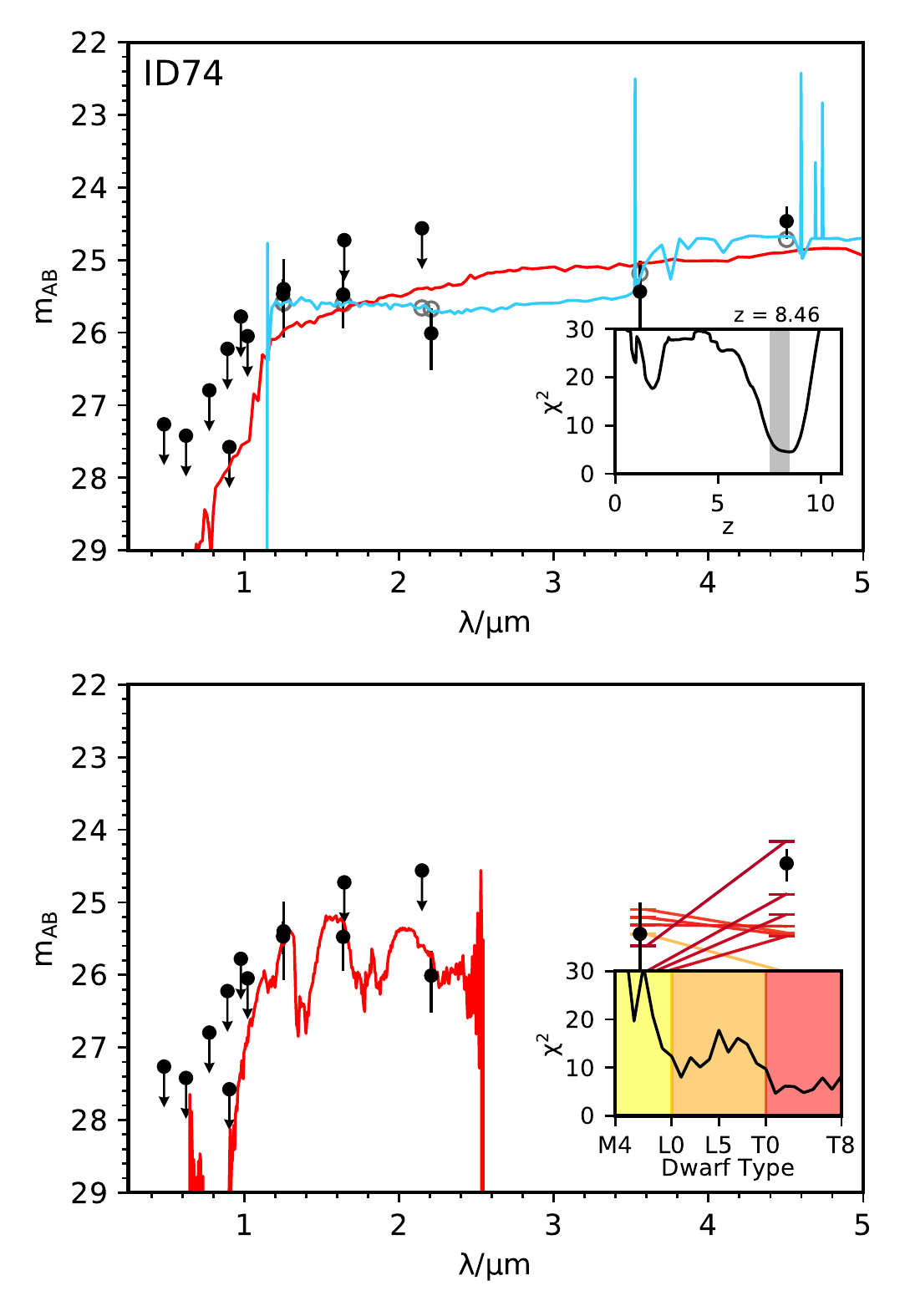}
\caption{Continued.}\label{fig:xmmz8sedii}
\end{figure*}

\begin{figure*}

\includegraphics[width = 0.3\textwidth]{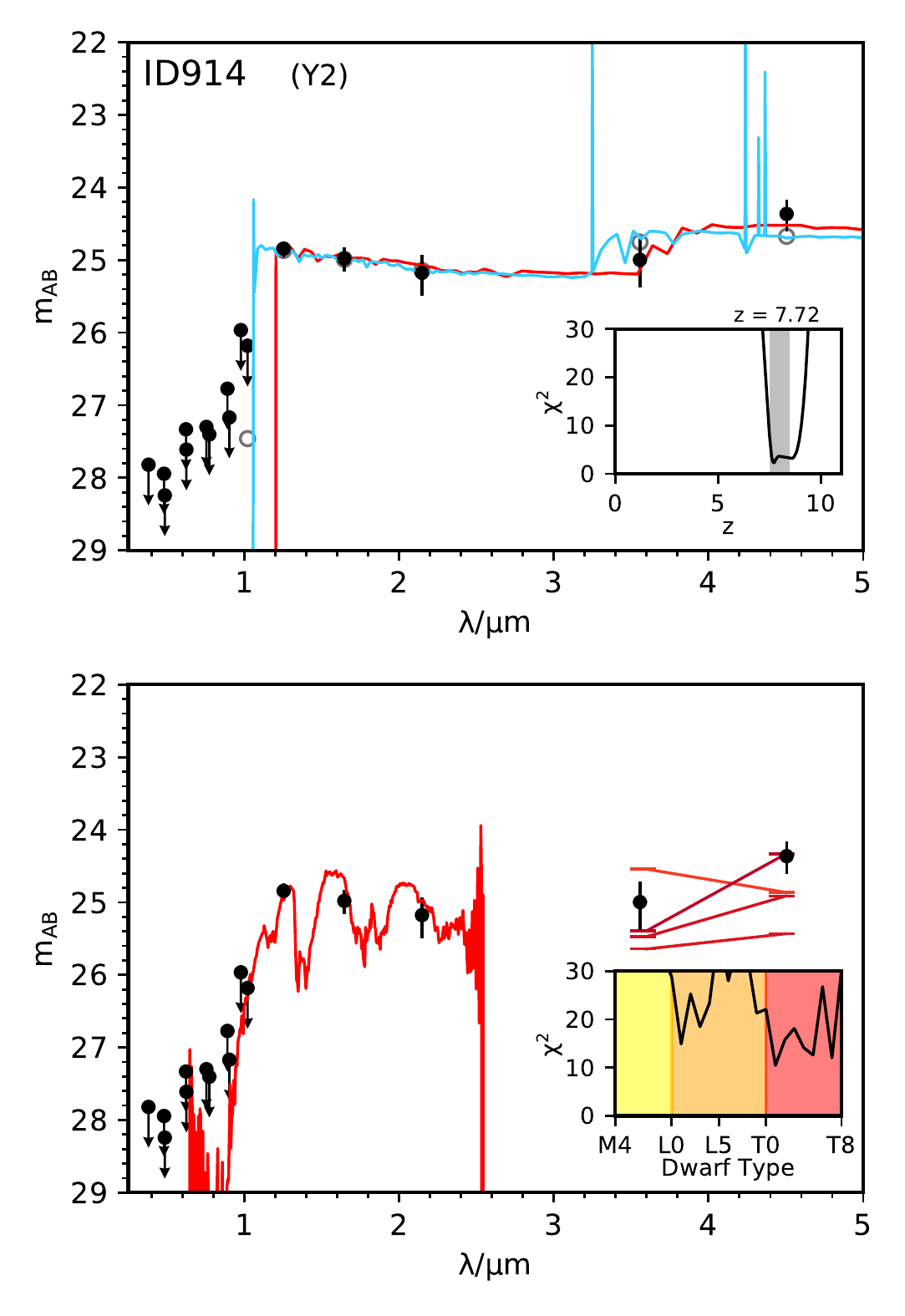}
\includegraphics[width = 0.3\textwidth]{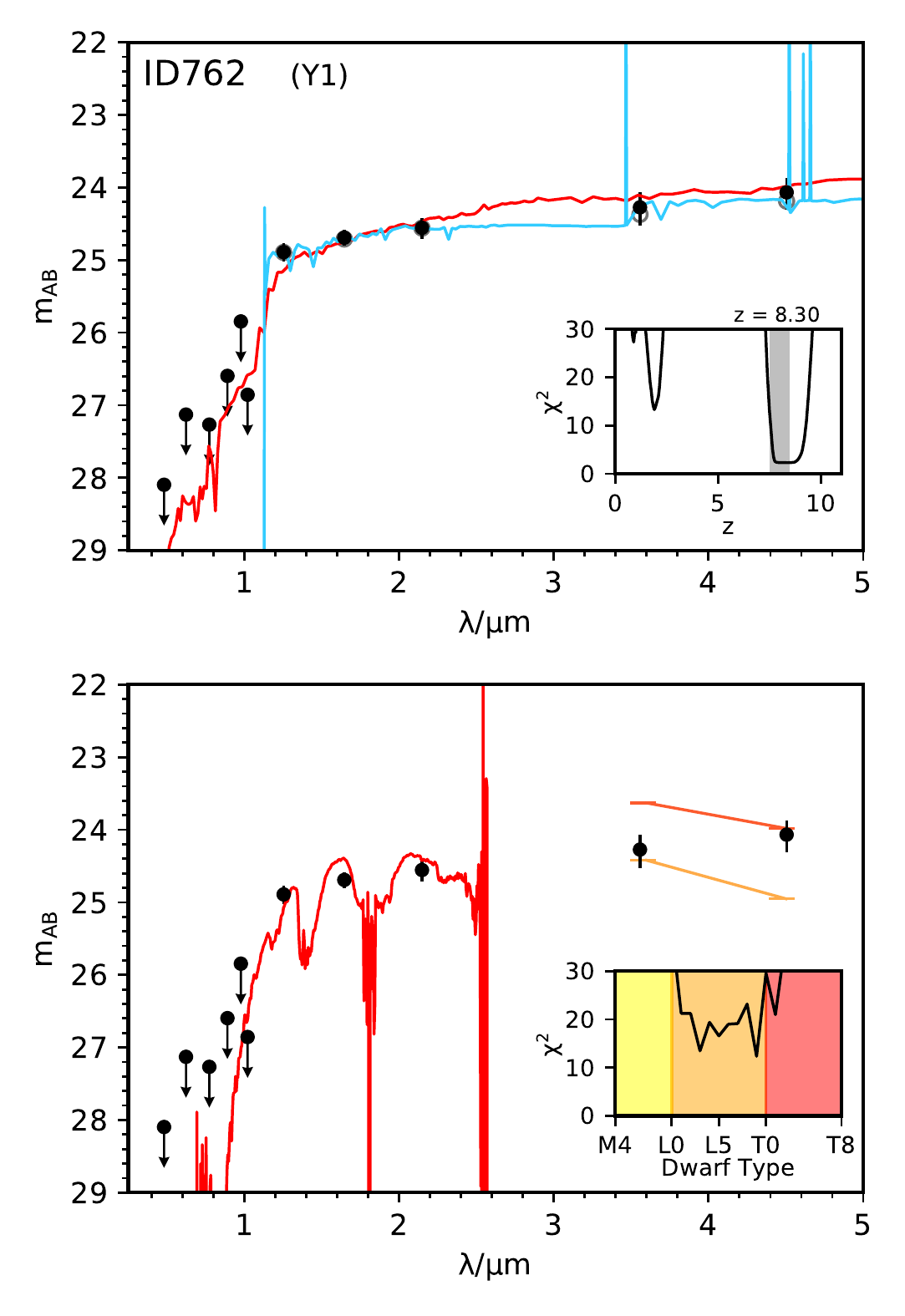}
\includegraphics[width = 0.3\textwidth]{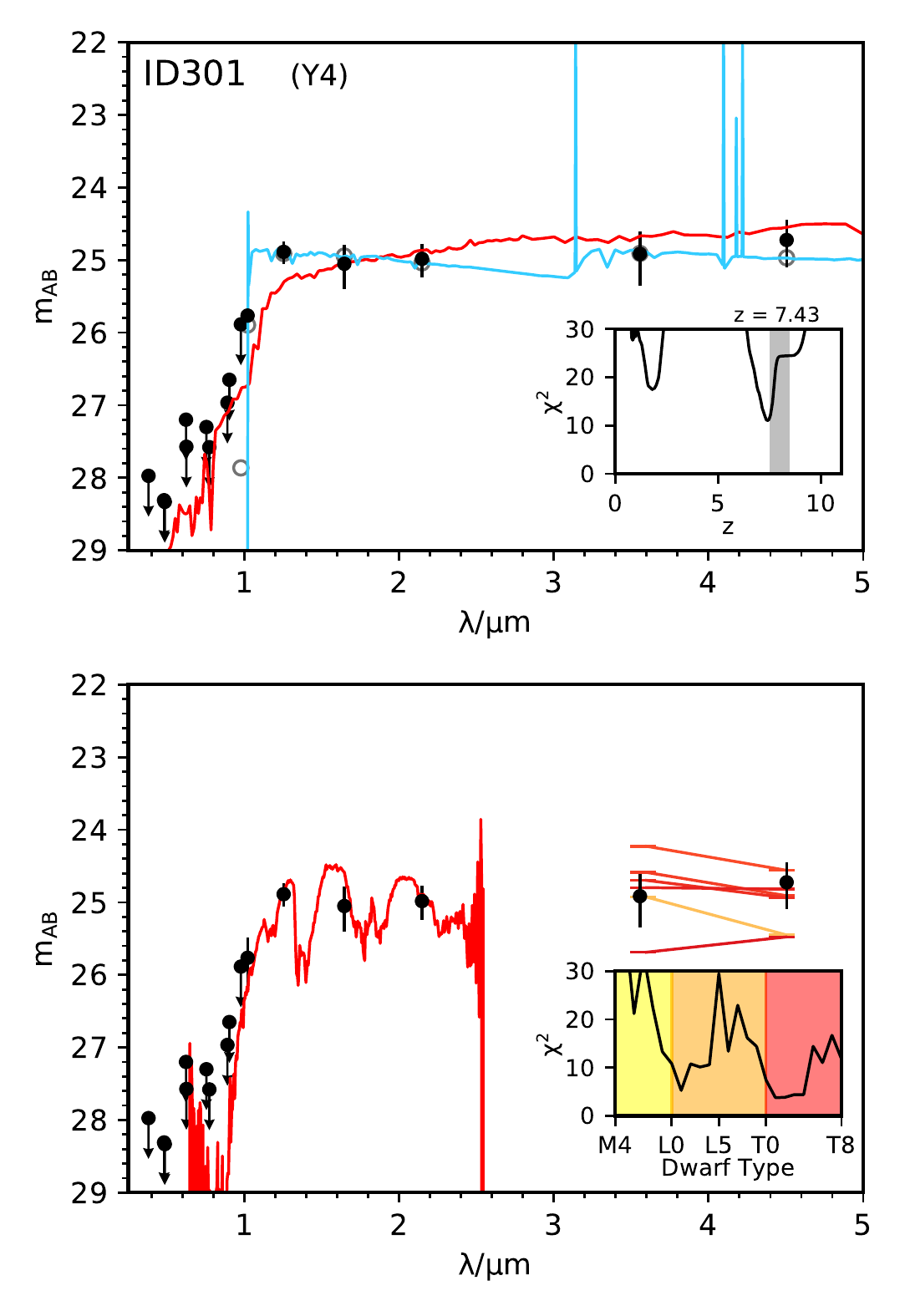}\\
\includegraphics[width = 0.3\textwidth]{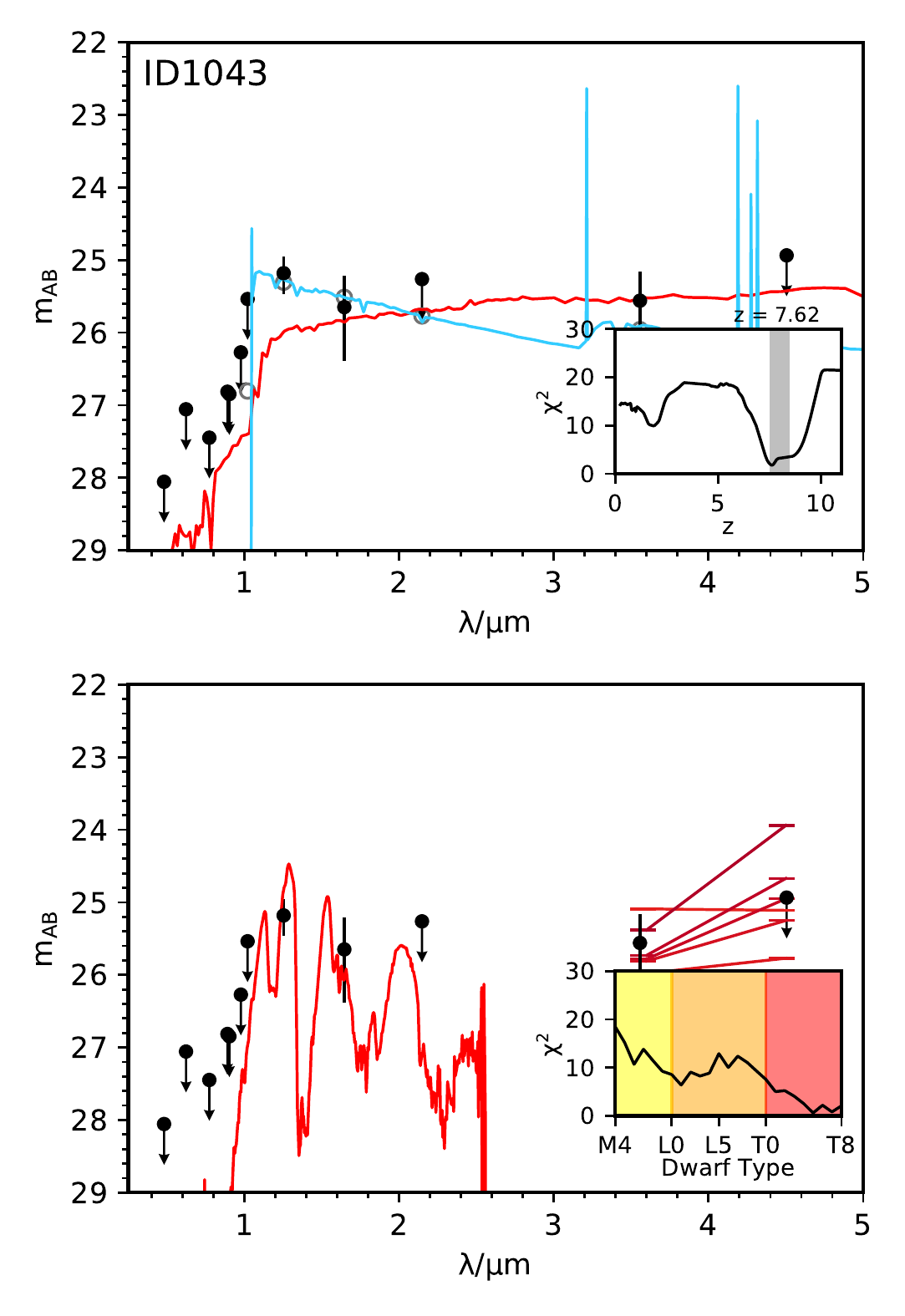}
\includegraphics[width = 0.3\textwidth]{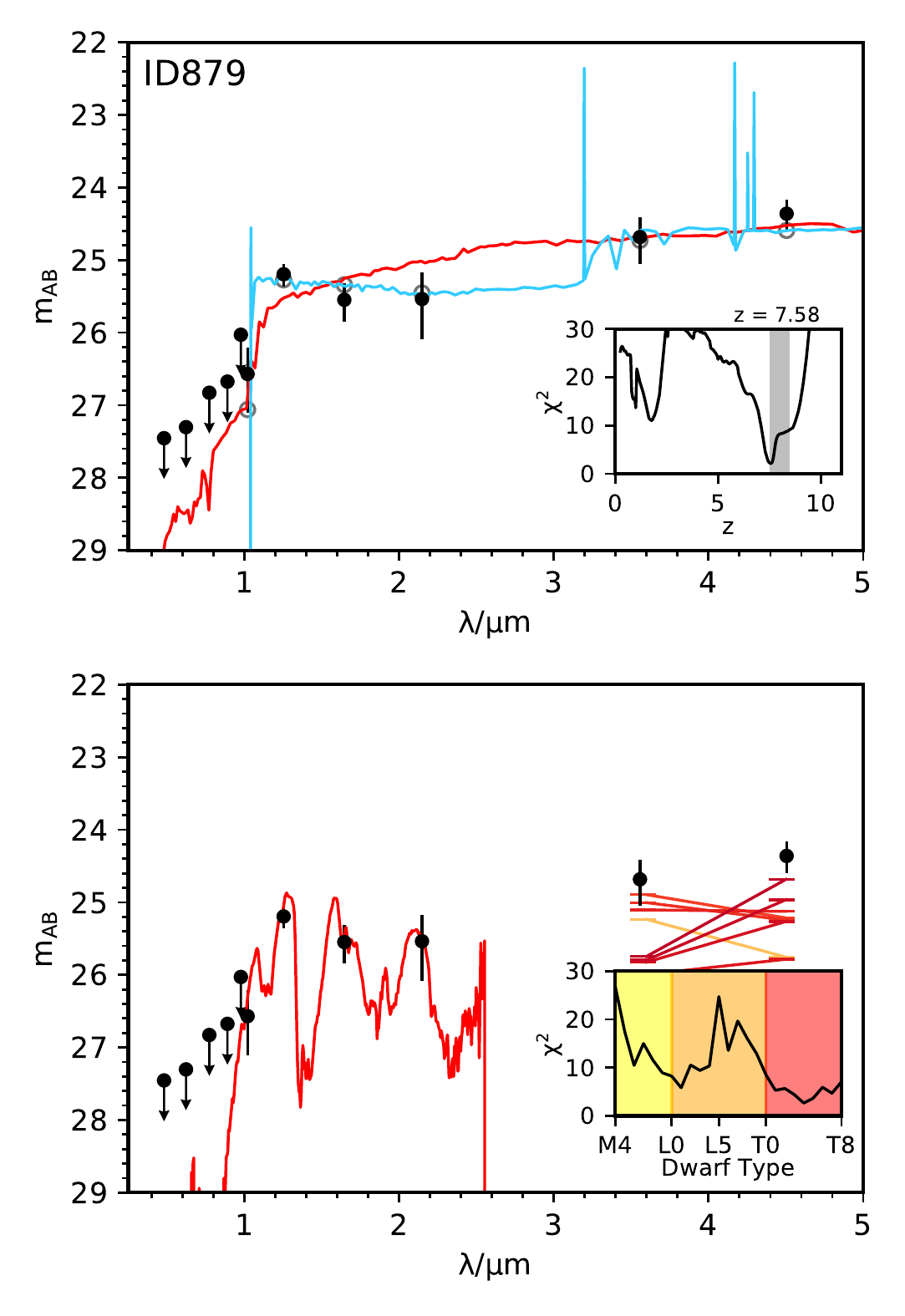}
\includegraphics[width = 0.3\textwidth]{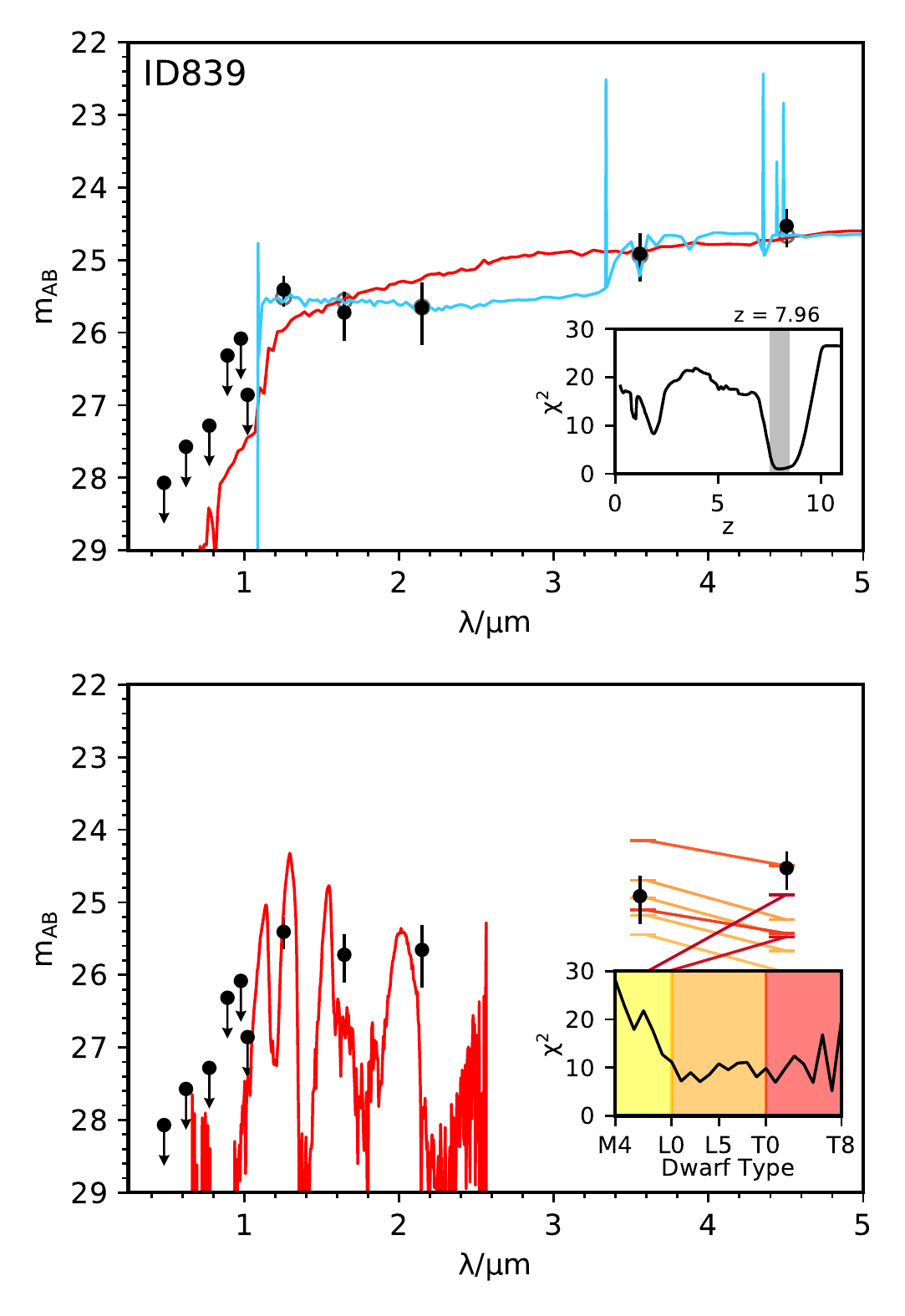}\\

\caption{The observed photometry and best-fitting SED templates for the $z \simeq 8$ candidates in the COSMOS field.
The plot format is described in the caption to Fig.~\ref{fig:xmmz8sed}.
}
\label{fig:cosz8sed}

\end{figure*}

\begin{figure*}
\includegraphics[width = 0.3\textwidth]{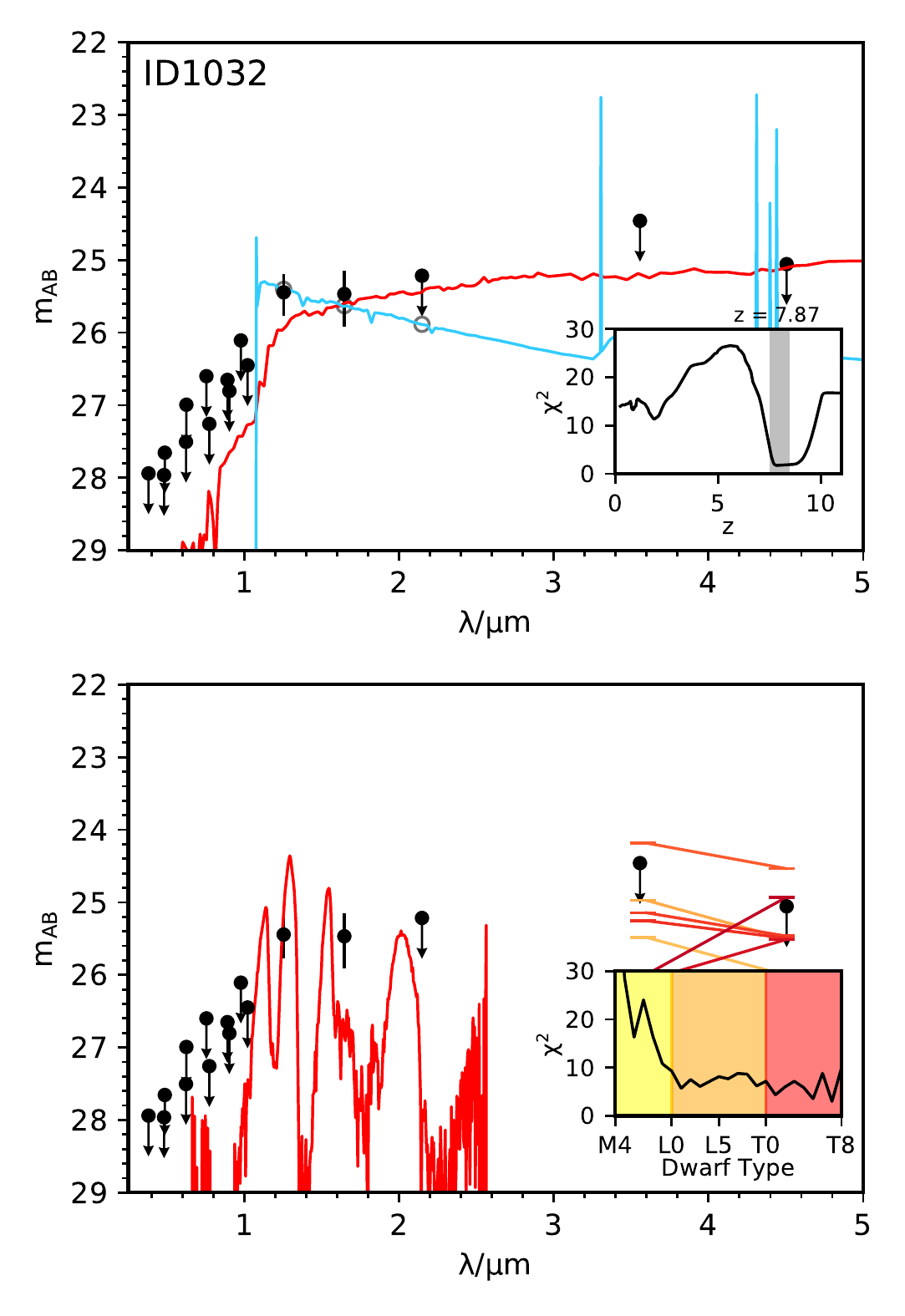}
\includegraphics[width = 0.3\textwidth]{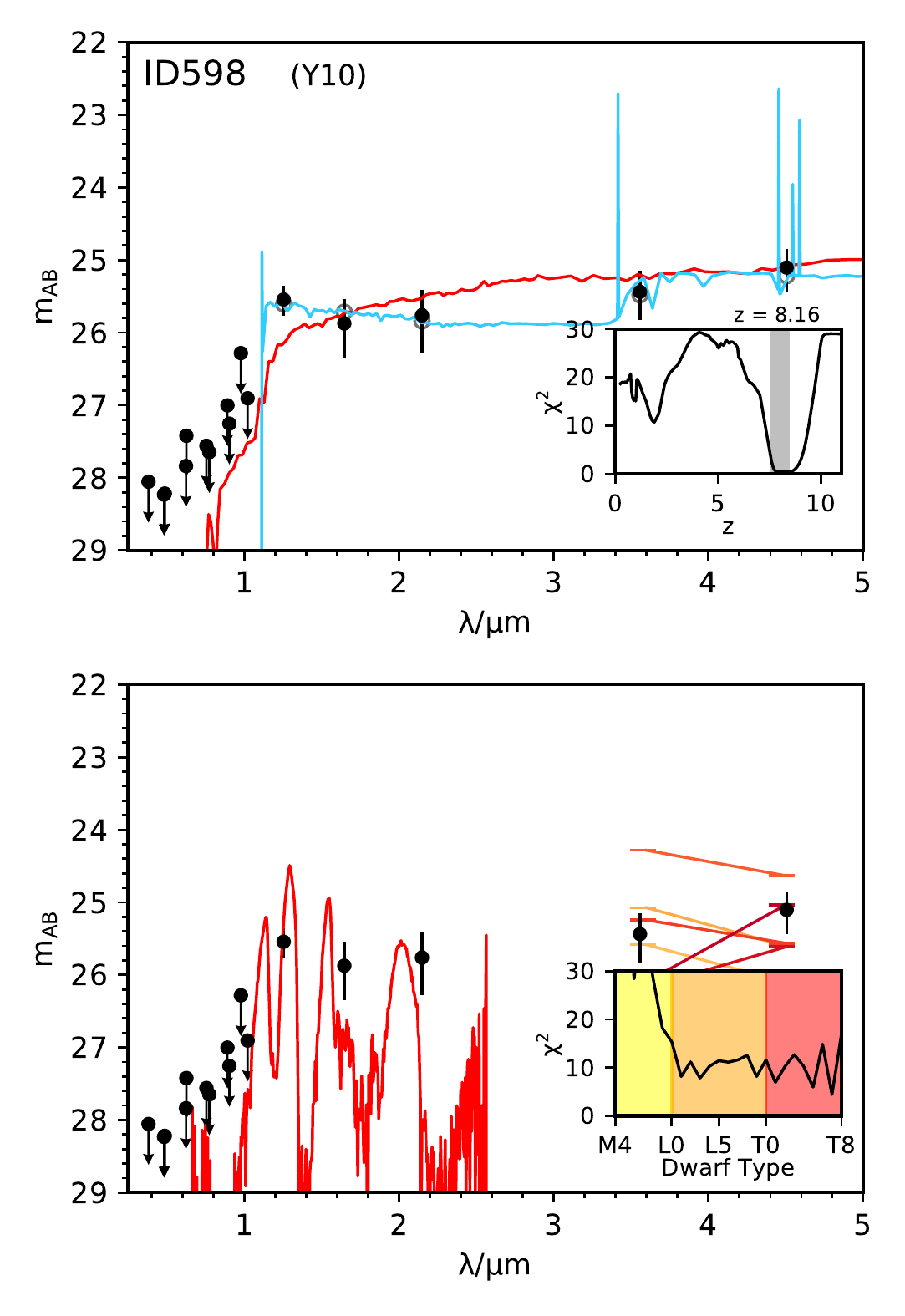}
\includegraphics[width = 0.3\textwidth]{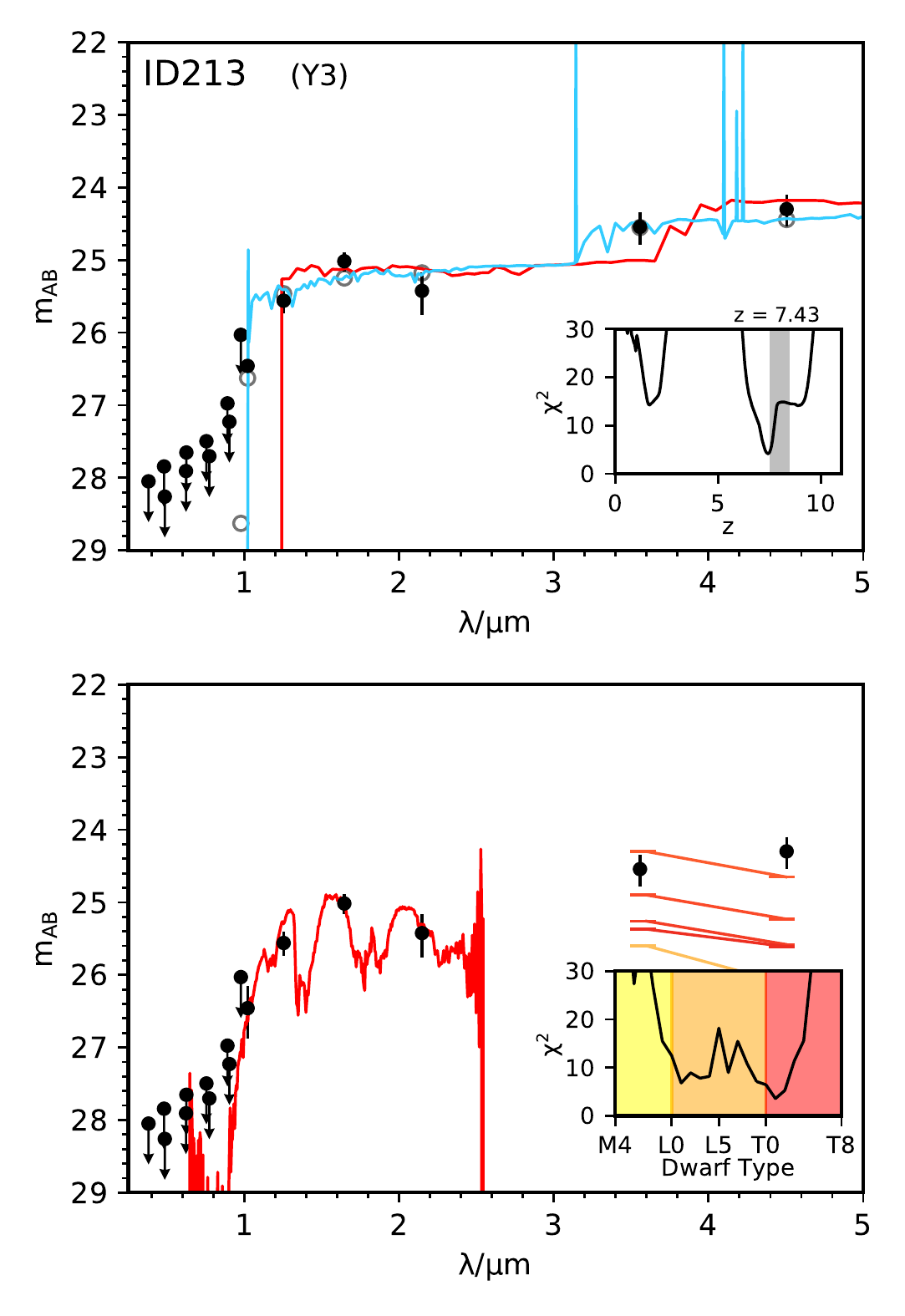}\\
\includegraphics[width = 0.3\textwidth]{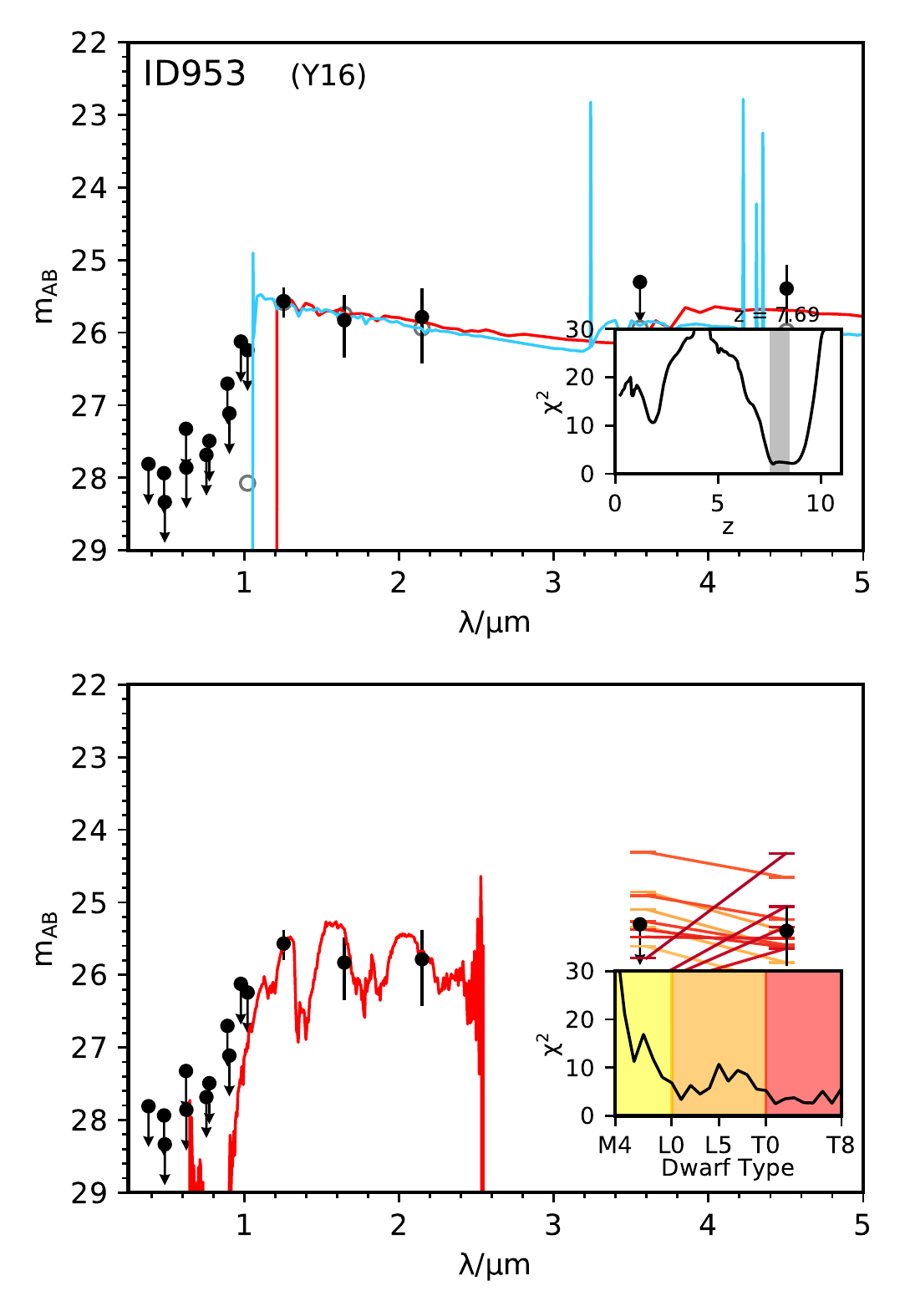}
\includegraphics[width = 0.3\textwidth]{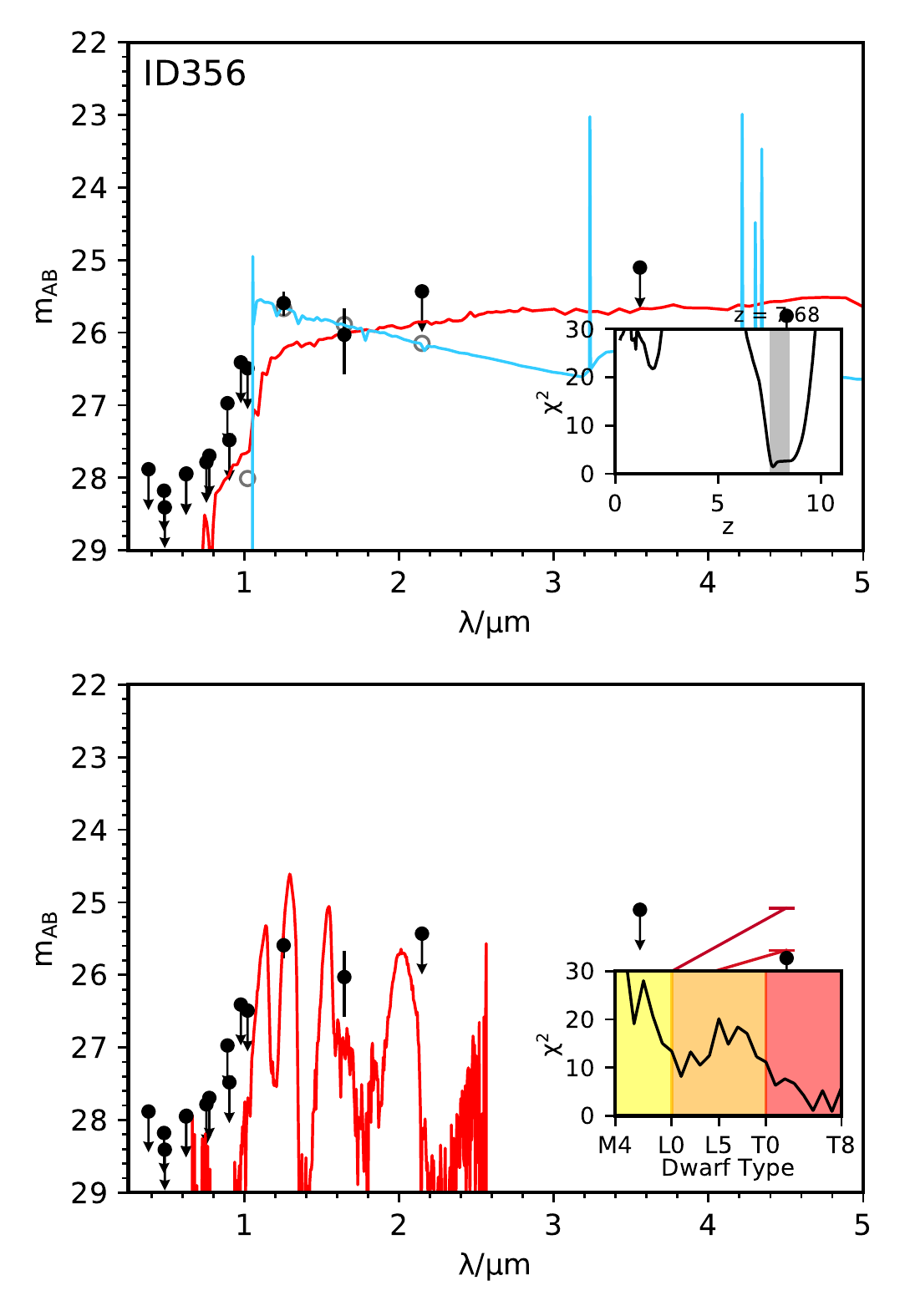}
\includegraphics[width = 0.3\textwidth]{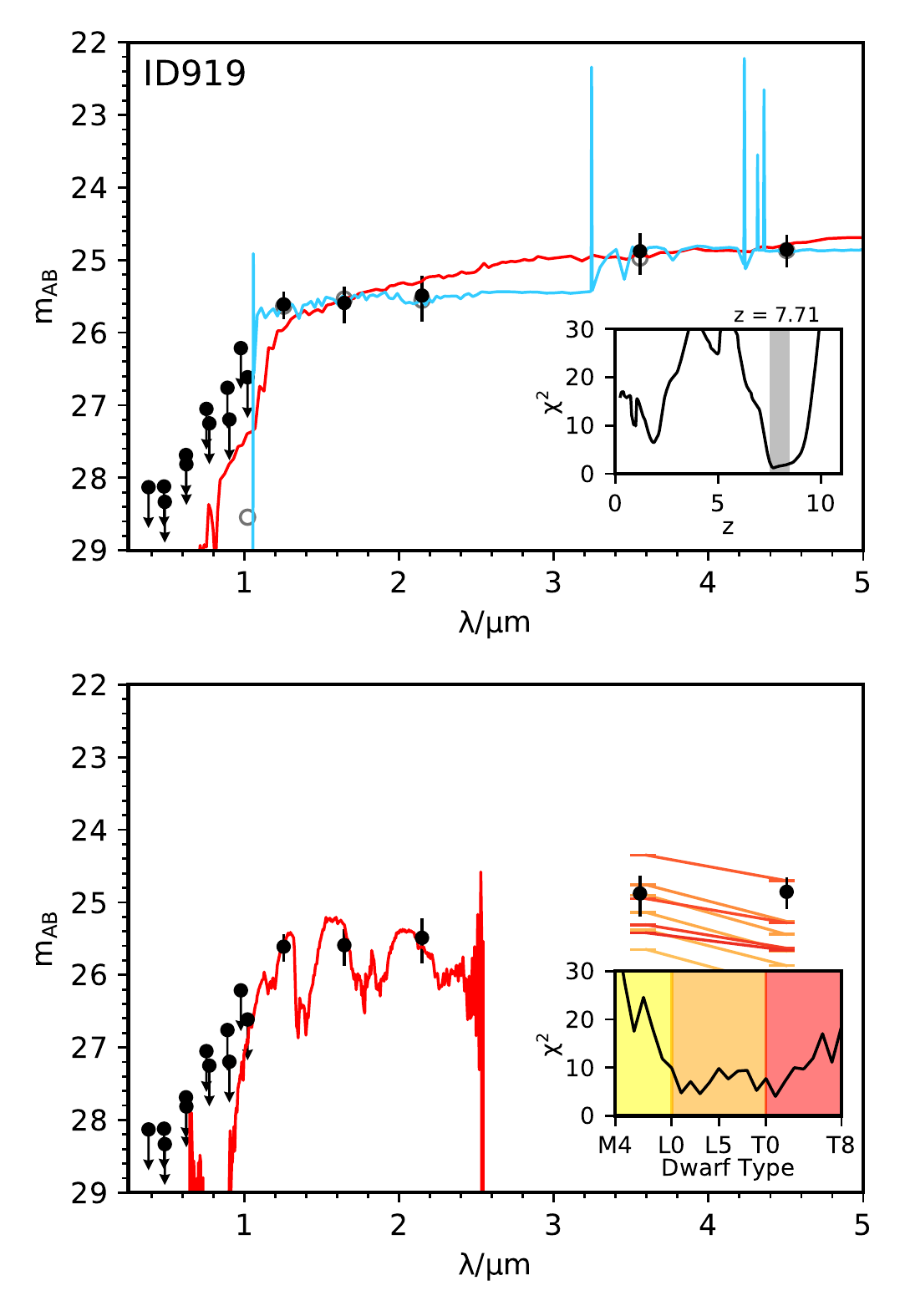}\\
\includegraphics[width = 0.3\textwidth]{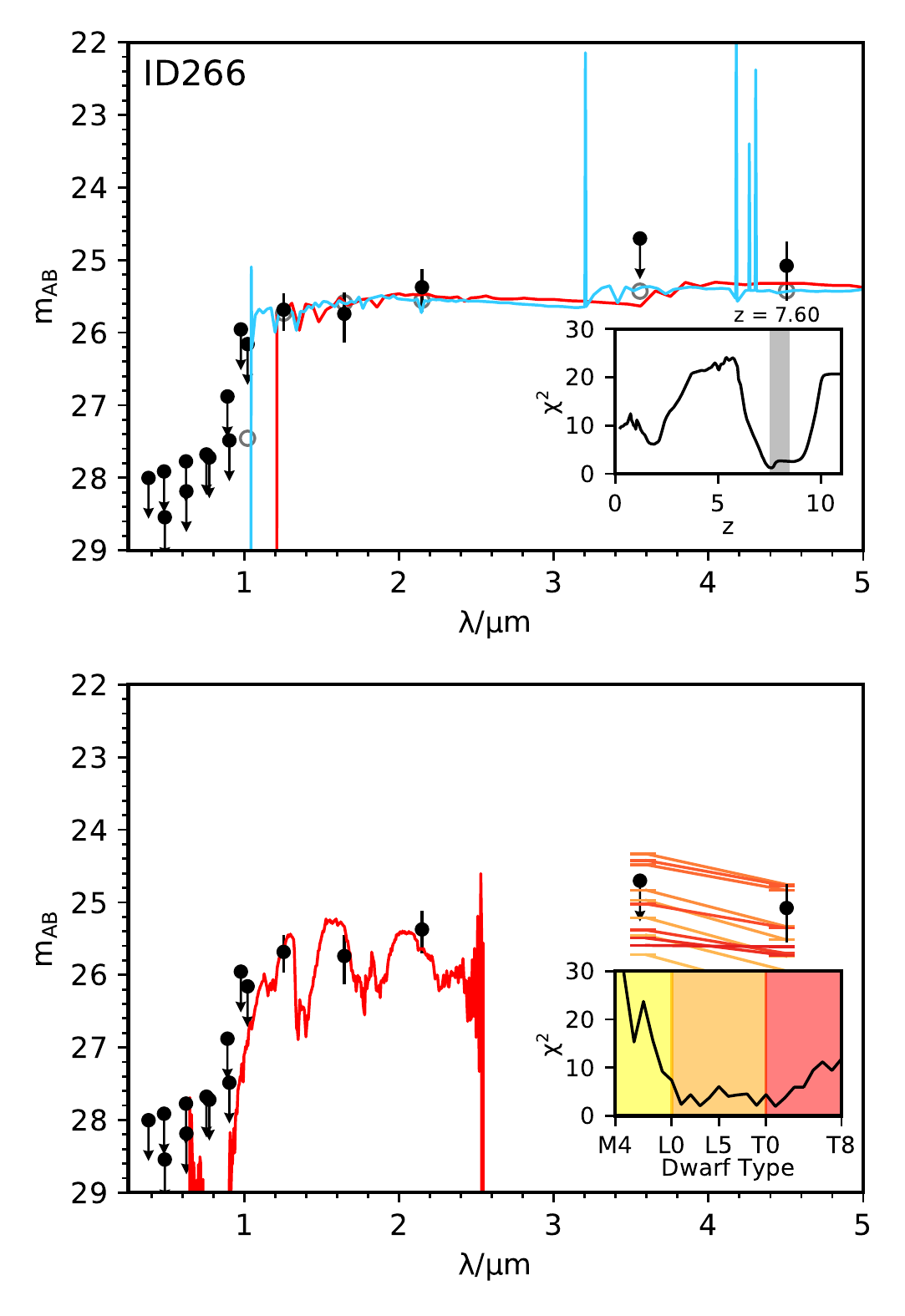}
\includegraphics[width = 0.3\textwidth]{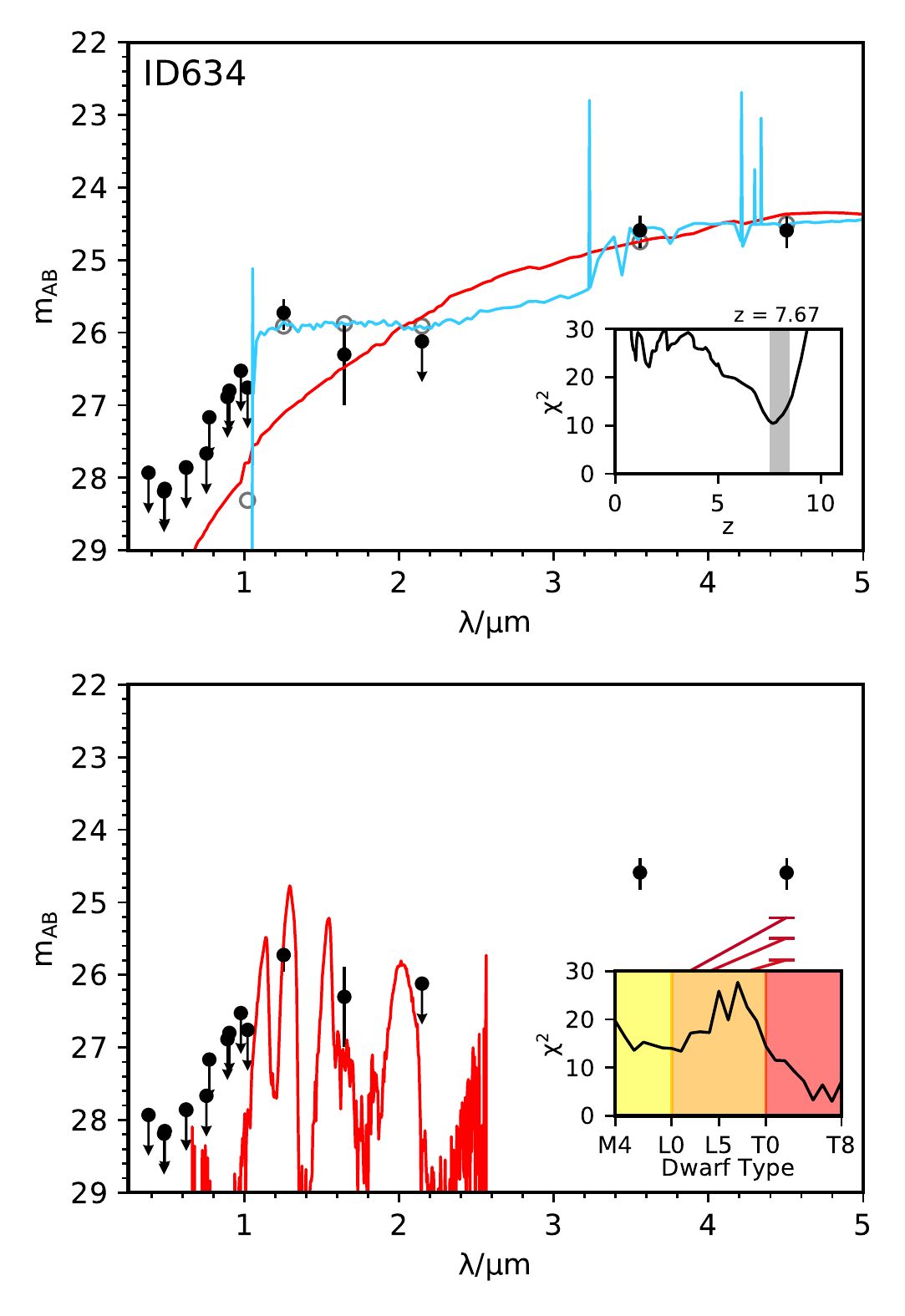}\\
\caption{Continued.}\label{fig:cosz8sedii}
\end{figure*}

\begin{figure*}
\includegraphics[width = 0.3\textwidth]{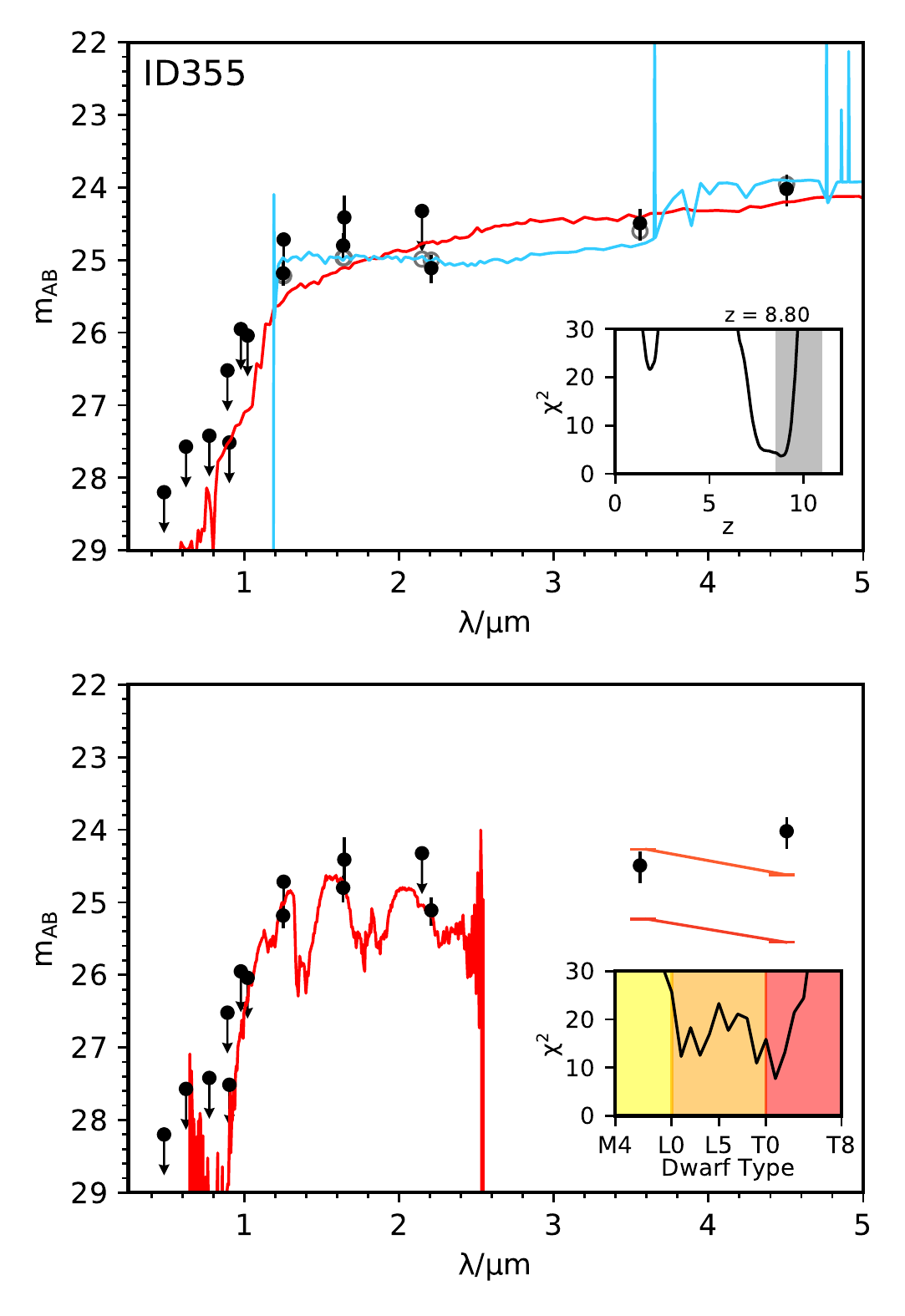}
\includegraphics[width = 0.3\textwidth]{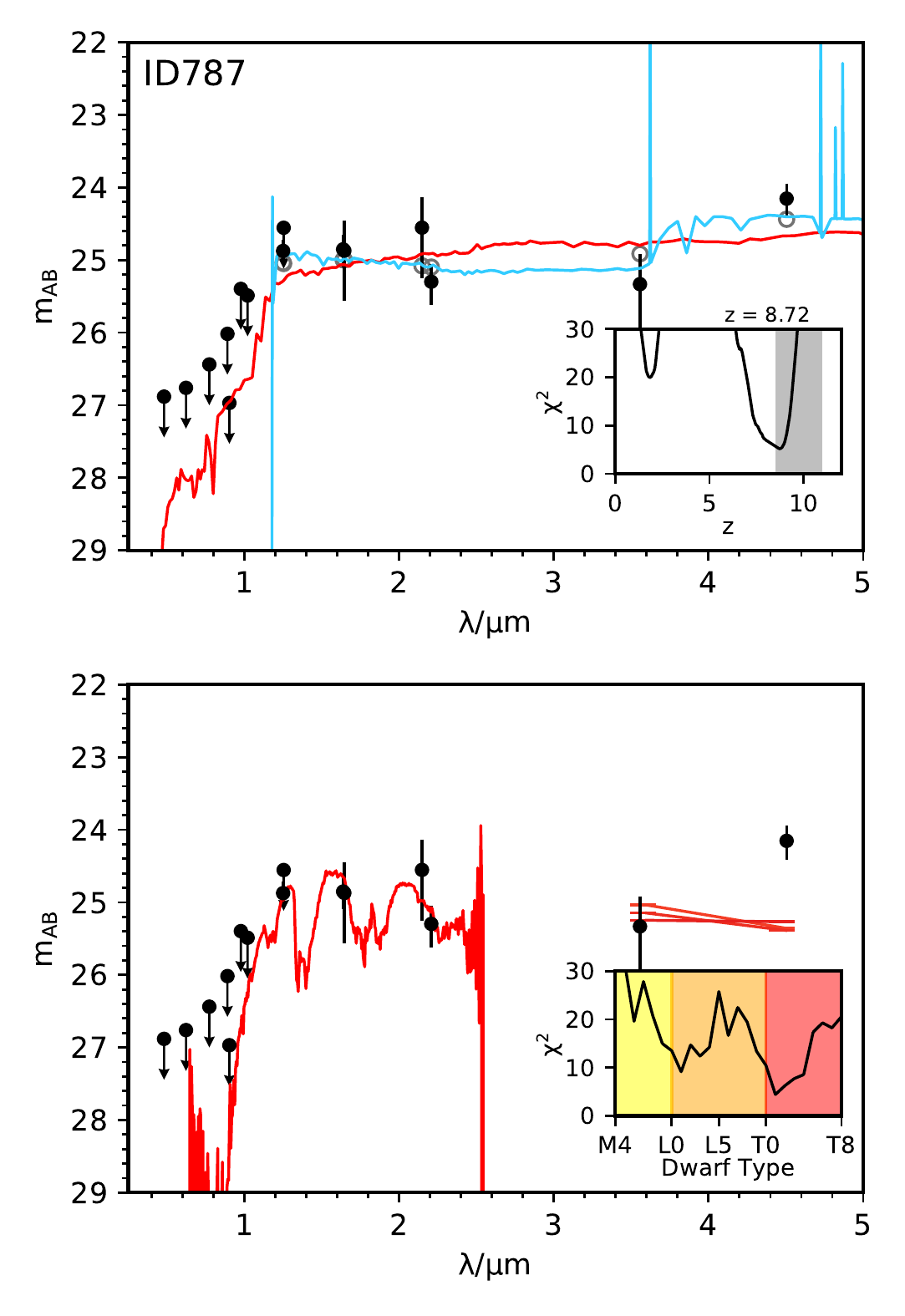}
\includegraphics[width = 0.3\textwidth]{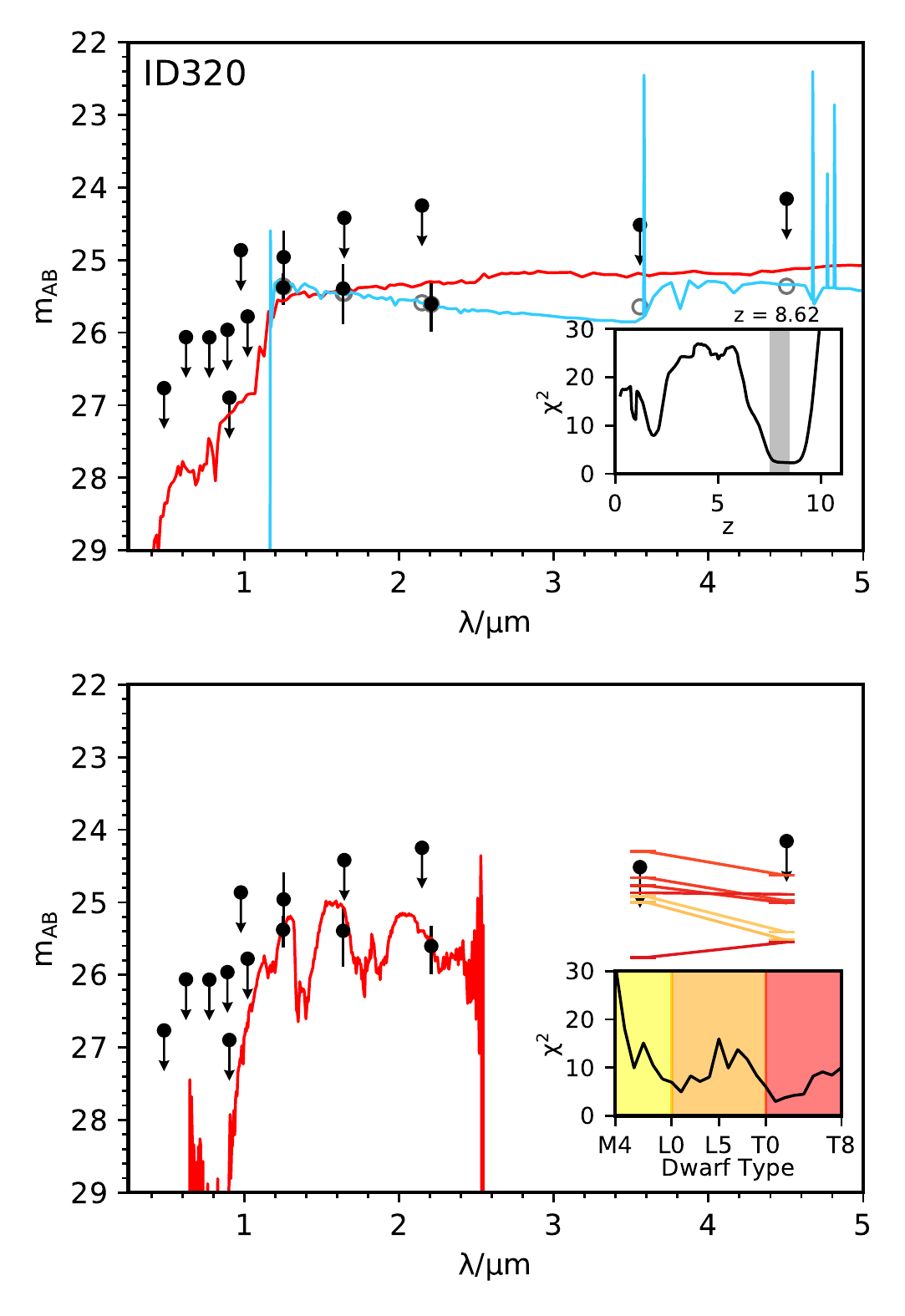}
\includegraphics[width = 0.3\textwidth]{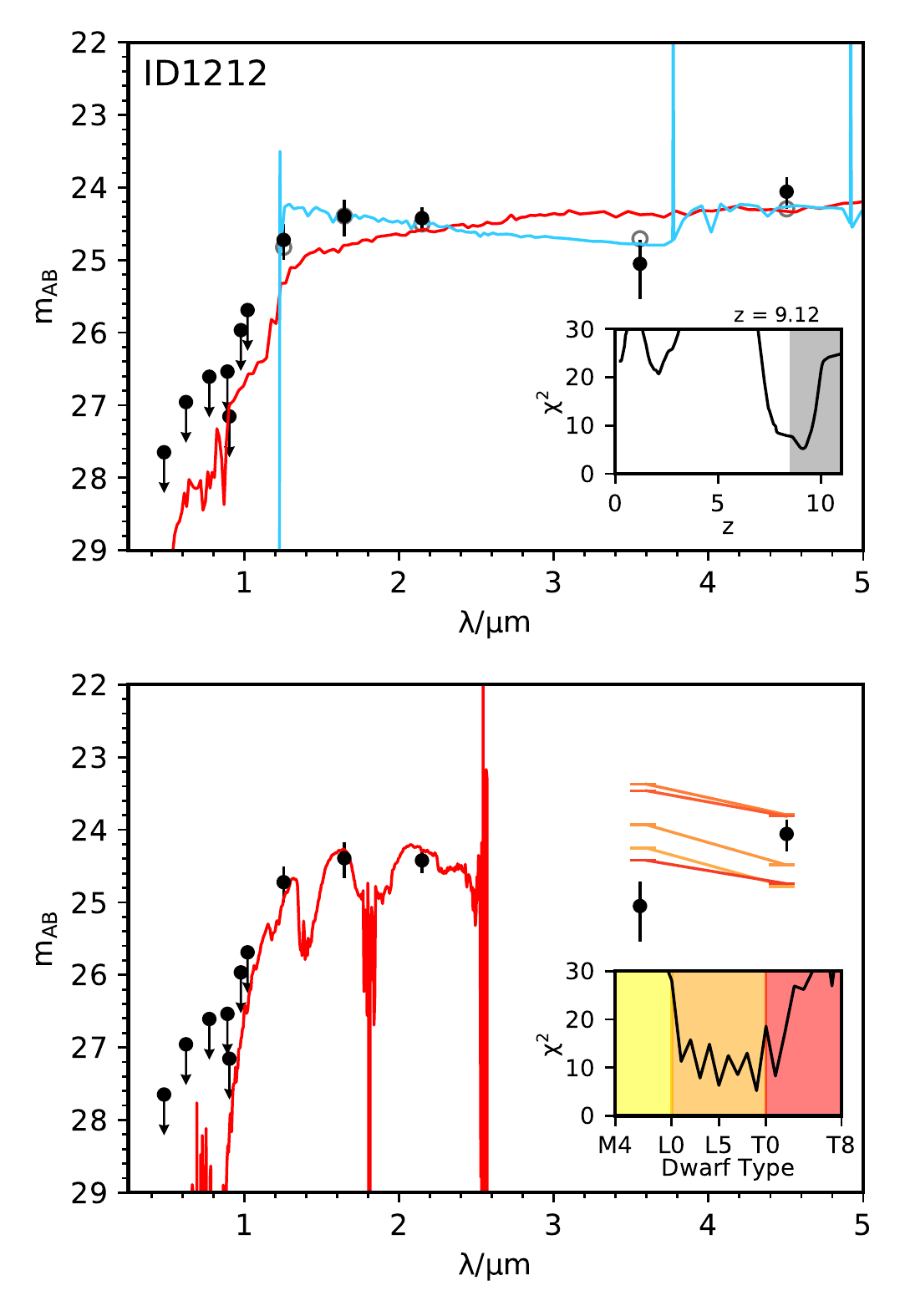}
\includegraphics[width = 0.3\textwidth]{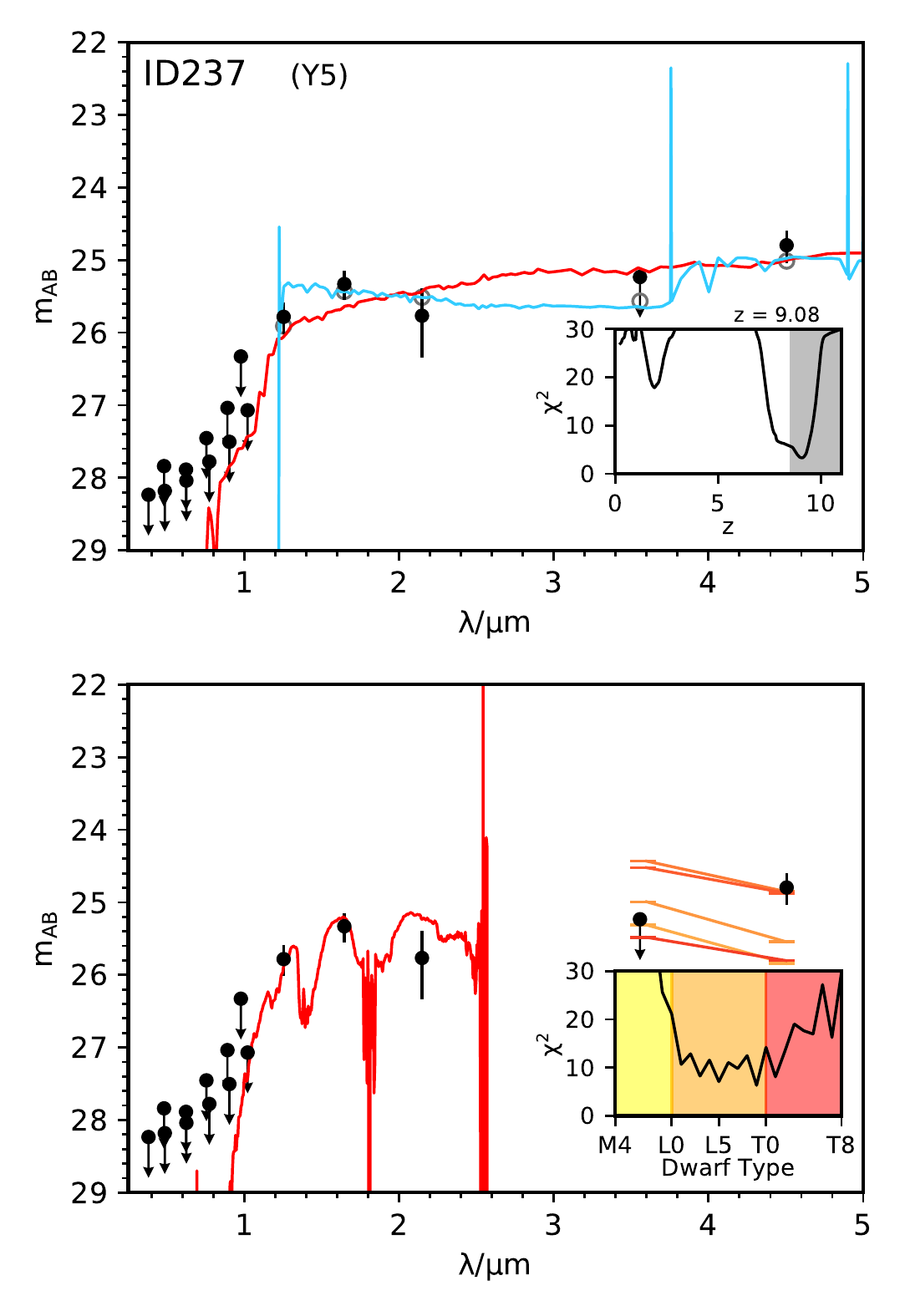}\\

\caption{The observed photometry and SED fitting results for the $z \simeq 9$ candidates.
The three XMM-LSS sources are shown on the top row, followed by the two COSMOS sources in the bottom row.
The plot format is described in the caption to Fig.~\ref{fig:xmmz8sed}.}
\label{fig:z9sed}
\end{figure*}


\bsp	
\label{lastpage}
\end{document}